\documentclass[11pt]{article}
\usepackage{jheppub}

\usepackage{epsfig}
\usepackage{amssymb}
\usepackage{amsfonts}
\usepackage{amsbsy}
\usepackage{amsmath}
\usepackage{dsfont}
\usepackage{bbm}
\usepackage{upgreek}
\usepackage{amssymb,amscd}
\usepackage{graphicx}
\usepackage{mathrsfs}
\usepackage{amsmath,amsthm}
\usepackage{slashed} 
\usepackage{dsfont}
\usepackage{tikz}
\usepackage[utf8]{inputenc}

\makeatletter
\DeclareFontFamily{OMX}{MnSymbolE}{}
\DeclareSymbolFont{MnLargeSymbols}{OMX}{MnSymbolE}{m}{n}
\SetSymbolFont{MnLargeSymbols}{bold}{OMX}{MnSymbolE}{b}{n}
\DeclareFontShape{OMX}{MnSymbolE}{m}{n}{
    <-6>  MnSymbolE5
   <6-7>  MnSymbolE6
   <7-8>  MnSymbolE7
   <8-9>  MnSymbolE8
   <9-10> MnSymbolE9
  <10-12> MnSymbolE10
  <12->   MnSymbolE12
}{}
\DeclareFontShape{OMX}{MnSymbolE}{b}{n}{
    <-6>  MnSymbolE-Bold5
   <6-7>  MnSymbolE-Bold6
   <7-8>  MnSymbolE-Bold7
   <8-9>  MnSymbolE-Bold8
   <9-10> MnSymbolE-Bold9
  <10-12> MnSymbolE-Bold10
  <12->   MnSymbolE-Bold12
}{}

\let\llangle\@undefined
\let\rrangle\@undefined
\DeclareMathDelimiter{\llangle}{\mathopen}%
                     {MnLargeSymbols}{'164}{MnLargeSymbols}{'164}
\DeclareMathDelimiter{\rrangle}{\mathclose}%
                     {MnLargeSymbols}{'171}{MnLargeSymbols}{'171}
\makeatother

\def\be{ \begin{equation} }
\def\ee{ \end{equation}}

\newcommand{\eq}[1]{\begin{align}\begin{split}#1\end{split}\end{align}}

\def\tildeB{{\widetilde{B}}}

\def\exp{{\rm exp}}

\def\Ind{{\rm Ind}}

\def\half{\frac{1}{2}}

\def\itwopi{\frac{i}{2\pi}}

\def\one{{\hbox{ 1\kern-.8mm l}}}
\def\CA{{\cal A}}

\def\CC {{\cal C}}

\def\CH {{\cal H}}
\def\CI {{\cal I}}

\def\CL {{\cal L}}

\def\CN {{\cal N}}

\def\CZ {{\cal Z}}

\def\CH {{\cal H}}
\def\CI {{{\cal I}}}

\def\CT {{\cal T}}

\def\CZ{{\cal Z}}

\def\IC{\mathbb{C}}

\def\IP{\mathbb{P}}
\def\ICP{\mathbb{CP}}

\def\IR{{\mathbb{R}}}

\def\IZ{{\mathbb{Z}}}

\def\fg{\mathfrak{g}}

\def\fn{\mathfrak{n}}
\def\fo{\mathfrak{o}}

\def\fs{\mathfrak{s}}

\def\fs{\mathfrak{s}}

\def\fI{\mathfrak{I}}

\def\rmk#1{\bigskip\noindent{\bf Remark} }
\def\cnj#1{\bigskip\noindent{\bf Conjecture:} }

\DeclareMathAlphabet{\mathpzc}{OT1}{pzc}{m}{it}

\def\Sym{ \textrm{Sym} }
\def\Tr{ \, \textrm{Tr}  }

\title{Anomalies of 4d $Spin_G$ Theories}

\author[a]{T.~Daniel Brennan}
\author[a]{Kenneth Intriligator}

\affiliation[a]{Department of Physics, University of California, San Diego}

\emailAdd{tbrennan@ucsd.edu, keni@ucsd.edu}

\abstract{We consider 't Hooft anomalies of four-dimensional gauge theories whose fermion matter content admits $Spin_G(4)$ generalized spin structure, with $G$ either gauged or a global symmetry.   We discuss methods to directly compute $w_2\cup w_3$ 't Hooft anomalies involving Stiefel-Whitney classes of gauge and flavor symmetry bundles that such theories can have on non-spin manifolds, e.g. $M_4=\mathbb{CP}^2$.    
Such anomalies have been discussed for $SU(2)$ gauge theory with adjoint fermions, where they were shown to give an effect that was originally found in the Donaldson-Witten topological twist of ${\cal N}=2$ SYM theory.  We directly compute these anomalies for a variety of theories, including general $G$ gauge theories with adjoint fermions, $SU(2)$ gauge theory with fermions in general representations, and $Spin(N)$ gauge theories with fundamental matter.    We discuss aspects of matching these and other  't Hooft anomalies in the IR phase where global symmetries are spontaneously broken, in particular for general $G_{\rm gauge}$ theory with $N_f$ adjoint Weyl fermions. For example, in the case of $N_f=2$ we discuss anomaly matching in the IR phase consisting of $h^\vee _{G_{\rm gauge}}$ copies of a $\mathbb{CP} ^1$ non-linear sigma model,  including for the $w_2w_3$ anomalies when formulated with  $Spin_{SU(2)_{\rm global}}(4)$ structure.

}

\begin{document}

\maketitle

\section{Introduction}

Global symmetries, including approximate and spontaneously broken symmetries, yield powerful tools, constraints, and insights for progress in physics.  See e.g. ~\cite{McGreevy:2022oyu,Cordova:2022ruw,Schafer-Nameki:2023jdn,Brennan:2023mmt,Bhardwaj:2023kri,Shao:2023gho} for reviews and references on the broadened notions and applications of generalized symmetries in quantum field theory.   In quantum theories, symmetries can be  projectively realized on the Hilbert space, as happens with anomalous symmetries.  't Hooft anomalies for global symmetries are especially useful: because they are invariant under continuous, symmetry-preserving deformations, including renormalization group (RG) flow, they are often exactly computable and usefully constrain the dynamics and the IR theory.   E.g. non-zero 't Hooft anomalies for continuous symmetries cannot be matched by a symmetry-preserving, gapped phase -- and that is also the case for some discrete symmetries (whereas other discrete symmetry 't Hooft anomalies can be matched by a symmetry-preserving, gapped TQFT)~\cite{Garcia-Etxebarria:2017crf, Cordova:2019bsd, Cordova:2019jqi, Brennan:2023kpo,Brennan:2023ynm}.
Anomalies of  $d$-dimensional QFTs on spacetime $M_d$ are usefully described via inflow from bulk $(d+1)$-dimensional topological theories on spacetime $N_{d+1}$ with boundary $M_d=\partial N_{d+1}$, see e.g.~\cite{Witten:1983tw, Alvarez-Gaume:1983ihn, Alvarez-Gaume:1984zlq, Callan:1984sa, Freed:2016rqq}.  For anomalies involving fermions, this relates the anomaly to the $\eta$-invariant of the $(d+1)$-dimensonal Dirac operator via the APS index theorem~\cite{Witten:1985xe,Dai:1994kq,Witten:2019bou} which, if the perturbative anomalies vanish, is classified by cobordism groups $\Omega _{d+1}$~\cite{Kapustin:2014tfa,Freed:2014eja,Freed:2016rqq}.  We will only discuss $d=4$ dimensional QFTs in this paper. 

Theories that are bosonic, or more generally admit a generalized spin structure, can be placed on non-spin manifolds, e.g. $M_4=\ICP ^2$, which can lead to additional 't Hooft anomalies.  In this paper we will especially focus on such `t Hooft anomalies of 4d QFTs, which involve Stiefel-Whitney classes of the gauge bundle, the background fields for global symmetries, and/or that of the spacetime geometry.  We focus on ``$w_2 w_3$" 't Hooft anomalies, i.e. where the associated 5d anomaly theory terms are $\sim \int w_2\cup w_3$.  Such anomalies have been discussed in e.g.~\cite{Brennan:2023kpo,Kapustin:2014tfa,Thorngren:2014pza, Cordova:2018acb, Wang:2018qoy,Wan:2018djl} and we further explore and illustrate them here.  

An example of such a $w_2w_3$ anomaly is the 4d purely gravitational 't Hooft anomaly, which is $\IZ _2$-valued and given by the 5d anomaly theory term~\cite{Kapustin:2014tfa,Thorngren:2014pza} 
\eq{\label{gravanomaly}
\CA\supset  i \pi  \kappa _{T,T} \int _{N_5} w_2(TN)\cup w_3(TN)~ \qquad \qquad \kappa _{T,T}\cong \kappa _{T,T}+2}
where the boundary of $N_5$ is the spacetime 4-manifold $M_4=\partial N_5$, and $TN$ refers to the tangent manifold of $N$; in what follows we will often denote $w_3(TN)$ by $w_3(TM)$ as the anomalous phase is only dependent on the choice of $M_4$.  The anomaly coefficient $\kappa _{T,T}$ is the $\IZ _2$ valued, theory-dependent purely gravitational 't Hooft anomaly; the gravitational anomaly is non-trivial if the theory has  $\kappa _{T,T}=1$ mod 2.  The Stiefel-Whitney classes $w_2(TN)$ and $w_3(TN)$  are obstructions to the existence of a spin- and spin$_c$-structure respectively on $N_5$.  All 4-manifolds $M_4$ have $w_3(M_4)=0$ and the anomaly ~\eqref{gravanomaly} trivializes if there is a spin structure on $M_4$, since then $w_2(TN)=w_2(TM)=0$.   The anomaly~\eqref{gravanomaly} is a cobordism invariant and is classified by the cobordism group for bosonic theories as $\Omega ^{SO}_5\supset  \IZ_2$, where the $SO$ refers to the global structure of the Lorentz group. On the other hand, in fermionic theories with spin structure the vanishing of the anomaly is reflected by the fact that the relevant cobordism group vanishes:
 $\Omega ^{Spin}_5=0$.  Two choices for 5-manifolds $N_5$ that generate the nontrivial  $\Omega _5^{SO}=\IZ_2$, with $\int _{N_5}w_2(TN)\cup w_3(TN)\neq 0$ mod 2, are called the Wu and Dold manifolds\footnote{As (foot)noted in~\cite{Lee:2020ojw}, $SU(3)/SO(3)$ is called the Wu manifold, but it was Calabi who found it as a generator of $\Omega _5^{SO}=\IZ _2$.  Wu had $\ICP ^2\rtimes _{\widehat \varphi}S^1$, which is now called the Dold manifold. The space $SU(3)/SO(3)$ also arises as the pion target space in $SU(3)\to SO(3)$ spontaneous symmetry breaking.} $N_5=SU(3)/SO(3)$ and $N_5=\ICP ^2\rtimes _{\varphi}S^1$; the latter arises in the mapping torus construction of the anomaly in~\cite{Wang:2018qoy} and in what follows here.     The 't Hooft anomaly~\eqref{gravanomaly} is non-zero e.g. in all fermion electrodynamics~\cite{Thorngren:2014pza, Kravec:2014aza} on $\ICP ^2$.

More generally, theories that (classically) admit a generalized $Spin_G(4)$ structure can be formulated on non-spin manifolds such as $M_4=\ICP ^2$, with non-zero $w_2(TM)$.\footnote{We can and will take $\ICP ^2$ as the representative 4-manifold with $w_2(TM)\neq 0$, to activate the new anomalies.}   A generalized $Spin _G(4)$ structure is possible only if all dynamical matter fields satisfy a generalized spin charge relation, with $(-1)^F$ acting the same as a $\IZ _2$ central element of the gauge or global symmetry $G$, i.e. with all bosons having even charge under the $\IZ _2$ and all fermions having odd charge.     The $Spin_{G}(4)$ structure requires the gauge fields to satisfy a flux constraint correlated with the geometry, $w_2(G)=w_2(TM)$.   For the case of gauge $G=G_{\rm gauge}$, this is a constraint on the gauge field functional integral and $(-1)^F$ is effectively gauged so there are no fermionic gauge invariant operators in the spectrum; the theory is actually bosonic.  For the case of $G=G_{\rm global}$,  $w_2(G)=w_2(TM)$ is a constraint on the background gauge fields and the interpretation is somewhat different.  The theory could have gauge invariant fermionic operators but, as with {\it twisting} of supersymmetric theories, the stress-tensor can be consistently modified by an improvement term involving the $G_{\rm global}$ currents so that activating the $w_2(G)=w_2(TM)$ background gauge field flux effectively twists the spins and converts the fermionic operators to be bosonic.
  
The $w_2(G)=w_2(TM)$ condition of generalized $Spin_G(4)$ structures relates the gravitational anomaly~\eqref{gravanomaly} to a mixed $w_2w_3$ type 't Hooft anomaly between $G$ and gravity
\eq{\label{anomalyGT}
\CA \supset  i\pi \kappa _{G,T} \int w_2(G)\cup w_3(TN)~.}
Irrespective of whether $G$ is gauged or global, the interpretation of~\eqref{anomalyGT} is that of an 't Hooft anomaly in the generalized $Spin_G(4)$ structure rather than an  inconsistency of the theory.  It does not spoil the consistency of the theory with ordinary spin structure on spacetimes with $w_2(TM)=0$, and even with the generalized $Spin_G(4)$ structure on a spacetime with $w_2(TM)\neq 0$ one can always choose local counterterms to represent a mixed anomaly as preserving the gauged symmetry, with the global symmetry projectively realized.

The new $SU(2)$ anomaly of~\cite{Wang:2018qoy} (hereafter ``WWW") is an example of a $w_2w_3$ 't Hooft anomaly.  It arises in $SU(2)$ gauge theory with $N_f=1$ Weyl fermion in the $j={3\over 2}$ representation.  The $\IZ _2$ center of the gauge group acts on all fields as $(-1)^F$, so the theory is {\it bosonic}:  all gauge invariant operators are bosons. The theory thus classically admits a $Spin_{SU(2)_{\rm gauge}}(4)$ structure.  As shown in~\cite{Wang:2018qoy}, the quantum theory with generalized $Spin_{SU(2)}(4)$ structure has the gravitational anomaly~\eqref{gravanomaly}, and consequently also the 't Hooft anomaly~\eqref{anomalyGT}.  

For general 4d $SU(2)$ gauge theory, the cobordism classification gives $\Omega _5^{Spin_{SU(2)}}=\IZ _2\oplus \IZ _2$~\cite{Freed:2016rqq}, corresponding to the original $SU(2)$ Witten anomaly~\cite{Witten:1982fp} and the WWW new $SU(2)$ anomaly~\cite{Wang:2018qoy}.   The original Witten anomaly  is a gauge anomaly (or 't Hooft anomaly if $SU(2)$ is a global rather than gauge symmetry) arises if there are an odd number of Weyl fermions in the $D=(2j+1)$-dimensional representation of $SU(2)$ for $D\in 2+4\IZ$; it can be seen from the fact that an $SU(2)$ instanton then has an odd number of fermion zero modes so its corresponding 't Hooft operator violates $SU(2)$ and Lorentz symmetry.  

The WWW~\cite{Wang:2018qoy} anomaly is a 't Hooft anomaly in the  $Spin_{SU(2)}(4)$ structure, which is non-trivial if the number of fermion zero modes of the 4d Dirac operator in a $Spin_{SU(2)}(4)$ flux background is not a multiple of 4; this is the case if there are an odd number of Weyl fermions in representations of dimension $D\in 4+8\IZ$~\cite{Wang:2018qoy}.  In particular,  the $Spin_G$ 't Hooft anomaly is non-zero for an odd number of chiral fermions in the ${\bf 4}$ (i.e $j=3/2$) of $SU(2)$ when placed on $M_4=\ICP ^2$ which has $w_2(TM)\neq 0$.  The theory with ordinary spin structure on $M_4$ with spin structure $w_2(TM)=0$ is perfectly healthy (e.g. the instanton 't Hooft vertex has an even number of fermion zero modes) \footnote{This was also noted e.g. in~\cite{Intriligator:1994rx}, where the ${\cal N}=1$ version of this theory was considered.}.   Deforming the $SU(2)$ theory by adding an adjoint-valued scalar field (preserving the spin-charge relation) and Higgsing $SU(2)\to U(1)$ leads to all fermion electrodynamics with the same gravitational anomaly \eqref{gravanomaly} ~\cite{Wang:2018qoy,Brennan:2022tyl}.

The present paper is on a class of  $\mathbb{Z}_2$-valued $w_2w_3$-type 't Hooft anomalies that can arise in theories with $Spin_{G_{\rm global}}(4)$ structure on $M_4=\ICP ^2$ in the presence of a non-zero background gauge field $B_2$ for a $\mathbb{Z}_2$ subgroup~\footnote{As we will discuss, there is no such anomaly if e.g. $\Gamma _g^{(1)}=\IZ _{N_g}^{(1)}$ with odd $N_g$: it can then be cancelled. } of a one-form global symmetry $\Gamma _g^{(1)}$. Such 't Hooft anomalies were first discussed in~\cite{Cordova:2018acb}
in the context of $SU(2)_{\rm gauge}$ theory with $N_f=2$ Weyl fermions in the adjoint of $SU(2)_{\rm gauge}$, which has a generalized spin structure is associated with a $SU(2)_\text{global}$ (rather than $SU(2)_\text{gauge}$ symmetry).  The  't Hooft anomaly is activated when a non-zero background gauge field $B_2$ is turned on for the $\Gamma _g^{(1)}=\IZ _2^{(1)}$ one-form global symmetry (the center symmetry of $SU(2)_{\rm gauge}$) when the $Spin_{SU(2)_{\rm global}}$ structure is used to put the theory on a spacetime with $w_2(TM)\neq 0$, e.g. $M_4=\ICP ^2$~\cite{Cordova:2018acb}.  Then WWW~\cite{Wang:2018qoy} discussed a relation to their anomaly.  Such anomalies will be generalized and applied here.

Generalizing~\cite{Cordova:2018acb}, the new  $\mathbb{Z}_2$-valued 't Hooft anomaly has an anomaly theory of the form 
\eq{\label{anomalyGammaT}\CA \supset  i\pi \kappa _{\Gamma _g, T} \int _{N_5} B_2 \cup w_3(TN)~, \qquad \qquad \kappa _{\Gamma _g, T}\cong \kappa _{\Gamma _g, T}+2~.}
The background $B_2$ activates a $\IZ _2$-valued subgroup of the $\Gamma _g^{(1)}$ one-form global symmetry associated with the center of $G_{\rm gauge}$, modifying the gauge fields to have $w_2(G_{\rm gauge})=B_2$, so~\eqref{anomalyGammaT} is indeed a type of $w_2\cup w_3$ anomaly.\footnote{There are also possible `t Hooft anomaly terms on non-spin manifolds of the form $\CA\supset \frac{\pi i  \kappa}{N}\int_{N_5}A_1\cup B_2\cup w_2(TN)$ where $A_1$ is the background gauge field for a discrete 0-form symmetry. We will not discuss such anomalies in this paper, see \cite{Anber:2020gig,Anber:2019nze} for examples and details. See also~\cite{Davighi:2019rcd,Davighi:2020bvi,Davighi:2023luh} on anomalies and~\cite{Wang:2021hob,Wang:2021vki} for analysis of $w_2w_3$ anomalies in BSM-inspired theories.}
The interpretation of the anomaly, as discussed in~\cite{Cordova:2018acb}, is in terms of the $\IZ$-valued integer lifts, $\tilde B_2$ and $\tilde w_3(TN)$, of the $\IZ _2$-valued quantities $B_2$ and $w_3(TN)$.  As in~\eqref{anomalyGammaT}, we will often suppress the tildes on the integer lifts to avoid clutter, when their presence should be understood from the context.  The integer lifts arise from defining differentials of cyclic-valued elements such as $B_2$ and $w_2(TN)$ via the Bockstein map; the cyclic-valued quantities are recovered via modding out by $\tilde B_2\to \tilde B_2+2x$ and $\tilde w_2(TN)\to \tilde w_2(TN)+2y$, with $x$ and $y$ integral-valued. 
The mixed anomaly~\eqref{anomalyGammaT} implies that the quantum partition function picks up a minus sign under either the $x$ or the $y$ shift above (i.e. either $Z\mapsto (-1)^{\int x\cup w_2(TM)}Z$, or $Z\mapsto (-1)^{\int B_2\cup y}Z$); the two options for how the anomaly shows up differ by a local counterterm in the backgrounds~\cite{Cordova:2018acb}.

 For the example of $SU(2)_{\rm gauge}$ with $N_f=2$ adjoint Weyl fermions, ref~\cite{Cordova:2018acb} argued that the theory must have $\kappa _{\Gamma _g, T}\neq 0$ via a  RG flow from  twisted ${\cal N}=2$ supersymmetric Yang-Mills on $\ICP^2$.  The 't Hooft anomaly~\eqref{anomalyGammaT} of that theory was presented in~\cite{Cordova:2018acb}  as a reinterpretation of an effect that arises in the context of Donaldon theory~\cite{Donaldson:1983wm,Donaldson:1990kn} and correspondingly in the twisted ${\cal N}=2$ supersymmetric gauge theory that leads to the Donaldson-Witten partition function, $Z_{DW}$~\cite{Witten:1994ev,Witten:1994cg,Moore:1997pc}.  To compute $Z_{DW}$, the $\IZ _2$ valued quantities $B_2$ and $w_2(TM)$ must be lifted to $\IZ$-valued integer quantities $\tildeB_2$ and $\tilde w_2(TM)$, and as above one recovers $B_2$ and $w_2(TM)$ by modding out $\tildeB_2$ and $\tilde w_2(TM)$ by $\IZ _2$ valued shifts,  $\tildeB_2\to \tildeB_2+2x$ or $\tilde w_2(TM) \to \tilde w_2(TM)+2y$ with $x$ and $y$ integral.   The fact that $Z_{DW}$ picks up a minus sign under one or the other shift, either $(-1)^{\int _{M_4} x\cup w_2(TM)}$ or $(-1)^{\int _{M_4} B_2\cup y}$  was found in ~\cite{Donaldson:1983wm,Donaldson:1990kn}  as coming from an orientation change of instanton moduli space, and it was shown to be reproduced in the twisted ${\cal N}=2$ $SU(2)$ Yang-Mills or the $U(1)$ (monopole) calculation of the partition function $Z_{DW}$, from the fermions in the functional integral in~\cite{Witten:1994ev,Witten:1994cg,Witten:1995gf,Moore:1997pc}.   
  
We will discuss and calculate $Spin_{G}(4)$ 't Hooft anomalies~\eqref{anomalyGammaT} for wide classes of theories.   We show how the anomaly coefficients $\kappa _{\Gamma _g, T}$ can often be simply directly computed by an adaptation of the method that was used in WWW~\cite{Wang:2018qoy} to compute the $\kappa _{T,T}$ 't Hooft anomaly of $SU(2)$ with a Weyl fermion in the ${\bf 4}$.  This method computes the anomaly from the index of the 5d mod-2 Dirac operator on the mapping torus $\ICP ^2\rtimes _{\widehat \varphi}S^1$, with appropriate fluxes for the  $Spin_G(4)$ background and also additional fluxes to activate the anomaly~\eqref{anomalyGammaT}.  Here $\widehat \varphi$ is a $\IZ _2$ valued isometry of $\ICP ^2$ and the flux background.  The 5d mod-2 Dirac index then reduces to a 4d Dirac index calculation in the flux background, with the fermion zero modes all in 2d representations of an $O(2)$ group~\cite{Wang:2018qoy}.  The upshot is a  simple way to compute $\kappa _{\Gamma _g, T}$: it is half (mod 2) of the number of fermion zero modes $\fI$ of the 4d Dirac operator in a flux background that activates the anomaly: 
\eq{\label{kappais}
\kappa_{\Gamma_g,T} = \half \fI=\frac{1}{4} \sum _i  \left(q_i^2-\frac{1}{4}\right)~, \qquad \text{mod}_2~,
}
where the Atiyah-Singer index theorem yields the last equation. The $q_i$ are flux-weighted charges of fermion species $i$, associated with an appropriately chosen $Spin^c$ connection.
 
 This simple method is generally applicable when the matter content consists of Weyl fermions in symplectic representations of the $G_\text{gauge}\times G_\text{global}$ group. This allows for non-zero 't Hooft anomalies, as there is no symmetry-preserving mass term.  E.g. for $SU(2)_\text{gauge}$ with $N_f=2$ adjoint Weyl fermions,  the $SU(2)_\text{gauge}$ and Lorentz invariant fermion bilinear is in the adjoint of $SU(2)_\text{global}$.    
 
We will consider and directly compute 't Hooft anomalies of the general form~\eqref{anomalyGammaT} that can arise in theories with a $Spin_{G_{\rm global}}(4)$ structure on $M_4=\ICP ^2$, generalizing the original example~\cite{Cordova:2018acb} to wide classes of theories.    The examples include e.g. general $G_{\rm gauge}$ theory with $N_f=2$ Weyl fermions in the adjoint, formulated with $Spin_{SU(2)}(4)$ structure on $\ICP ^2$, with background gauge fields $B_2$ for the one-form global symmetry $\Gamma _g$ (the center of the gauge group $G_{\rm gauge}$).  These same 't Hooft anomalies would arise in ${\cal N}=2$ SUSY Yang-Mills with gauge group $G_{\rm gauge}$ when formulated with  $Spin_{SU(2)_R}(4)$ structure on $\ICP ^2$, as in the Donaldson-Witten topological twist, where $SU(2)_R\equiv SU(2)_{\rm global}$ is the R-symmetry of ${\cal N}=2$, which acts on the two adjoint fermions as in the non-SUSY theory.

The outline of this paper is as follows.  In Section~\ref{sec:review} we review aspects of $w_2w_3$-type anomalies and their computation from~\cite{Cordova:2018acb} and~\cite{Wang:2018qoy}, and generalize the methods to be applicable to more general $Spin_G(4)$ theories. 
In Section~\ref{sec:examples} we consider a variety of classes of gauge theories that admit generalized $Spin _G(4)$ structures, and compute the $w_2 w_3$ anomaly coefficients of the theory on $\ICP ^2$.  We find:
\begin{itemize}
\item \textbf{Section \ref{sec:su2janomaly}:} $SU(2)$ gauge theory with $N_f=2n_f$ flavors of matter fields in the general isospin $j\in \IZ$ representation  
has a $w_2w_3$ anomaly coefficient $\kappa _{\Gamma _g, T}=\half N_f$ for $N_f\in 4\IZ +2$ if $j\in 4\IZ +1$  
or $j\in 4\IZ +2$; $\kappa _{\Gamma_g,T}=0$ for $j\in 4\IZ$ and $j\in 4\IZ +3$
\item  \textbf{Section \ref{sec:AQCDanom}:} $SU(N_c)$ with $N_f$ chiral fermions in the adjoint representation has a $w_2w_3$ anomaly coefficient $\kappa _{\Gamma _g, T}$ is non-zero iff both $N_c=2$ mod 4 and $N_f=2$ mod 4
\item  \textbf{Section \ref{sec:spinoddvec}:} $Spin(2n_c+1)$ gauge theory with $N_f=2n_f$ chiral fermions in the fundamental vector representation has anomaly coefficient is $\kappa _{\Gamma _g, T}=n_f$ mod 2.  
\item \textbf{Section \ref{sec:SpinvecQCD}:} $Spin(2n_c)$ gauge theory with $N_f=2n_f$ chiral fermions in the vector representation has anomaly coefficient  $\kappa _{\Gamma _g, T}=n_f$ mod 2.  
\item \textbf{Section \ref{sec:spinaqcd}:} $Spin (N_c)$ gauge theory with $N_f=2n_f$ chiral fermions in the adjoint representation  has $\kappa _{\Gamma _g, T}=n_fN_c$ mod 2. 
\item \textbf{Section \ref{sec:otherexamples}:} $Sp(N_c)$ for $N_c=1,2 $ mod${}_4$ (but not $N_c=0,3$ mod${}_4$) and $E_7$ gauge theories with $N_f=2n_f$ adjoint Weyl fermions have  $\kappa _{\Gamma _g, T}=n_f$ mod 2. 
\end{itemize}
In Section \ref{sec:abelian}, we discuss how to modify the $\widehat \varphi$ mapping torus method when it needs to include charge conjugation, as in cases with $U(1)$ factors and for anomaly matching when the gauge group is Higgsed.  In Section \ref{sec:CP1model}, we consider anomaly matching if the global symmetries are spontaneously broken, e.g. in theories with a global $SU(2)_R\equiv SU(2)_\text{flavor}$ symmetry that is spontaneously broken to a $U(1)_R\subset SU(2)_R$, resulting in a $\ICP ^1\cong SU(2)_R/U(1)_R$ non-linear sigma model.  We discuss anomaly matching in the sigma model, generalizing the analysis in~\cite{Cordova:2018acb} to other theories.  In Section \ref{sec:SMG}, we discuss a class of chiral gauge theory examples that are motivated by examples considered in \cite{Razamat:2020kyf,Tong:2021phe,Smith:2021vbf} for possibly exhibiting symmetric mass generation in 4d; we double the spectrum to obtain an extra $SU(2)_R$ global symmetry that can be used for $Spin_{SU(2)_R}(4)$ twisting. We note that the resulting theory has a $w_2w_3$ anomaly, so it cannot be completely trivially gapped in the IR.  In Appendix \ref{app:fractionalization} we will recall the fractionalization technique of \cite{Brennan:2022tyl,Wang:2018qoy} and provide details of its application to compute discrete anomalies for a few examples quoted in the main text.

\section{Computing Generalized $Spin_G$ Structure 't Hooft Anomalies}
\label{sec:review}

In this section we will review and generalize aspects of the new 't Hooft anomalies that can arise for theories with generalized $Spin _G(4)$ structure on $\ICP ^2$.   The review will be based on discussion and methods from~\cite{Cordova:2018acb, Wang:2018qoy}.  For a generalized $Spin_G(4)$ structure, all dynamical fields in the theory are in  faithful representations of  
\eq{\label{GGamma}
G_{\rm total}=\frac{\frac{G_{\rm gauge}}{\Gamma_g}\times G_{\rm global}\times Spin(4)}{\Gamma}
}
where $Spin(4)$ is the Euclidean Lorentz group, $\Gamma _g$ is a subgroup of the center of $G_{\rm gauge}$, and $\Gamma$ is a subgroup of the centers of all of the groups.  The $\Gamma _g$ quotient implies a $\Gamma _g^{(1)}$ one-form global symmetry acting on the $G_{\rm gauge}$ Wilson lines.  A $Spin_G(4)$ structure is possible if $\Gamma$ identifies $(-1)^F\in Spin(4)$ with a central element $-1$ in the center of $G=G_\text{gauge}$ or $G=G_\text{global}$.  In the case of $Spin_{G_\text{gauge}}$, the path integral over the $G_\text{gauge}$ fields effectively gauges $(-1)^F$ and the gauge invariant spectrum of the theory is bosonic; there is a constraint $w_2(G_\text{gauge})=w_2(TM)$ in the path integral.   In the case of $Spin_{G_\text{global}}$, the interpretation is instead that of twisting and there are background fields constrained to satisfy  $w_2(G_\text{global})=w_2(TM)$.   We will here be especially interested in the new 't Hooft anomalies that arise for $Spin_{G_\text{global}}$ theories on $M_4=\ICP ^2$ in the presence of background gauge fields for the $\Gamma _g^{(1)}$ one-form global symmetry.   Anomalies can be exhibited by considering background gauge fields and geometry, subject to the $\Gamma$ identifications, and arise from the variation of the path integral measure.   

Anomalies arise in the path integral description from the transformation of the functional integration measure, which often reduces to an anomalous variation of zero modes in an appropriate background.   
 These zero modes can be read off from the index $\Ind[\slashed{D}]$ of the 4d Dirac operator on $M_4$.\footnote{The more general APS index theorem, including the $\eta$ invariant contribution~\cite{Witten:1985xe,Dai:1994kq,Witten:2019bou} is not needed here.}   
 
 For example, the $ U(1)_A \times G_\text{gauge}^2$ ABJ anomaly can be read off from the fermion zero modes of $\slashed{D}$ in the background of a $G_{\rm gauge}$ instanton\footnote{For  $U(1)_{\rm gauge}$ there is no instanton and the anomaly leads to a non-invertible symmetry category \cite{Cordova:2022ieu,Choi:2022jqy}.}.  The 't Hooft vertex's fermion zero modes have net $U(1)_A$ charge, explicitly breaking $U(1)_A$ to a discrete subgroup even in a trivial background. The Atiyah Singer index theorem implies that a 
Weyl chiral fermion $\psi _r$ in representation $r$ has $N_r=I(r)$ fermion zero modes in the $k_\text{inst}=\frac{1}{8\pi ^2} \int \Tr F\wedge F=1$ instanton sector where $\Tr _{r}T^aT^b =2I(r)\delta ^{ab}$ is the quadratic index (normalized to be an integer\footnote{E.g. for the adjoint representation $N_{adj}=2h^\vee _{G_\text{gauge}}$, with $h^\vee _{G_\text{gauge}}$ the dual Coexter number.}).  The $U(1)_A$ that assigns charge $q_i=1$ to all fermions is broken explicitly by the instanton to a $\IZ_{N_0}$ subgroup where $N_0=\sum _i I(r_i)$ is the total number of fermion zero modes in the instanton background.  For consistent theories without gauge anomalies, $N_0$ is always even and the generator $g$ of $\IZ _{N_0}$ has $g^{\half N_0}=(-1)^F$.  See e.g.~\cite{Hsieh:2018ifc,Delmastro:2022pfo} for discussion of the 't Hooft anomalies of such a $\IZ _{N_0}$ symmetry.  
As another example, gauge anomalies from fermions come from an index theorem for the 6d Dirac operator but can nevertheless often be detected from the 4d index in an instanton background, if the product of the fermion zero modes in the instanton's Hooft vertex is not Lorentz and/or gauge invariant, e.g. because of having an odd number of fermion zero modes.    
 
\subsection{$w_2w_3$ 't Hooft anomalies  
}
\label{sec:w2w3gen}

If the one-form group $\Gamma _g^{(1)}$ in~\eqref{GGamma} is $\Gamma _g^{(1)}\supseteq \IZ _{N_g}^{(1)}$, and $(-1)^F\in \Gamma$ so the theory admits a $Spin_G(4)$ structure, we can consider a $\IZ_{N_g}$ version of the $w_2w_3$ 't Hooft anomaly  in~\eqref{anomalyGammaT}
\eq{\label{anomalyGammaTgen}
\CA \supset  \kappa _{\Gamma _g, T} \frac{2\pi i }{2N_g} \int _{N_5} B_{N_g} \cup \delta w_2(TN) \qquad \hbox{or}\qquad \CA \supset \kappa _{\Gamma _g, T} \frac{2\pi i }{2N_g} \int _{N_5} \delta B_{N_g} \cup w_2(TN)~.
}
Here $B_{N_g}\in H^2(N_5, \IZ_{N_g})$ is the $\Gamma _g^{(1)}$ background gauge field for the $\IZ_{N_g}^{(1)}\subset \Gamma_g^{(1)}$ subgroup and $w_3(TN)=\half \delta w_2(TN)$ and ${1\over N_g}\delta B_{N_g}=\beta_{N_g}(B_{N_g})$ is the $\IZ_{N_g}$ Bockstein of $B_{N_g}$. 
The %background $B_{N_g}$ is $\IZ _{N_g}$ valued and the anomaly is understood 
anomaly encodes a  dependence of the path integral on the choice of integer lift 
%in terms of the integer lifts 
$\tilde B_{N_g}$ or $\tilde w_2(TN)$  where different choices of lift are related by the shifts 
%as a mixed anomaly associated with  simultaneously imposing 
$\tilde B_{N_g} \to \tilde B_{N_g}+N_gx$ and $\tilde w_2(TN)\to \tilde w_2(TN)+2y$ with integral-valued $x$ and $y$.  This generalizes the $N_g=2$ discussion in~\cite{Cordova:2018acb}.  The two expressions for the anomaly in~\eqref{anomalyGammaTgen} differ by an integration by parts, corresponding to modifying the 4d action by a local counterterm 
\eq{\label{sct}
S_{c.t.} = \frac{i \pi}{N_g} k\int _{M_4} \tilde B_{N_g}\cup \tilde w_2(TM)~,
}
with $k=\kappa_{\Gamma_g,T}$.  The two presentations of the anomaly in~\eqref{anomalyGammaTgen} lead to a variation of the path integral under shifts of the integer lifts $\tilde B_{N_g} \to \tilde B_{N_g}+N_gx$ or $\tilde w_2(TN)\to \tilde w_2(TN)+2y$ respectively:
\eq{Z\to Z\cdot (-1)^{\kappa _{\Gamma _g, T} \int _{M}x \cup w_2(TM)} \qquad \hbox{or} \qquad Z\to Z \cdot e^{\frac{2\pi i}{N_g} \kappa _{\Gamma _g, T}\int _M B_{N_g} \cup y }~,
}
with integral-valued $x$ and $y$. 
This variation can be cancelled if $\kappa_{\Gamma_g,T}$ is even, so we restrict to the case where $\kappa_{\Gamma_g,T}$ is odd. 
Additionally, for $N_g$ odd we can use the counterterm \eqref{sct} with $k=N_g-\kappa_{\Gamma_g,T}$ to completely trivialize the anomaly, so we take also $N_g$ even.   The first presentation shows that the anomaly is $\IZ _2$ valued for all even $N_g$.  In the second presentation, anomaly appears to be $\IZ _{N_g}$-valued -- but it really is not: we can shift $\kappa _{\Gamma _g, T}\to \kappa _{\Gamma _g, T}+2$ by the counterterm with $k=2$. This is why the anomaly is trivializable when $N_g$ is odd: $2$ is a divisor of -1 in $\IZ_{2n+1}$. 

Because the anomaly is $\IZ _2$ valued and vanishes for $N_g$ odd, it suffices as in~\eqref{anomalyGammaT} to activate a flux in a $\IZ _2$ subgroup of $\IZ _{N_g}$ as $B_{N_g}=\frac{N_g}{2}B_2$ where $\tilde B_2\to \tilde B_2+2x$. 
  The method discussed in the next section only probes the $\kappa _{\Gamma _g, T}$ anomaly for flux in a $\IZ _2\subset \Gamma _g$.  We will see that this anomaly trivializes\footnote{The Bockstein $\beta = \half \delta$ for a $\IZ_2$-valued cochain trivializes if the co-chain has a $\IZ _4$ lift.  Nevertheless, theories can have other types of non-trivial, mixed $\IZ _4$, $\IZ _2$ 't Hooft anomalies, e.g. via the discrete gauge version of the ${\cal A}=\frac{i}{2\pi} \int B^{(e)} \wedge B^{(m)}$ mixed 't Hooft anomaly.  E.g. if we start with a theory with a $\IZ _8$ global symmetry and gauge a $\IZ _2$ subgroup, then the  anomaly of~\cite{Tachikawa:2017gyf} will give a mixed $\IZ _4$, $\IZ _2$ 't Hooft anomaly.}  if $N_g\in 4\IZ$. The $w_2 w_3$ anomalies can more generally also be computed via the symmetry fractionalization techniques of \cite{Wang:2018qoy,Brennan:2022tyl}, as illustrated in the appendix.   

\subsection{The $\widehat \varphi$ Mapping Torus and its  
 Anomaly Indicator}
\label{sec:MapTor}

As discussed in  \cite{Wang:2018qoy}, the global anomalies associated with gravitational 't Hooft anomaly coefficient can be computed from a mod-2 index of the 5d Dirac operator on the mapping torus $N_5=\ICP ^2\rtimes _{\widehat \varphi} S^1$associated with the theory on $M_4=\ICP ^2$, where $\widehat \varphi$ includes the operation $\varphi _{c.c.}$ that acts as complex conjugation on $\ICP ^2$ along with a Weyl transformation to preserve the flux backgrounds.   This is natural, since the $\Omega _5$ bordism class with  $\int _{N_5} w_2w_3\neq 0$ can be generated by the Dold manifold $N_5=\ICP ^2\rtimes _{\varphi _{c.c.}}S^1$.   We will here generalize this method to directly compute other $w_2w_3$ type anomalies, and apply the method to a variety of examples.  

As in~\eqref{GGamma}, consider a general theory where the fermions form a faithful representation of a total group that involves the gauge group, global group, and Lorentz group, modded out by $\Gamma$.  We divide such theories into two cases: (1) when $(-1)^F\in Spin(4)$ is not in the identification, i.e. $(-1)^F\notin \Gamma$;  and (2) when $(-1)^F\in \Gamma$.  If $(-1)^F \notin \Gamma$, the theory is fermionic, and does not admit a generalized spin structure so it cannot be put onto a non-spin manifold.  For such cases, a basic $M_4$ with spin structure that exhibits the possible anomalies is $M_4=\ICP^1\times \ICP^1$.  Our interest here is in the case where $(-1)^F \in \Gamma$, so the theory classically admits a $Spin_{G}$ structure; we can then consider the theory on the non-spin manifold $M_4=\ICP ^2$ with discrete flux in $G$ such that $w_2(TM)=w_2(G)$. We generally refer to the process of elevating a $G$-connection to a $Spin_G$ connection  or reducing a $G$-connection to a $G/Z$ connection with obstruction class $w_2(G)$ as ``turning on a discrete flux'' in the group $G$.  We will be interested in cases where there is a one-form global symmetry $\Gamma _g^{(1)}$, coming from the center of the gauge group (or subgroup thereof), and turning on an associated 2-form background gauge field $B_2$ is also a type of discrete flux, which is unconstrained by the $Spin_G$ structure but must be activated to see the associated 't Hooft anomaly.  For reasons we will discuss shortly, we will only consider activating background fluxes for $\IZ_2\subset \Gamma_g^{(1)}$, so we will restrict to $B_2\in H^2(BG_{\rm gauge}/\Gamma_g;\IZ_2)$.  

To activate the needed background fluxes, both for the $Spin_G$ structure and for $\Gamma _g^{(1)}$ background gauge fields, it suffices to consider a group $U(1)_Q$ that is embedded in the gauge group and global symmetries. 
 More precisely, the fluxes are in a discrete subgroup of $U(1)_Q$ and, if the discrete flux is associated with an obstruction class, then $U(1)_Q$ is actually a $Spin^c$ connection.  For example, if the gauge group is $SU(N_g)$ and the theory has a $\Gamma^{(1)}_g=\IZ_{N_g}^{(1)}$ one-form center symmetry, then turning on a generic $\IZ_{N_g}$-valued non-zero background $B_2^{(N_g)}$ gauge field is achieved by choosing $U(1)_{\Gamma_g}\supset  
 Z(SU(N_g))$  (which itself embeds in $U(1)_Q$) 
and turning on a discrete  
flux  $B_2^{(N_g)}$ along this $U(1)_{\Gamma_g}$ by elevating the $U(1)_{\Gamma_g}$ connection to a $Spin^c$ connection.   More generally, we can turn on generic fluxes for $\Gamma=\IZ_N$ and one-form group $\Gamma_g=\IZ_{N_g}$ in~\eqref{GGamma},  and write fluxes associated with non-trivial $w_2(\Gamma)$ and $w_2(\Gamma _g)$ by embedding $\Gamma \subset U(1)_\Gamma$ and $\Gamma _g\subset U(1)_{\Gamma _g}$ which are generated by some charges $Q_{\Gamma},Q_{\Gamma_g}$ respectively. We can then write the flux backgrounds as an explicit 
$Spin^c$ connection using $U(1)_Q^c\equiv {Q}_\Gamma U(1)_\Gamma^c + {Q}_{\Gamma_g}U(1)_{\Gamma_g}^c$:
\eq{\label{Qflux}
A_{Q}={a}_{\IZ_N}{Q}_\Gamma+{a}_{\IZ_{N_g}}{Q}_{\Gamma_g} \quad ,\quad \oint\frac{{f}_{\IZ_n}}{2\pi}=\frac{1}{n}~,
}
where ${f}_{\IZ_n}$  is the field strength for the $Spin^c$ connection ${a}_{\IZ_n}$.

The  number of fermion zero modes of the Dirac operator in the $U(1)_Q$ flux background  is then given by the Atiyah Singer index theorem as $\Ind[\slashed{D}]=\int _{M_4} \widehat A(TM)\wedge\Tr _\psi  \exp (F_Q/2\pi)$, with $\widehat A(TM)=1-p_1(TM)/24+\dots$, where $\Tr_\psi$ runs over the Weyl fermions.  We decompose the fermions into representations of $U(1)_Q$ as $\psi=\sum_{i} \psi_i\, v_i$, where $v_i$ is a basis of eigenvectors of the $A_Q$ flux,  $F_Q\cdot v_i=q_i \,v_i$; our notation will be to put the $1/n$ fractional fluxes in~\eqref{Qflux} into the normalization of the charges $q_i$.  The index theorem gives  the number of fermion zero modes  $\fI$ to be\footnote{Equation \eqref{OGindex} is the formula for the index of the Dirac operator from the Atiyah Singer index theorem. As mentioned in \cite{Wang:2018qoy}, the index of the Dirac operator on $\ICP^2$ is equivalent to the number of zero modes. This can be shown by a Lichnerowicz-type argument using the fact that $\ICP^2$ has positive curvature and that we are coupling to only self-dual gauge fields. 
}
 \eq{\label{OGindex}
\fI(R)=\begin{cases}
\sum_{i}q_i^2&M=\ICP^1\times \ICP^1\\
\sum_{i}\frac{1}{2}(q_i^2-\frac{1}{4})&M=\ICP^2 
\end{cases}
}

The $-\frac{1}{8}$ comes from the $-p_1(TM)/24$ term in $\widehat A$ evaluated on $\ICP ^2$, and the overall relative factor of $\half$ between the two cases is because the 2-cycle in $\ICP ^2$ has intersection pairing 1 whereas in $\ICP ^1\times \ICP ^1$ the intersection pairings are 2.

As in \cite{Wang:2018qoy},  we exhibit $w_2w_3$ anomalies by considering the Dirac operator on the 5d mapping torus associated with a classical symmetry $\widehat \varphi$ that acts on $\ICP ^2$ as complex conjugation.  To be a symmetry of the flux background, $\widehat \varphi$ generically needs to be augmented with a $G$ Weyl group operation.   Although $\widehat \varphi$ is then a symmetry of the bosonic backgrounds, the fermion zero modes of $\slashed{D}$ can transform non-trivially under $\widehat \varphi$ which may give rise to an anomaly.   We will show, as in~\cite{Wang:2018qoy}, that the fermion zero modes of the 4d Dirac operator all form 2d representations of $O(2)$ under $U(1)_Q$ and $\widehat \varphi$, and that this implies that the $\IZ _2$ valued $w_2w_3$ anomaly coefficient is given by half of the number of zero modes counted in~\eqref{OGindex}. 

We can apply these considerations to both the $M_4=\ICP ^1\times \ICP ^1$ and the $M_4=\ICP ^2$ cases.  The standard mapping torus construction of anomalies involves a 5th dimension that is an interval where the ends are identified with a bosonic symmetry transformation.  
For $w_2w_3$ anomalies, the transformation is given by 
\eq{
\widehat\varphi=\varphi_{c.c.}\circ W~,
}
where $\varphi _{c.c.}: (z_1, z_2)\to (z_1^*, z_2^*)$ which is a symmetry of $M_4$.  The Weyl group element $W$ compensates for the fact that $\varphi _{c.c.}$  acts as complex conjugation on the fluxes associated with the $\Gamma,\Gamma_g$-fluxes, where $\Gamma$ and $\Gamma _g$ are as in ~\eqref{GGamma}, as $\varphi_{c.c.}:w_2(\Gamma),w_2(\Gamma_g)\longmapsto -w_2(\Gamma),-w_2(\Gamma_g)$.  The $W$ group element restores the fluxes to obtain a classical symmetry of the background, since $W$ also acts as charge conjugation on $U(1)_\Gamma,U(1)_{\Gamma_g}$. 
For example, given an embedded $U(1)\subset SU(2)$, the Weyl generator of $SU(2)$ acts as charge conjugation on the embedded $U(1)$.  This construction from ~\cite{Wang:2018qoy} generalizes immediately for theories with fermions in irreducible real or pseudo-real representations of $G_{\rm gauge}\times G_{\rm global}$. 

Note that such $W\in G_{\rm gauge}\times G_{\rm global}$ only exist when $w_2(\Gamma),w_2(\Gamma_g)$ are $\IZ_2$-valued fluxes. The reason is that $G$ acts trivially on $Z(G)$ so the only way that there can exist $W$ that undoes the action of $\varphi_{c.c.}$ on the fluxes $w_2(\Gamma),w_2(\Gamma_g)$ is if $\varphi_{c.c.}$ only acts to change their representative -- i.e. if they are $\IZ_2$-valued. For vector-like matter representations, there is a similar construction with $W$ an outer automorphism charge conjugation symmetry, as we will discuss and illustrate in  Section \ref{sec:AnomalyCSymmetry}.     

As discussed in~ \cite{Wang:2018qoy}, this $\widehat \varphi$ satisfies $\widehat \varphi ^2=1$, although $\varphi _{c.c.}^2= W^2=(-1)^F$, as seen from the fact that $\varphi _{c.c.}$ acts  as a $\pi$-rotation on $M_4=\IC\IP^2$ \cite{Wang:2018qoy} (and similarly for $M_4=\ICP^1\times \ICP^1$): if we parameterize $M_4$ in real coordinates $(z_1,z_2)=(x_1+ix_2,x_3+ix_4)$, complex conjugation takes $(x_1,x_2,x_3,x_4)\mapsto (x_1,-x_2,x_3,-x_4)$, which is a $\pi$-rotation in the $x_2-x_4$ plane.  Likewise, $W$ that acts a $\pi$-rotation in the Lie algebra $\fg_{\rm gauge}\times \fg_{\rm global}$ of the fermions.   Since $W$ acts as complex conjugation on $U(1)_Q$, the fermions transform in reps of $O(2)=U(1)_Q\rtimes \IZ_2^{\widehat\varphi}$.  The group $O(2)$ has only one- and two-dimensional representations.  The nontrivial action of  the generators $Q$ of  $U(1)_Q$, and $\widehat\varphi$ of $O(2)$, on the fermions implies that the  $\fI$ fermion zero-modes must reside in $\half \fI$ copies of the $2d$ representations of $O(2)$, and the generators act as   
 \eq{
Q=\left(\begin{array}{cc}q_i &0 \\ 0 &-q_i\end{array}\right)\quad, \quad 
\widehat\varphi=\left(\begin{array}{cc}0&1\\1&0\end{array}\right)~. 
}
Thus $\widehat \varphi$ acts on the zero-modes, and thus on the fermion measure of the path integral, as:
\eq{\label{PIvariation}
\int [d\psi ]\big(\dots \big) \stackrel{\widehat\varphi}{\xrightarrow{\hspace{2cm}}} \int [d\psi]\times {\rm det}[\widehat\varphi] \big(\dots\big)=(-1)^{\fI /2}\int [d\psi] \big(\dots \big)~.
}
Here $\half \fI$ is the number of 2d $O(2)$ representations, with $\fI$ the total number of fermion zero modes as given by~\eqref{OGindex}.
There is a $\widehat\varphi$ anomaly if there is a non-zero (odd) value of the $\IZ_2$-valued anomaly indicator
\eq{
\sigma_\fI:=\frac{\fI}{2}~{\rm mod}_2~.
}
Since the anomaly is $\IZ _2$ valued,  it is always trivialized upon doubling the matter content.

The anomaly exhibited by~\eqref{PIvariation} can be described in terms of a $5d$ anomaly polynomial/anomaly SPT phase given by a $w_2\cup w_3$ anomaly theory
\eq{\label{genanom}
\CA\subset\pi i \,\sigma_\fI\int _{N_5} w_2 \cup w_3^\prime~,
}
where $w_2,w_2^\prime$ are discrete fluxes associated with $\Gamma,\Gamma_g$ and $w_3^\prime$ is the Bockstein of $w_2^\prime$: $w_3^\prime={\rm Bock}[w_2^\prime]$. 
The anomalous phase can be realized by considering the SPT/anomaly polynomial on the mapping torus $N_5=\ICP^2\rtimes_{\widehat\varphi} S^1$ which is twisted by the action of $\widehat\varphi$.  Then  
\eq{
(w_3^\prime)_N=(w_3^\prime)_{\ICP^2}\oplus (w_2^\prime)_{\ICP^2}\cup (w_1^\prime)_{S^1}~. 
}
Since $\oint_{S^1}w_1=1$ on the mapping torus, the anomaly SPT  reduces to the phase on $\ICP^2$:
\eq{\label{anomSPT}
S_{5d}=\pi i \sigma _\fI \int_N w_2\cup w_3^\prime=\pi i\sigma _\fI  \int_{\ICP^2} w_2\cup w_2^\prime~,
}
which implies the anomalous phase~\eqref{PIvariation} of the path integral under $\widehat\varphi$.

\subsection{Review of the WWW $w_2w_3$ ``New $SU(2)$ Anomaly''  
}
\label{sec:NewSU2}

We here illustrate the the discussion of the previous section by  reviewing the original case from~\cite{Wang:2018qoy}: an $SU(2)_\text{gauge}$ theory with $N_f=1$  Weyl fermion $\psi _\alpha$ in the $j=3/2$ representation of $SU(2)_\text{gauge}$.  This is a chiral theory in that   
it is impossible to write a mass term for $\psi _\alpha$ as there is no quadratic, gauge invariant, Lorentz scalar operator.  The classical, ABJ-anomalous  $U(1)_A=U(1)_\psi$ global symmetry is broken by instantons to $\IZ _{10}$: the Dirac operator in flat $\IR ^4$ in an instanton background has $N_{0} =I_2(j=3/2)=10$ fermion zero modes\footnote{For a Weyl fermion in the $2j+1$ dimensional representation, $N_0=I(j)=\frac{2}{3}j(j+1)(2j+1)$.}; the fact that this is even shows that the theory does not have the original $SU(2)_\text{gauge}$ anomaly. 
The generator $g$ of the $\IZ _{10}$ global symmetry has $g^5=(-1)^F$ and the $\IZ _{10}$ 't Hooft anomalies obstruct a symmetry preserving gapped phase~\cite{Cordova:2019jqi}; this symmetry will not play a role here.  

All dynamical fields form faithful representations of the symmetry group 
\eq{
G_{\rm total}=\frac{SU(2)_{\rm gauge}\times Spin(4) \times \IZ _{10}}{\IZ_2\times \IZ_2}.
}
One $\IZ _2$ identifies $g^5=(-1)^F$ and the other identifies $(-1)^F=-\mathds{1}_{SU(2)_\text{gauge}}$.  So this theory (unlike the SUSY version in~\cite{Intriligator:1994rx}) satisfies a spin-charge relation  and thus can be classically formulated with a $Spin_{SU(2)}(4)$ structure on $M=\ICP ^2$ via a twist of the $SU(2)_{\rm gauge}$-bundle to have discrete flux  
$w_2(SU(2))\in H^2(M;\IZ_2)$ equal to the second Stiefel-Whitney class of $M$:
\eq{
w_2(SU(2))=w_2(TM)~. 
}
For $\ICP ^2$, $w_2(TM)$ is non-trivial and thus the $SU(2)_{\rm gauge}^c$ bundle is an $SO(3)$ bundle with fixed $w_2(SO(3))=w_2(TM)$.  Here we write the gauge bundle as $SU(2)^c_{\rm gauge}$ rather than $SO(3)_{\rm gauge}$ because the gauge group is not actually $SO(3)$, and is more analogous to a $Spin^c$ connection -- moreover we will often simply write the gauge bundle as $SU(2)_{\rm gauge}$ despite the non-zero $w_2(SU(2))$ obstruction from being an $SU(2)$ bundle.  Restricting to a $U(1)_Q\subset SU(2)_{\rm gauge}$, for the theory on $\ICP ^2$ we turn on a $Spin^c$ connection along   
\eq{
 A_{SU(2)}=\left(\begin{array}{cc}
\widehat{a}_{\IZ_2}&0\\0&-\widehat{a}_{\IZ_2}
\end{array}\right)\quad, \quad \int_H \frac{\widehat{f}_{\IZ_2}}{2\pi}=\half~,
}
corresponding to $Q=T^3$.  
This background is preserved by the $\widehat \varphi$ map 
\eq{
\widehat\varphi=\varphi_{c.c.}\circ W_{SU(2)}\quad, \quad W_{SU(2)}=\left(\begin{array}{cc}
0&-1\\1&0
\end{array}\right)~.
}
$W_{SU(2)}$ acts as the generator of the Weyl group acting on the Cartan generated by $Q$.  

To determine the fermion zero modes and $\widehat \varphi$ action on them,  we decompose the fermion into $U(1)_Q$ representations, which results in four fermions with flux $\times$ charges given by
\eq{
q_i=\left(\frac{3}{2},\half,-\half,-\frac{3}{2}\right)~. 
}
The $U(1)_Q$ charges of the fermions are odd integers, consistent with the spin-charge relation of a $Spin^c$ connection.  Our notation is that the charges $q_i=\half \cdot 2 \cdot T^3_{ii}$ includes a factor from the half-unit of flux,  $\int _{\Sigma _2}\frac{F}{2\pi}=\half \int _{\Sigma _2}w_2(TM)=\half$, that cancels a factor of $2$ from rescaling the $U(1)\subset SU(2)$ charges to be integers (with the $T^\pm$ generators of charge $\pm 2$); the upshot is that the $q_i$ are simply the $T^3$ eigenvalues with the usual normalization for $T^\pm$ charges. The number of zero modes of the Dirac operator in this flux background on $\ICP ^2$ is
\eq{\label{WWWOGindex}
\fI(R_{j=3/2})=\sum_{\psi_i}\frac{4q_i^2-1}{8}\quad\Longrightarrow \quad \fI(\mathbf{4})=2~. 
} 

From this we can read off the anomaly indicator for the chiral fermion $\psi$ in the ${\bf 4}$ rep:
\eq{
\sigma_{\fI ({\bf 4})}= \frac{1}{2} \fI ({\bf 4})\ \rm{mod}_2=1~.
}
The theory thus has a $\widehat \varphi$ anomaly, corresponding to the gravitational 't Hooft anomaly \eqref{genanom} with $w_2,w_2^\prime=w_2(TM)$:
\eq{
\CA\subset\pi i \kappa _{T,T} \int _{N}w_2(TN)\cup w_3(TN)~ 
} 
with $\kappa _{T,T}=\sigma_{\fI ({\bf 4})}=1$.  Since $w_2(TM)=w_2(SU(2))$ and this anomaly can also be interpreted as a mixed gauge-gravity anomaly
\eq{
\CA\subset\pi i \int  _Nw_2(SU(2))\cup w_3(TN)~.
}
Generalizing to $SU(2)_{\rm gauge}$ with $N_f$ fermions in the $\mathbf{2j+1}$-rep, the index with $w_2(SU(2))=w_2(TM)$ on $M=\ICP ^2$ is  
\eq{
\fI(R)=N_f\sum _{f=-j}^j \frac{4f^2-1}{8}=\frac{N_f}{24}(4j^2-1)(2j+3)~.}
The new $SU(2)$ anomaly  $\kappa _{T,T}=\sigma_\fI(R)=\half \fI$ is odd only if $N_f$ is odd and $j={3\over 2}+4r$ with $r\in \IZ$~\cite{Wang:2018qoy}. 
  The $Spin_{SU(2)}(4)$ structure implies that $\kappa _{T,T}=\kappa _{SU(2),T}$.

\subsection{Application to $Spin_{G_\text{global}}(4)$ Structure Theories}

Our main interest here are the $w_2w_3$ anomalies of theories with $Spin_{G_\text{global}}(4)$ structure, i.e. those where all matter couples to bundles $G_{\rm total}$ of the form:
\eq{
G_{\rm total}=\frac{G_{\rm gauge}}{\Gamma_g}\times \frac{G_{\rm global}\times Spin(4)}{\Gamma}~,
} 
where $\Gamma=\IZ _2$ that identifies $(-1)^F$ with a $\IZ _2$ in the center of $G_{\rm global}$.  The generalized $Spin_{G_\text{global}}(4)$ structure involves using background gauge fields for the $G_\text{global}$ symmetry to twist the theory, allowing it to be put on non-spin manifolds like $M_4=\ICP ^2$.   There is also the option to introduce a background gauge field  $B_2$  for the $\IZ_2^{(1)}\subset \Gamma ^{(1)}_g$ one-form symmetry.  The possible $w_2w_3$ 't Hooft anomalies are the gravitational one, with coefficient $\kappa _{T,T}=\kappa _{G_\text{global},T}$, and a mixed 't Hooft anomaly between gravity and the global one-form symmetry $\Gamma _g^{(1)}$, with coefficient $\kappa _{\Gamma _g,T}$.\footnote{\label{WuRelation} 
The reason there is no independent $\kappa_{\Gamma_g,\Gamma_g}$ anomaly is that it can equivalently be written in terms of $\kappa_{\Gamma_g,T}$ because $w_2(G_{\rm gauge})\cup w_3(G_{\rm gauge}) \sim w_2(G_{\rm gauge}) \cup w_3(TN)$ for $\IZ _2$ valued $B_2\equiv w_2(G_{\rm gauge})$.  This follows by a Steenrod identity for any $\IZ _2$ valued cohomology  class $x_2\in H^2(N_5;\IZ_2)$
\eq{\label{appendixrelation} \int_{N_5} x_2\cup \beta(x_2)=\int_{N_5} \nu_2(TN)\cup \beta(x_2)~.}
 where $\nu_2(TN)$ is the $2^{nd}$ Wu class, and $\beta$ is the $\IZ_2$ Bockstein map. For orientable $N$, $\nu_2 (TN)=w_2(TN)$.  See e.g. \cite{Lee:2020ojw} for further details.  }  
 Both anomalies can be probed by turning on appropriate $U(1)_Q^c$ background fluxes as in~\eqref{Qflux}; for $\kappa _{T,T}$ the $\Gamma _g$ flux $B_2$ can be set to zero, and for $\kappa _{\Gamma _g,T}$ the $U(1)^c_Q$ should include both the $\Gamma _g$ and $\Gamma$ fluxes.  The anomaly coefficients are then determined from the 5d Dirac operator on the $\ICP ^2\rtimes _{\widehat \varphi} S^1$ mapping torus with appropriate flux backgrounds, with $\widehat \varphi$ that preserves the bosonic backgrounds, leading to the result~\eqref{kappais} in terms of the charges $q_i$. 

In the original example of~\cite{Cordova:2018acb}, i.e, $SU(2)$ gauge theory with $N_f=2$ Weyl fermions in the adjoint representation, the one-form global center symmetry is $\Gamma _g^{(1)}=\IZ _2^{(1)}$ and the $Spin_G$ structure involved $G_{\rm global}\equiv SU(2)_R$, with $-\mathds{1}_{SU(2)_R}\sim(-1)^F$ by $\Gamma = \IZ _2$.   In that example, all of the discrete symmetries are $\IZ _2$, and the $\kappa _{\Gamma _g, T}$ 't Hooft anomaly is $\IZ _2$ valued: the partition function can pick up a minus sign $Z\mapsto (-1)^{x\cup w_2(TM)}Z$ or $Z\mapsto (-1)^{B_2\cup y}Z$ under $\IZ _2$ trivial shifts $\tildeB_2\to \tildeB_2+2x$ and $\tilde w_2(TM)\to \tilde w_2(TM)+2y$ of the integer lifts of the $\IZ _2$-valued $B_2$ and $w_2(TM)$~\cite{Cordova:2018acb}.   We discuss some general aspects in this subsection, and will then illustrate the general discussion with the original example of~\cite{Cordova:2018acb} in a later subsection.    

Suppose that the one-form global symmetry is $\Gamma _g^{(1)}=\IZ _{N_g}^{(1)}$, with 2-form background gauge field $B_2$ that is $\IZ _{N_g}$ valued.  The integer lift of $B_2^{(N_g)}$ is then $\tildeB_2^{(N_g)}$, which reduces to $B_2^{(N_g)}$ if we mod out by $\tildeB_2^{(N_g)}\to \tildeB_2^{(N_g)}+N_gx$ with integer-valued $x$.  The $\kappa _{\Gamma _g, T}$ anomaly only arises for $N_g$ even and it suffices to turn on a background gauge field $B_2$ for $\IZ_2\subset \Gamma_g$, setting $B_2^{(N_g)}=\frac{N_g}{2}B_2$. An integer lift of $B_2$, $\tildeB_2$ also induces an integer lift of $B_2^{(N_g)}$: $\tildeB_2^{(N_g)}=\frac{N_g}{2}\tildeB_2$. 
 The $\kappa _{\Gamma _g,T}$ mixed anomaly also involves the $\IZ _2$ valued $w_2(TM)$, which has integer lift $\widetilde w_2(TM)$ that reduces to $w_2(TM)$ if we mod out by $\widetilde w_2(TM)\to \widetilde w_2(TM)+2y$.  As in~\cite{Cordova:2018acb} a non-zero $\kappa _{\Gamma _g,T}$ mixed anomaly will imply an anomalous phase transformation of the partition function under either of these shifts, and the two options differ by a choice of local counteterm.

The global symmetry $G$ in the $Spin_G(4)$ structure must have a $\IZ _2$ subgroup that can be identified with $(-1)^F$ by $\Gamma$.  The $\IZ _2$ subgroup of $G$ can be viewed as the center of an $SU(2)_{\rm global}\subset G_{\rm global}$ and the $U(1)_\Gamma$ in~\eqref{Qflux} is the Cartan of this $SU(2)_{\rm global}$.  There is some freedom in the choice of this  $SU(2)_{\rm global}\subset G_{\rm global}$, but a non-trivial $\kappa _{\Gamma _g,T}$ anomaly can be misleadingly trivialized if there is a factor of two in the index of embedding of $SU(2)_{\rm global}$ in $G_{\rm global}$; we will discuss this further in the next subsection.  

We will generally assume that the fermions are in reps $R_{\rm gauge}\otimes R_{\rm global}$ of $G_{\rm gauge}\times G_{\rm global}$, and that upon 
restricting $G_{\rm global}\to SU(2)_{\rm global}$ the fermions transform as $R_{\rm global}\to \mathbf{2}^{\oplus n_f}$,  where the Cartan of $SU(2)_{\rm global}$ can be identified with $U(1)_\Gamma$ and the $w_2(G_{\rm global})=w_2(SU(2)_{\rm global})=w_2(TM)$.\footnote{For gauge theories with $\Gamma=\IZ_2$, we generally have $G_{\rm global}\subset SU(2n_f)$ where $\IZ_2\subset Z(SU(2n_f))\mapsto \IZ_2\subset Z(G_{\rm global})$. We can then restrict $SU(2n_f)\mapsto SU(2)$ where $\mathbf{2n_f}\mapsto \mathbf{2}^{\oplus n_f}$ while preserving $\IZ_2\subset Z(SU(2n_f))\mapsto \IZ_2\subset Z(SU(2)_{\rm global})$, and we can also generally restrict $G_{\rm global}\mapsto SU(2)$ so that $R_{\rm global}\mapsto \mathbf{2}^{\oplus n_f}$ factors through the lift to $SU(2n_f)$.} 
The $U(1)^c_\Gamma$ flux $\times$ charge pairing  of each $SU(2)_{\rm global}$ fundamental is then given by ${\bf 2}\to \half \oplus -\half$.   In the $B_2=0$ background, the index of the Dirac operator is proportional to $\fI=\half \sum _i (q_i^2-\frac{1}{4})$ with $q_i=\pm \half $, so $\fI=0$. Since this reduction $G_{\rm global}\to SU(2)_{\rm global}$ preserves the $Spin_G$ structure, our computation shows that 
 $\kappa _{T,T}=0$ in gauge theories with $Spin_{G_{\rm global}}$ structure (assuming $R_{\rm global}\mapsto \mathbf{2}^{\oplus n_f}$).  This is to have been expected given that the WWW new $SU(2)$ anomaly only arises for $SU(2)$ representations $j=\frac{3}{2}+4r$ for $r\in \IZ$~\cite{Wang:2018qoy} as reviewed in the previous subsection.

To compute the $\kappa _{\Gamma _g,T}$ anomaly, we need to additionally turn on non-zero $B_2$ background gauge field for the $\IZ_2\subset \Gamma _g$ center symmetry, leading to the $\IZ_2$-valued obstruction class $B_2=w_2(G_{\rm gauge})$, and the $Spin _{G_{\rm global}}$ structure requires flux $w_2(G_{\rm global})=w_2(TM)$.  The total flux can be embedded in a $U(1)_Q^c=U(1)^c_{\Gamma _g}+U(1)^c_\Gamma$ where the $U(1)_{\Gamma _g}^c\subset G_{\rm gauge}^c$ connection activates the center flux $B_2=w_2(G_{\rm gauge})$ and the $U(1)^c_\Gamma\subset G_{\rm global}^c$ connection activates the  $w_2(G_{\rm global})=w_2(TM)$ flux on $\ICP ^2$.  The $U(1)_{\Gamma _g}$ has generator that we can write as 
\eq{
Q_{\Gamma_g}=\sum_I Q^IH_I~,
}
for some flux charges $Q^I$ associated with $B_2=w_2(G_{\rm gauge}^c)$, where $H_I$ are $G_{\rm gauge}$ simple co-roots generating a Cartan torus containing $U(1)_{\Gamma_g}$.    The fermion representations $R_{\rm gauge}$ decompose to have $Q_{\Gamma _g}$ charges $q^g_i$.  The fermions are in the fundamental of $SU(2)_{\rm global}$ and the $Spin_{G_\text{Global}}(4)$ structure on $\ICP ^2$ shifts the $U(1)_{\Gamma _g}$ charges $q_i ^g$ as 
\eq{(q_i ^g, q_i^g)\longmapsto \left(q^g_i+\half\right)\oplus \left(q^g_i-\half\right)\equiv q_i ~.
}
The index theorem for the number of zero modes in this flux background with $B_2=w_2(G_{\rm gauge}^c)$ and $w_2(G_{\rm global}^c)=w_2(\ICP ^2)$ is then given by the index theorem to be 
\eq{
\label{index}
\fI(R_{\rm gauge}\otimes R_{\rm global})&=\half \sum_{\psi_i}\left(q_i^2-\frac{1}{4}\right) 
=\frac{\fn}{2}N_f\, I_2(R_{\rm gauge})\Tr_{R_{\rm def}}[Q_{\Gamma_g}^2]
\\&=\frac{\fn}{2}N_f \,I_2(R_{\rm gauge})\sum_a Q^IC_{IJ}Q^J~.
} 
Here we write $\Tr_R[T^aT^b]=\mathfrak{n}\,I_2(R)\, \Tr_{\rm def}[T^aT^b]$ with $\fn=\frac{1}{I_2(R_{\rm def})}$ (e.g. for $SU(N)$ groups the factor $\frac{\fn}{2}=1$).
The anomaly can thus be characterized by $U(1)_{\Gamma _g} \supset Z(G_{\rm gauge})$ and matching the $Z(G_{\rm gauge})$-ality of fermion matter representations with the divisors of their Dynkin indices.

Here we see that if $\Gamma_g^{(1)}\supset \IZ_{4N}^{(1)}$, then the index will be a multiple of 4 when coupled to a $\IZ_2^{(1)}\subset \IZ_{4N}^{(1)}\subset \Gamma_g^{(1)}$ background gauge field $B_2$. The reason is that the index $\fI$ must be an integer for the minimal $\IZ_{4N}^{(1)}$ background as well as for the $\IZ_2^{(1)}\subset \IZ_{4N}^{(1)}$ background. These are related by $Q^I_{\IZ_2}=2NQ^I_{\IZ_{4N}}$ so that $\fI_{\IZ_2}=4N^2 \fI_{\IZ_{4N}}$. This implies that for the backgrounds with $\IZ_2^{(1)}\subset \IZ_{4N}^{(1)}\subset \Gamma_g^{(1)}$, there will be no anomalous phase of the partition function under the action of $\widehat\varphi$.

Recall from our discussion in Section \ref{sec:w2w3gen} that a $w_2w_3$ anomaly betwen $\Gamma_g^{(1)}$ and $Spin_G(4)$ structure implies that the partition function depends on the choice of integer lift of $B_2$ or $w_2(TM)$, so that under a shift of integer lift $\tildeB_2\mapsto \tildeB_2+2x$ or $\tilde{w}_2(TM)\mapsto \tilde{w}_2(TM)+2y$, partition function varies by
\eq{
Z\to Z\cdot (-1)^{\kappa _{\Gamma _g, T} \int _{M}x \cup w_2(TM)} \qquad \hbox{or} \qquad Z\to Z \cdot (-1)^{ \kappa _{\Gamma _g, T}\int _M B_{2} \cup y }~.
}
As we have discussed in Section \ref{sec:w2w3gen}, this anomaly only depends on the $\kappa_{\Gamma_g,T}$ mod 2 and the phase can be generically activated for $\IZ_2^{(1)}\subset \Gamma_g^{(1)}$ backgrounds. Therefore, since this anomalous phase vanishes for any $\IZ_2^{(1)}\subset \IZ_{4N}^{(1)}\subset \Gamma_g^{(1)}$, it must be that $\kappa_{\Gamma_g,T}=0$ mod 2 and there is no $w_2w_3$ anomaly between $\IZ_{4N}^{(1)}$ and $Spin_G(4)$ structures. 

\subsection{Determining Anomalies via Reduction to WWW ``New $SU(2)$'' Anomaly}

As discussed in the previous subsections, the $w_2w_3$ anomalies in theories with $Spin_G(4)$ structure, 
where $G$ can be a gauge or  global symmetry, can be understood in terms of a  $Spin _{U(1)_Q}(4)$ structure for some appropriate $U(1)_Q$ subgroup of the gauge and global symmetry group: $U(1)_Q=U(1)_{\Gamma _g}+U(1)_\Gamma$, as in the previous subsection.  Here the  $U(1)_{\Gamma _g}^c$ connection activates a $\IZ_2$-flux in $G_{\rm gauge}$ that is determined by $w_2(G_{\rm gauge})=B_2$, and the  $U(1)_\Gamma^c$ connection activates a flux in $G_{\rm global}$ determined by $w_2(G_{\rm global})=w_2(TM)$.  The $U(1)_{\Gamma _g}$ can often be embedded in an $SU(2)_{\rm gauge}\subset G_{\rm gauge}$ and the $U(1)_\Gamma$ can often be embedded in an $SU(2)_{\rm global}\subset G_{\rm global}$.  Then $U(1)_Q$ can be embedded in a diagonal $SU(2)_{\rm diag}\subset G_{\rm gauge}\times G_{\rm global}\subset G_{\rm gauge}\times G_{\rm global}$.   In such cases, the computation of the $\kappa_{\Gamma_g,T} $ 't Hooft anomaly coefficient reduces  WWW anomaly calculation \cite{Wang:2018qoy} for this $SU(2)_{\rm diag}$ subgroup.   This method was already mentioned and used in 
 \cite{Wang:2018qoy} to reproduce the $w_2w_3$ anomaly of $SU(2)$ with $N_f=2$ Weyl adjoints~\cite{Cordova:2018acb} from the anomaly in the $SU(2)_{\rm diag}$ of a matter field in the $j=3/2$ representation.   We will review and further discuss this method here, and also a caveat is that this method can fail to exhibit a  possibly non-trivial $w_2w_3$ anomaly if the index of the embedding of $SU(2)_{\rm diag}\subset G_{\rm gauge}\times G_{\rm global}$ is even. 

As in the discussion above, we turn on background gauge field $B_2$ for the one-form center symmetry $\IZ_2\subset \Gamma _g$ and consider a $Spin_{G_{\rm global}}(4)$ structure with spin-charge relation $\Gamma$ that includes  $-\mathds{1}_{G_{\rm global}}\sim (-1)^F$.  We consider the anomaly 
\eq{\label{WeylAnomaly}
\CA\subset\pi i \kappa _{\Gamma _g,T}\int   B_2\cup w_3(TM)~. 
} 
Since such anomalies are $\IZ _2$ valued, they can only be non-trivial if there is a $\IZ_2\subset \Gamma_g$.   If we restrict   to  $SU(2)_{\rm gauge}\subset G_{\rm gauge}$ and $SU(2)_{\rm global}\subset G_{\rm global}$ such that $\IZ_2\subset Z(G_{\rm gauge/global})\mapsto Z(SU(2)_{\rm gauge/global})$,  
then the anomaly will be matched by identifying the fluxes:
\eq{
w_2\big(SU\big(2\big)_{\rm gauge}\big)={B}_2 \quad, \quad w_2\big(SU\big(2\big)_{\rm global}\big)=w_2\big(G_{\rm global}\big)=w_2(TM)~.
}
One can now exhibit anomalies by reducing $SU(2)_{\rm gauge}\times SU(2)_{\rm global}$ to the diagonal $SU(2)_{\rm diag}$ subgroup: the original theory has an anomaly~\eqref{WeylAnomaly} if the reduced theory has a WWW in $SU(2)_{\rm diag}$ with ${B}_2\neq 0$ which vanishes when ${B}_2=0$.   This follows from the fact that when we restrict to the $SU(2)_{\rm diag}\subset SU(2)_{\rm gauge}\times SU(2)_{\rm global}$ theory with $ B_2\neq 0$, we identify:
\eq{
w_2\big(SU\big(2\big)_{\rm diag}\big)=w_2\big(SU\big(2\big)_{\rm gauge}\big)=w_2\big(SU\big(2\big)_{\rm global}\big)=w_2(TM)~,
}
so that the mixed anomaly~\eqref{WeylAnomaly}  reduces to  the WWW $SU(2)$ anomaly \eqref{gravanomaly}:
\eq{
\CA\subset\pi i \kappa _{\Gamma _g,T}\int B_2 \cup w_3(TM)\longmapsto \CA\subset\pi i \kappa _{\Gamma _g,T}\int w_2(TM)\cup w_3(TM)~.
}
So if the original theory had no WWW anomaly for $G_{\rm global}$, and the $SU(2)_{\rm diag}\subset G_{\rm gauge}\times G_{\rm global}$ theory does have a WWW anomaly, then the original theory has $\kappa _{\Gamma _g,T}\neq 0$~\eqref{WeylAnomaly}.

Thus, rather than computing $\kappa _{\Gamma _g,T}$ from the anomaly indicator $\sigma _{\fI}$ by the mod-2 index $\fI$ of the Dirac operator in a $U(1)_Q$ flux background, we can simply use the results of \cite{Wang:2018qoy} which classified which $SU(2)$ representations yield $w_2w_3$ anomalies for fermions coupled to an $SU(2)$ gauge field: when $N_f$ fermions transform under a representation $R$ which pulls back to the $SU(2)$ representation $R\to \bigoplus \mathbf{d}^{\oplus r_d}$, the anomaly indicator is given by 
\eq{
\sigma_\fI(R)=N_f\sum_{n\geq 0} r_{8n+4}~{\rm mod}_2~. 
}
This provides an easy way to determine whether or not a theory has the anomaly \eqref{WeylAnomaly}.

A caveat is that this reduction method can fail to exhibit the anomaly if the embedding of $SU(2)_{\rm gauge/ global}\hookrightarrow G_{\rm gauge / global}$ where $R_{\rm gauge/global}\mapsto \bigoplus_i R^{(i)}_{\rm gauge/global}$ has an even index of imbedding\footnote{The index of embedding also determines if there are instantons in partially broken groups \cite{Intriligator:1995id,Csaki:1998vv}.}, $\mu_{G_{\rm gauge/global}}$:  
\eq{\mu_{G_{\rm gauge/global}}=\frac{\sum_i I_2\left(R^{(i)}_{\rm gauge/global}\right)}{I_2\left(R_{\rm gauge/global}\right)}~, 
} 
where $I_2(R)$ is the Dynkin index. In this scenario, the index of embedding determines the 
minimal Chern (i.e. instanton) number of the $SU(2)_{\rm gauge}\subset G_{\rm gauge}$ sub-bundle, with minimal $k_{G}$ related by 
$k_{SU(2)_{\rm gauge}}=\mu_{G_{\rm gauge}}\, k_{G_{\rm gauge}}$.   
So the index of embedding determines the proportionality constant in $\fI$ as shown in \eqref{index}:
\eq{
\fI(SU(2)_{\rm gauge}\times SU(2)_{\rm global})=\mu_{G_{\rm gauge}}\fI(G_{\rm gauge}\times SU(2)_{\rm gauge})~. 
} 
The anomaly indicator $\sigma_\fI(R)$ in \eqref{OGindex} will be a multiple of $\mu_{G_{\rm gauge}}$ and hence the restriction $G_{\rm gauge}\to SU(2)_{\rm gauge}$ fails to compute the anomaly if the index of embedding is even\footnote{
For non-simply laced $G$, a factor in restricting $G\to SU(2)$ can compensate for an even index of embedding, $\mu_G$, so that the anomaly is not trivialized. This occurs in the $Sp(N_c)$ theories considered in Section \ref{sec:otherexamples}. 
}
\eq{
\mu_{G_{\rm gauge}} \in 2\IZ: \qquad\hbox{anomaly trivialized}.
}

The restriction $G_{\rm global}\to SU(2)_{\rm global}$ where $R_{\rm global}\to \mathbf{2}^{\oplus n_f}$, if  $G_{\rm global}\subset SU(2n_f)$ where again $\mathbf{2n_f}\to R_{\rm global}$, would not trivialize the anomaly. Under this embedding, $\mu_{G_{\rm global}}=n_f$ which is only even when the anomaly identically vanishes.

\subsection{Review of  the $w_2w_3$ Anomaly of $SU(2)$ with 2 Adjoint Weyl Fermions  
}

Consider $SU(2)_{\rm gauge}$ theory with $N_f=2$ massless Weyl fermions in the adjoint representation, i.e. in the $({\bf 3, 2})$ rep of $SU(2)_\text{gauge}\times SU(2)_\text{global}$.   The fields are in faithful representations of
\eq{\label{su2aqcd2}
G_{\rm total}=\frac{SU(2)_{\rm gauge}}{\IZ_2^{(1)}}\times\frac{ SU(2)_{\rm global}\times Spin(4)\times \IZ _{8}}{\IZ_2\times \IZ_2}~,
}
where $\Gamma_g=\IZ ^{(1)}_2$ is the 1-form center global symmetry of $SU(2)_{\rm gauge}$ with matter in only $j\in \IZ$ representations, and the $\IZ_2\times \IZ_2$ quotient is because $-\mathds{1}_{\rm SU(2)_{\rm global}}=(-1)_{\IZ _8}=(-1)^F$.  The  $U(1)_A$ global symmetry that acts on both fermions is broken by instantons to a $\IZ _8$ subgroup, with generator $g$ satisfying $g^4=(-1)^F$.  This theory admits a $Spin _{SU(2)_{\rm global}}(4)$ twisting, which can be used to put the theory on a non-spin manifold via background gauge field flux with $w_2(SU(2)_{\rm global})=w_2(TM)$; this is also used in the Donaldson-Witten twist of ${\cal N}=2$.  The various 't Hooft anomalies of the theory are discussed in detail in  \cite{Cordova:2018acb}.

The $Spin_{SU(2)_{\rm global}}(4)$ theory on $M_4=\ICP ^2$ does not have a purely gravitational 't Hooft anomaly, $\kappa _{T,T}=0$, since $SU(2)_{\rm global}$  has fermions in the $j=\half$ representation whereas only $j=\frac{3}{2}+4r$ with $r\in \IZ$ has the 
$\kappa _{T,T}$ anomaly of \cite{Wang:2018qoy}.   The flux background for $\kappa _{T,T}$ is the minimal $w_2(SU(2)_{\rm global})=w_2(\ICP ^2)$ background of the  $Spin_{SU(2)_{\rm global}}(4)$ theory on $\ICP ^2$, and one can then take $\widehat\varphi = \varphi _{c.c.}\circ W_{\rm global}$, which preserves these fluxes.   The $U(1)_Q^c$ flux charges are $q_i=\pm \half$ (the $SU(2)_{\rm gauge}$ charges do not enter in this $U(1)_Q$) so there are $\fI=0$ fermion zero modes of the Dirac operator, and the anomaly indicator for $\kappa _{T,T}$ is $\sigma _\fI=0$.  

The $\kappa _{\Gamma _g,T}=1$ mixed 't Hooft anomaly between the 1-form center symmetry $\IZ _2^{(1)}$ and geometry was deduced in \cite{Cordova:2018acb}  by interpreting its relation to an anomaly found earlier in \cite{Witten:1995gf} and in the context of the twisted ${\cal N}=2$ theory of the Donaldson-Witten partition function $Z_{DW}$ \cite{Donaldson:1983wm,Donaldson:1990kn,Witten:1988ze,Witten:1994cg,Moore:1997pc}.   The $\Gamma _g ^{(1)}=\IZ _2^{(1)}$ global symmetry of the $SU(2)$ gauge theory is coupled to a background gauge field $B_2$ and the choice $\text{spin}^c$ structure is a lift of $w_2(TM)$ to an integral (vs mod 2) cohomology class $\tilde w_2(TM)$, with $w_2(TM)$ recovered by mod 2 reduction under shifts $\tilde w_2(TM)\to \tilde w_2(TM)+2y$ with $y$ an integral cohomology class.    Under this transformation, however, $Z_{DW}$ is not invariant but transforms as $Z_{DW}\to Z_{DW}(-1)^{B_2\cup y}$ \cite{Witten:1995gf}.  As discussed in  \cite{Cordova:2018acb}, one can add a local counterterm to remove the $y$ dependence, at the expense of introducing dependence on the integral lift $\tildeB_2$ of $B_2$, with $\tildeB_2\to \tildeB_2+2x$ with $x$ an integer 2-cochain leading to $Z_{DW}\to Z_{DW}(-1)^{x\cup w_2(TM)}$; as usual for mixed anomalies, one can preserve one or the other symmetry, shifting between them by a local counterterm, but not both.  As noted in \cite{Cordova:2018acb}, this anomaly is described by inflow from the 5d action
\eq{\label{B2w3} S_5=i\pi \int _{{\cal M}_5} B_2\cup w_3({\cal M}_5)~.
}
We now illustrate using the methods of the previous subsections to compute this anomaly. 

The $Spin_{SU(2)_{\rm global}}$ theory on $\IC\IP^2$ with $\IZ _2^{(1)}$ background $B_2 = w_2(SU(2)_{\rm gauge})$ has both $\Gamma: (-1)^F \sim -\mathds{1}_{SU(2)_{\rm global}}$ flux, $w_2(SU(2)_{\rm global})=w_2(\ICP ^2)$, and $\Gamma _g^{(1)}$ flux for $B_2\neq 0$.  The former can be activated by  a $U(1)_{\rm global}^c\subset SU(2)_{\rm global}^c$ connection 
\eq{
A_{\rm global}=\left(\begin{array}{cc}
\hat{a}_{\IZ_2}&0\\0&-\hat{a}_{\IZ_2}
\end{array}\right)\quad, \quad \frac{\hat{f}_{\IZ_2}}{2\pi}=\half [H]^\vee~,
}
where $H$ is the generator $[H]\in H_2(\ICP^2;\IZ)$.   To exhibit the $\kappa _{\Gamma _g,T}$ anomaly, we activate the $\Gamma ^{(1)}_g$ flux of the background gauge field $B_2=w_2(SU(2)_{\rm gauge})$ via the $U(1)^c_{\Gamma_g}$ connection 
\eq{
A_{\rm gauge}=\left(\begin{array}{cc}
\hat{a}_{\IZ_2}&0\\0&-\hat{a}_{\IZ_2}
\end{array}\right)=A_{\rm global}~.
}
The total flux is along a diagonal $U(1)_Q^c\subset SU(2)_{\rm gauge}^c\times SU(2)_{\rm global}^c$.   The $\kappa _{\Gamma _g,T}$ anomaly is exhibited by a mapping torus with $\widehat \varphi$ acting on the $S^1$ that preserves this total flux:
\eq{
\widehat\varphi=\varphi_{c.c.}\circ W_{\rm gauge}\circ W_{\rm global}~,
}
with $\varphi_{c.c.}$ complex conjugation on $\IC\IP^2$.  Again, $\varphi _{c.c.}$ reverses the orientation of the gauge and global fluxes and acts as a $\pi$ rotation, so $\varphi _{c.c.}^2=(-1)^F$.  The Weyl transformations are 
\eq{
W_{\rm gauge/global}=\left(\begin{array}{cc}
0&1\\-1&0
\end{array}\right)_{\rm gauge/global}~.  
} 
These act as $\pi$-rotations in $SU(2)_{\rm gauge/global}$ around the $\hat{y}$ axis, so $W_{\rm global}^2=(-1)^F$ since the fermions are in the ${\bf 2}$ and $W_{\rm gauge}^2=1$ since the fermions are in the ${\bf 3}$; the upshot is $\widehat \varphi ^2=(-1)^F \cdot (-1)^F=1$, so $\widehat \varphi $ is $\IZ _2$-valued as needed for a mod 2 anomaly $\kappa _{\Gamma _g,T}$.   

The fermion zero modes transform as doublets under the $O(2)$ symmetry group generated by flat $U(1)_Q\subset U(1)_{\rm gauge}\times U(1)_{\rm global}$ transformations and $\widehat \varphi$ and consequently,  $\kappa _{\Gamma _g,T}=\sigma _{\fI}$ $mod_2$ will be given by half the number of zero modes of the 4d Dirac operator in the $U(1)_Q^c$ flux background. To compute this, we will apply the Atiyah-Singer index theorem. Under the restriction  of $SU(2)_{\rm gauge/global}\to U(1)_{\rm gauge/global}$, normalized to give integer charges, the fermions have charges under $U(1)_{\rm gauge}\times U(1)_{\rm global}\mapsto U(1)_Q$ given by
\eq{
\mathbf{3}\otimes \mathbf{2}\longmapsto (2,0,-2)\otimes (1,-1)=\left(\begin{array}{crc}
3&1&-1\\
1&-1&-3
\end{array}\right)
}
Since we have $\half$-integral background fluxes in the $U(1)$s, we apply the $\ICP ^2$ flux background Atiyah Singer index theorem $\fI =\sum _i (4q_i^2-1)/8$ with $\half$ of the above charges, i.e.  
\eq{
q_i=
\left(\begin{array}{crc}
\frac{3}{2}&\half&-\half\\
\half&-\half&-\frac{3}{2}
\end{array}\right)
}
The result is $\fI(\mathbf{3}\otimes \mathbf{2})=2$.  We can also obtain this result using \eqref{index} with $I_2(\mathbf{3})=2$, $Q^1=\half$ and $N_f=2$: $\fI(\mathbf{3}\otimes \mathbf{2})= N_f\, I_2(\mathbf{3})\,\Big(2(Q^1)^2\Big)=2\times 2\times \left(2\times \frac{1}{4}\right)=2~$.  The anomaly indicator is thus non-trivial, $\sigma_\fI=\frac{\fI(\mathbf{3}\otimes \mathbf{2})}{2}~{\rm mod}_2=1$, 
re-deriving the anomaly of~\cite{,Cordova:2018acb} 
\eq{
\CA\subset\pi i \int B_2\cup w_3(TM)~,
}
that was obtained via the Donaldson-Witten anomaly in the shifts of the integer lifts of $B_2$ or $w_2(TM)$~\cite{Donaldson:1983wm,Donaldson:1990kn,Witten:1988ze,Witten:1994cg,Moore:1997pc}.   
This can be easily generalized to $N_f=2n_f$ adjoint Weyl fermions, simply by restricting $G_{\rm global}=SU(N_f)\mapsto SU(2)_{\rm global}$.  The fermions then transform under $
(\mathbf{3}\otimes \mathbf{2})^{\oplus n_f}$ so the index is given by $\fI\left((\mathbf{3}\otimes \mathbf{2})^{\oplus n_f}\right)=2n_f=N_f~,$ which leads to the anomalous phase of the action $(-1)^{n_f}$.  The anomaly indicator is thus given by 
$\sigma_\fI=n_f~{\rm mod}_2~$.  So $SU(2)$ gauge theory with $N_f=2n_f$ adjoint fermions has the 't Hooft anomaly
\eq{\label{SU2AQCDAnom}
\CA\subset\pi i \,n_f \int B_2 \cup w_3(TM)~.
}
This anomaly was also obtained in \cite{Brennan:2022tyl} in connection with symmetry fractionalization. 

The anomaly found in \cite{Cordova:2018acb} for $SU(2)$ with $N_f=2$ adjoint fermions can also be derived by reducing to the WWW $SU(2)$ anomaly \cite{Wang:2018qoy}.   Generalizing to $N_f=2n_f$ adjoint fermions, the fields transform faithfully under 
\eq{
G_{\rm total}=\frac{SU(2)_{\rm gauge}}{\IZ_2}\times \frac{SU(2n_f)_{\rm global}\times Spin(4)\times \IZ _{8n_f} }{\IZ_2\times \IZ _2}~,
}
where there is a $\IZ _2^{(1)}$ global center symmetry and $\IZ _{8n_f}$ has generating group element $g$ that satisfies $g^4=C_{SU(2n_f)}$ where $C_{SU(2n_f)}$ is the generator of the $\IZ _{2n_f}$ center of the flavor group and $g^{4n_f}=C_{SU(2n_f)}^{n_f}=(-1)^F$ are the $\IZ_2$ quotients indicated in $G_{\rm total}$. Neither $SU(2)_{\rm gauge}$ nor $SU(2)_{\rm global}$ separately have a WWW $SU(2)$ anomaly since they have $j\notin \frac{3}{ 2}+4\IZ$  representations.  For $n_f=1$ the $SU(2)_{\rm global}$ has an 't Hooft anomaly analog of the original Witten anomaly of $SU(2)_{\rm global}$, since there are an odd number (3) of $SU(2)_{\rm global}$ doublets; for $n_f>1$, this becomes simply a perturbative, non-zero 't Hooft anomaly: $\Tr SU(2n_f)_{\rm global}^3=3$.    The anomaly~\eqref{SU2AQCDAnom} can be exhibited by reducing $SU(2)_{\rm gauge}\times SU(2n_f)_{\rm global}\to SU(2)_{\rm diag}$. In the restriction $G_{\rm global}\to SU(2)_{\rm global}$, we can choose the embedding $\mathbf{2n_f}\mapsto \mathbf{2}^{\oplus n_f}$ to not trivialize the anomaly, so the fermions transform under $SU(2)_{\rm gauge}\times SU(2)_{\rm global}$ as the $(\mathbf{3},\mathbf{2})^{\oplus n_f}$ representation.  We now couple the fermions to $Spin_{SU(2)_{\rm diag}}$ bundles where the fermions transform with respect to $SU(2)_{\rm diag}\subset SU(2)_{\rm gauge}\times SU(2)_{\rm global}$ as 
\eq{
(\mathbf{3}\otimes \mathbf{2})^{\oplus n_f}\longmapsto \mathbf{4}^{\oplus n_f}\oplus \mathbf{2}^{\oplus n_f}~.
}
When $n_f$ is odd, the theory has both a WWW $SU(2)$ anomaly, due to the odd number of fermions in the $\mathbf{4}$ representation of $SU(2)_{\rm diag}$, and an $SU(2)$ Witten anomaly~\cite{Witten:1982fp} (which is here a 't Hooft anomaly) from the odd number of fermions in the $\mathbf{2}$.   The WWW anomaly of $SU(2)_{\rm diag}$ implies that the $\kappa _{\Gamma _g,T}$ anomaly \eqref{SU2AQCDAnom} is indeed non-trivial for $n_f$ odd.

\subsection{Gauging $\Gamma _g^{(1)}$ subgroups, e.g.  $SO(3)_\text{gauge}$ with 2 Adjoint Weyl Fermions}
\label{sec:SO(3)}

In the following sections, we will mostly consider simply connected $G_{\rm gauge}$ theories.  One can obtain other cases by gauging the one-form global symmetry $\Gamma _g^{(1)}$ or subgroups, i.e. subgroups of the original $G_\text{gauge}$ symmetry.    For example, the 
$SO(3)_\pm$ gauge theories can be obtained from $SU(2)$ gauge theories by gauging the $\IZ _2^{(1)}$ one-form electric global symmetry~\cite{Kapustin:2014gua}.  If the original $SU(2)$ theory has partition function $Z_Q[B]$, one replaces  the background gauge field $B_2=w_2(SU(2))$ with a dynamical $\IZ _2$ gauge field, $B_2\to b_2$, to obtain the $SO(3)_+$ partition function with dual magnetic $\IZ _2^{(1)}$ global symmetry with background gauge field $B_2'$~\cite{Kapustin:2014gua,Lee:2021crt}
\eq{Z_{SQ}[B_2', \tilde w_2(TM)]=\sum _{b_2} (-1)^{\int B_2'\cup b_2}Z_Q[b_2, \tilde w_2(TM)].}
For $SO(3)_\text{gauge}$ theory with $N_f=2$ Weyl fermions in the adjoint representation, it suffices to consider the $SO(3)_+$ theory since the $SO(3)_\pm$ cases are related by $\theta \to \theta +2\pi$.  

Mixed anomalies of the original $Z_Q$ theory involving $B_2$ can lead to 2-group extensions upon gauging\footnote{Other mixed anomaly terms, e.g. $\kappa _{r\Gamma}$ in~\eqref{sect5anomalies}, lead to non-invertible symmetries see e.g.~\cite{Kaidi:2021xfk, Kaidi:2023maf, Bah:2023ymy, Apruzzi:2022rei, Choi:2022zal, Choi:2021kmx, Cordova:2022ieu,  Choi:2022jqy}}, as illustrated in e.g.~\cite{Tachikawa:2017gyf, Benini:2018reh, Hsin:2020nts, Lee:2021crt}.  Consider the effect of the anomaly term~\eqref{SU2AQCDAnom} of the $SU(2)_\text{gauge}$ theory in the  $SO(3)_\text{gauge}$ version.   By choice of counterterm, we can choose the presentation of the anomaly where $Z_Q[\tilde b_2, \tilde w_2(TM)]$ is invariant under $\tilde b_2\to \tilde b_2+2x$, so that we can unambiguously sum over the gauged $\mathbb{Z}_2$-valued $b_2$.  The anomaly affects the shift of $\tilde w_2(TM)$ as $Z_Q[b_2, \tilde w_2(TM)+2y]=(-1)^{\kappa _{\Gamma _g, T} b_2\cup y} Z_Q[b_2, \tilde w_2(TM)]$.  So the $SO(3)$ theory has $Z_{SQ}[\tilde B_2'+2x, \tilde w_2(TM)]=Z_{SQ}[\tilde B_2, \tilde w_2(TM)]$ and 
\eq{\label{Z4extension} Z_{SQ}[B_2', \tilde w_2(TM)+2y]=Z_{SQ}[B_2'+\kappa _{\Gamma _g, T}y, \tilde w_2(TM)].}
If the original theory had vanishing 't Hooft anomaly, $\kappa _{\Gamma _g, T}=0$, then the theory after gauging $\IZ _2^{(1)}$ also has vanishing 't Hooft anomaly.  If the original theory had 't Hooft anomaly $\kappa _{\Gamma _g, T}=1$ then the theory after gauging has a $\IZ _4$ extension of the $\IZ _2\times \IZ _2$ shift symmetries of $\tilde B_2'$ and $\tilde w_2$, as in~\eqref{Z4extension}, which implies $\IZ _4$ symmetry  $\tilde w_2(TM)\to w_2(TM)+4y$.

\section{Determining the $w_2w_3$ Anomalies of Other Theories}
\label{sec:examples}
In this section we exhibit the $w_2 w_3$ 't Hooft anomalies for a variety of gauge theories. 

\subsection{$SU(2)_\text{gauge}$ with $2n_f$ Weyl Fermions in the ${\bf (2j+1})$-Representation} 
\label{sec:su2janomaly}

in this family of theories, the case $j=1$, $N_f=2n_f=2$ is the adjoint representation case considered in~ \cite{Cordova:2018acb}, reviewed in the previous section.  We here generalize to other spin-$j$ $SU(2)$-representations for the fermions,\footnote{For general $N_f$ and $j$ the theory is not asymptotically free in the UV, but could arise as a low-energy effective field theory in the IR.} first for $N_f=2$.  For $j\in \IZ +\half$ there is a $SU(2)_{\rm gauge}\times SU(2)_{\rm flavor}$ symmetry-preserving mass term for the fermions, and thus there can be no anomaly.  We thus consider $j\in \IZ$, and then the gauge invariant fermion bilinear is in the adjoint of $SU(2)_{\rm flavor}$ so there is no symmetry-preserving mass term.   The fields couple to $G_{\rm total}$-bundles where:\footnote{We omit writing the $\IZ _{4n_f I_2(R_j)}\subset U(1)_A$, where  $2I_2(R_j)=2j(j+1)(2j+3)/3$ is the number of fermion zero modes for each flavor in representation $R_j$ in an instanton background.   Its generator $g$ has $g^{2I_2(R_j)}=C_{SU(2n_f)}$ where $C$ generates the $\IZ _{2n_f}$ center of $SU(2n_f)$, with $C^n_f=-1=(-1)^F$.}
\eq{
G_{\rm total}=\frac{SU(2)_{\rm gauge}}{\IZ_2}\times \frac{SU(2n_f)_{\rm global}\times Spin(4)}{\IZ_2}~. 
}
The theory classically admits a $Spin_{SU(2n_f)_{\rm global}}(4)$ structure, allowing us to consider the theory on $\ICP^2$ upon twisting by $SU(2n_f)$ flux, taking $w_2(SU(2N_f))=w_2(\ICP^2)$.  Since $j\in \IZ$ the theory has unbroken $\IZ _2^{(1)}$ one-form center symmetry, which we can couple to background field $B_2$. 

There is no purely gravitational anomaly, $\kappa _{T,T}=0$, and the $\kappa _{\Gamma _g}$ anomaly is non-zero if there are 2 mod 4 fermion zero modes of the  Dirac operator on $\ICP ^2$ in the flux background with  $w_2(SU(2N_f)) =w_2(\ICP^2)$ and $B_2\neq0$.  This can be computed from the Atiyah Singer index by decomposing the fermions under the diagonal $U(1)_Q\subset U(1)_{\rm gauge}\times U(1)_{\rm global}$, with $U(1)_{\rm gauge}\subset SU(2)_{\rm gauge}$ and $U(1)_{\rm global} \subset SU(2)_{\rm flavor}$ and half-integral flux in each for the non-zero $w_2(SU(2)_{\rm global})$ and $w_2(SU(2)_{\rm gauge})$.  The Dirac index is $\fI = \sum _i (4q_i^2-1)/8$, where for each $n_f$ the $q_i$ run over $m\pm \half$ with $m=j, j-1, \dots -j$, so $\fI = n_f\sum _{m=0}^j m^2=n_f T_2(R_j)
$ the Dynkin index of the $j$-representation:
\eq{
\fI(R_j\otimes \mathbf{2n_f})=n_f\times \frac{j(j+1)(2j+1)}{3}~. 
}
The $\IZ _2$ anomaly coefficient is $\sigma_\fI=\half \fI$ mod 2, i.e.
\eq{
\sigma_\fI= n_f\frac{j(j+1)(2j+1)}{6}~{\rm mod}_2~. 
}
It is easily seen that this implies that the anomaly indicator is given by 
\eq{
\sigma_\fI=\begin{cases}
n_f~{\rm mod}_2&j=1,2~{\rm mod}_4\\
0&j=0,3~{\rm mod}_4
\end{cases}
}
Thus the anomaly polynomial is non-trivial iff $n_f$ is odd and $j=1,2, ~{\rm mod}_4$:
\eq{\label{anomalysu2j} 
\CA\subset\begin{cases}
\pi i \,n_f\int B_2\cup  w_3(TM)&j=1,2~{\rm mod}_4\\
0 &j=0,3~{\rm mod}_4
\end{cases}
}

The result~\eqref{anomalysu2j} can also be derived by reducing to the  WWW $SU(2)$ anomaly, as discussed for the $j=1$ case in \cite{Wang:2018qoy} and reviewed in the previous section.  We project onto the diagonal subgroup 
$SU(2)_{\rm diag}\subset SU(2)_{\rm gauge}\times SU(2n_f)$ corresponding to $B_2=w_2(SU(2n_f))=w_2(TM)$.   The fermions decompose under the $SU(2)_{\rm diag}$ via the $SU(2)$ rep tensor product decomposition:  
\eq{\label{su2diagreps}
(\mathbf{2j+1})\otimes \mathbf{2n_f}\to \left( (\mathbf{2j+1})\otimes \mathbf{2}\right) ^{\oplus n_f}=(\mathbf{2j+2})^{\oplus n_f}\oplus (\mathbf{2j})^{\oplus n_f}~. 
}
The WWW anomaly  \cite{Wang:2018qoy} is non-zero  iff there are an odd number of Weyl fermions in the representations with $j=\frac{3}{2}$ mod 4, i.e. of dimension 4 mod$_8$.  The original $SU(2)_{\rm gauge}$ matter content is anomaly free and the $SU(2)_{\rm global}\subset SU(2n_f)_{\rm global}$ has an odd number of doublets for $n_f$ odd (since $2j+1$ is taken to be odd), so it has the original $SU(2)$ anomaly, but not the WWW anomaly.  On the other hand, $SU(2)_{\rm diag}$ has the WWW anomaly for $j=1,2$ mod$_4$ and $n_f$ is odd, reproducing the anomaly \eqref{anomalysu2j}.

\subsection{$SU(N_c)_{\rm gauge}$ theory with $2n_f$ Weyl Fermions in the Adjoint Representation}

\label{sec:AQCDanom}

In these theories, the fields couple to $G_{\rm total}$ bundles (we suppress the $\IZ _{2N_cN_f}\subset U(1)_A$ \footnote{Its generator $g$ has $g^{2N_c}=C$, the center generator of $SU(2n_f)$, and $g^{2n_cn_f}=C^{n_f}=(-1)^F$.}): 
\eq{
G_{\rm total}=\frac{SU(N_c)}{\IZ_{N_c}}\times \frac{SU(2n_f)\times Spin(4)}{\IZ_2}~. 
} 
We take $N_f=2n_f$ here for a generalized $Spin _{SU(2n_f)}(4)$ twisting, with non-zero $w_2(SU(2n_f))=w_2(TM)$ flux background for the theory on $\ICP ^2$.  There is a $\IZ_{N_c}^{(1)}=Z(SU(N_c))$ 1-form symmetry associated with the center of $SU(N_c)$. As we discussed, the $w_2w_3$ type 't Hooft anomalies under consideration can only be $\IZ _2$-valued and such anomalies involving the $\IZ_{N_c}^{(1)}$ and $Spin_{G_{\rm global}}$ structure  can only be non-trivial  if $\IZ_{N_c}^{(1)}$ has a $\IZ _2^{(1)}$ subgroup.  The $w_2w_3$ 't Hooft anomalies thus only arise if $N_c$ is even, so we write $N_c=2n_c$ and  
we denote the background gauge field coupled to the $\IZ_2^{(1)}\subset \IZ_{2n_c}^{(1)}$ as $B_2$.

To exhibit the $\IZ _2$-valued anomaly, let us turn on the  $B_2$ background gauge field. This can be activated by a $U(1)_{\Gamma_g}^c$ connection  
which is generated by 
\eq{\label{QsuN}
Q _{\Gamma_g}= 
\frac{1}{2} {\rm diag}(\underbrace{1,...,1}_{n_c},\underbrace{-1,...,-1}_{n_c})~,
}
acting on the fundamental representation. 
The Atiyah Singer index theorem in the $w_2(SU(2n_f))=w_2(TM)$ flux background on $\ICP ^2$ then gives for the number of adjoint fermion zero modes: $\fI=\sum _i (4q_i^2-1)/8=N_f N_c \Tr _\text{fund}Q^2$ where we used $I_2({\rm adj})=2N_cI_2({\rm fund})$.  The $w_2w_3$ anomaly indicator is thus 
\eq{\label{suadjind}
\sigma_\fI=\half \fI ~{\rm mod}_2=n_f n_c~{\rm mod}_2~. 
}
Hence adjoint $SU(2n_c)$ QCD has an anomaly term
\eq{\label{anomalysuN}
\CA\subset\pi i\,n_f n_c \int {B_2}\cup w_3(TM)~. 
} 
The anomaly is thus non-trivial only if $n_c$ and $n_f$ are both odd, i.e. if $N_c=4n+2$ for $n\in \IZ$ and $N_f=4m+2$ for $m\in \IZ$.

The $w_2w_3$ anomaly in the $Spin_{G_{\rm global}}(4)$ twisted theory can be seen from the WWW  $SU(2)$ anomaly \cite{Wang:2018qoy} in a $SU(2)_{\rm diag}\subset SU(2)_{\rm gauge}\times SU(2)_{\rm global}$ with appropriately chosen $SU(2)_{\rm gauge}\subset G_{\rm gauge}$ and $SU(2)_{\rm global}\subset G_{\rm global}$ 
so that the $\IZ_2\subset Z(G_{\rm gauge/global})\mapsto Z(SU(2)_{\rm gauge/global})$.  
Such embeddings will 
will only be non-trivial for $n_c$ odd. We choose the  embedding $SU(2)_{\rm gauge/global}\subset SU(2n_c)_{\rm gauge/global}$ where $\mathbf{2n_c}\mapsto \mathbf{2}^{\oplus n_c}$ (and similarly for $n_f$).  Note that the index of the embedding of $SU(2n_c)_{\rm gauge}\to SU(2)_{\rm gauge}$ is $\mu = n_c$, so this would trivialize the anomaly for $n_c$ even.  
 Restricting to $SU(2)_{\rm diag} \subset SU(2)_{\rm gauge}\times SU(2)_{\rm global}$ gives the restriction 
\eq{
B_2=w_2(SU(n_f))=w_2(TM)~. }
The $SU(2n_c)$ adjoint fermions decompose under the $SU(2)\subset SU(2n_c)$ as  
\eq{
\mathbf{adj}=\mathbf{2n_c}\otimes \overline{\mathbf{2n_c}}-\mathbf{1}\longrightarrow  \mathbf{2}^{\oplus n_c}\otimes  \mathbf{2}^{\oplus n_c}-\mathbf{1}=\mathbf{3}^{\oplus n_c^2}\oplus \mathbf{1}^{\oplus n_c^2-1}~.
}
So the fermions decompose under $SU(2)_{\rm diag}$ as 
\eq{
\mathbf{adj}\otimes \mathbf{2n_f}\longrightarrow\left[\left(\mathbf{3}^{\oplus n_c^2}\oplus \mathbf{1}^{\oplus n_c^2-1}\right)\otimes \mathbf{2}\right]^{\oplus n_f}=\mathbf{4}^{\oplus n_f n_c^2}\oplus \mathbf{2}^{\oplus n_f (2n_c^2-1)}~.
}
So $SU(2)_{\rm diag}$ has the $w_2w_3$ 't Hooft anomaly for $n_f n_c^2$ odd, i.e. for both $n_f$ and $n_c$ odd, reproducing the anomaly \eqref{anomalysuN}.  The odd number of $SU(2)_{\rm diag}$ doublets implies that the theory also has a $\IZ _2$ 't Hooft anomaly associated with the original $SU(2)_{\rm global}$ Witten anomaly.

\subsection{$Spin(2n_c+1)_{\rm gauge}$ with $2n_f$ Weyl Fermions in the Vector  Representation} 
\label{sec:spinoddvec}

In these theories, the fields couple to $G_{\rm total}$-bundles with (suppressing the $\IZ _{4n_f}\subset U(1)_A$)
\eq{
G_{\rm total}= \frac{Spin(2n_c+1)}{\IZ_2}\times \frac{SU(2n_f)\times Spin(4)}{\IZ_2}~. 
}
We can put the theory on $\ICP^2$ via twisting $w_2(SU(2n_f))=w_2(TM)$ and can turn on background field $B_2$ for the $\IZ_2^{(1)}$ center symmetry.  This $\IZ _2^{(1)}$ can be 
chosen to be generated by the Cartan element $Q_{\Gamma_g}=\half H_{n_c}=\alpha_{n_c}^\ast$ where $H_{n_c}$ is the co-root that is twice the dual of the short root of $\fs\fo(2n_c+1)$.   The Atiyah Singer index theorem in this flux background then gives\footnote{$\mathfrak{so}(2N+1)$ is a non-simply laced Lie algebra. This means that the formula relating the trace to inner products on the root space via the Cartan matrix differs along short roots. For $\mathfrak{so}(2N+1)$, this differs for the coroot $H_N^\vee=2\alpha_N$. So that $\Tr_{\rm vec}[H_N^2]=4 (\alpha_N,\alpha_N)=4C_{NN}=8$. } 
\eq{\label{SOodd}
\fI=n_f \Tr\left[\frac{1}{4}H_{n_c}^2\right]=n_f (\alpha _{n_c}, \alpha _{n_c})=2n_f \quad \Longrightarrow \quad \sigma_\fI=n_f~. 
}
This leads to the anomaly 
\eq{\label{sooddanomaly}
\CA\subset\pi i \, n_f\int B_2\cup  w_3(TM)~. 
}
For $n_c=1$, this matches the $SU(2)$ with adjoints case reviewed in Section \ref{sec:AQCDanom}.

We can alternatively demonstrate the anomaly by reducing to the WWW $SU(2)$ anomaly in some appropriately chosen $SU(2)_{\rm diag}$. 
 Let us restrict to $SU(2)_{\rm diag} \subset Spin(2n_c+1)\times SU(2n_f)$ where the vector representation of $Spin(2n_c+1)$ maps to the $(\mathbf{2n_c+1})$ representation of $SU(2)$ and $SU(2)$ is embedded diagonally in $SU(2n_f)$. This restriction identifies 
\eq{
w_2(\Gamma_g)=w_2(\Gamma)=w_2(TM)~.
}
 Under this restriction, the fermions decompose as 
\eq{
\left[(\mathbf{2n_c+1})\otimes \mathbf{2}\right]^{\oplus n_f}=\left[\mathbf{2n_c}\oplus (\mathbf{2n_c+2})\right]^{\oplus n_f}
}
which demonstrates that this theory has the WWW $SU(2)$ anomaly for $n_c=1,2$ mod$_4$ and $n_f$ odd.  The index of embedding, $\mu$, is the ratio of the Dynkin index in the $(\mathbf{2n_c+1})$ irrep of $SU(2)$ to the 
 $(\mathbf{2n_c+1})$ of $SO(2n_c+1)$: 
\eq{
\mu=\frac{n_c(n_c+1)(2n_c+1)}{6}~.
}
The anomaly is trivialized by the embedding if $\mu$ is even, i.e. if $n_c=0,3~{\rm mod}_4$.   To avoid trivializing the anomaly for $n_c=0,3$ mod$_4$, we can choose a different embedding of $Spin(2n_c+1)\to SU(2)$, e.g. we can take $\mathbf{(2n_c+1})\to (\mathbf{2n_c-3})\oplus \mathbf{1}^{\oplus 4}$.   This embedding effectively replaces $n_c\to n_c-2$ as compared with the previous one, and in particular the index of embedding differs from that above by this replacement.  So the anomaly with this new embedding is no longer trivialized for $n_c=0,3$ mod$_4$ (it is instead trivialized for $n_c=1,2$ mod$_4$).  Indeed, the fermions decompose under the new embedding as 
\eq{
\left[\left(\mathbf{2n_c-3}\oplus \mathbf{1}^{\oplus 4}\right)\otimes \mathbf{2}\right]^{\oplus n_f}=\left[\mathbf{(2n_c-4)}\oplus \mathbf{(2n_c-2)} \oplus \mathbf{2}^{\oplus 4} \right]^{\oplus n_f}~,
}
which exhibits WWW $SU(2)$ anomaly for $n_c=0,3$ mod$_4$ and $n_f$ odd. Together these two embeddings exhibit the anomaly of \eqref{sooddanomaly} for all $n_c$. 

\subsection{$Spin(2n_c)_{\rm gauge}$ with $2n_f$ Weyl Fermions in the Vector Representation} 

\label{sec:SpinvecQCD}

The fields in this case couple to $G_{\rm total}$-bundles where (suppressing the $\IZ _{4n_f}\subset U(1)_A$)
\eq{
G_{\rm total}= \frac{\frac{Spin(2n_c)}{\IZ^{(1)}_2}\times SU(2n_f)\times Spin(4)}{\IZ^F_2\times \IZ^C_2}~. 
}
The $\IZ_2^F\times \IZ_2^C$ identify $\IZ_2^F:~-\mathds{1}_{{2n_f}}\sim (-1)^F$ and  $\IZ_2^C:~(-1)^F\sim z\in Z(Spin(2n_c))$ where here $z^2=\mathds{1}_{Spin(2n_c)}$ for $n_c$ even and $z^2=-\mathds{1}_{Spin(2n_c)}\in \Gamma_g$ for $n_c$ odd.   The $\IZ _2^C$ identification implies that $(-1)^F$ is effectively gauged, so the theory is bosonic and admits a $Spin_{Spin(2n_c)}(4)$ structure.  The $\IZ _2^F$ identification implies that there is also a $Spin_{SU(2n_f)}(4)$ structure. The theory on $\ICP ^2$ can be realized with either of these generalized spin structures.  For odd $n_c$, the 
$\IZ _2^{(1)}$ center symetry is actually in a 2-group with $(-1)^F$ because a Wilson line $W$ in the spinor rep of the gauge group has $W^2$ which is charged under $\IZ_2^C$ (while being uncharged under $\IZ_2^F$), which can only be screened by the dynamical fermions at the expense of transforming non-trivially under $(-1)^F$; this implies that $\delta w_2(\Gamma _g)=\beta w_2(\IZ_2^C)$~\cite{Hsin:2020nts,Lee:2021crt}.\footnote{Alternatively, there is a 2-group between the $\IZ_2^{(1)}$ and $SU(2n_f) $ since $W^2$ is also charged under $-\mathds{1}_{SU(2n_f)}$: this is equivalent to the above-mentioned 2-group due to the quotient by $\IZ_2^F$ which relates $-\mathds{1}_{SU(2n_f)}\sim (-1)^F$. }

We first consider $Spin_{SU(2n_f)}(4)$ backgrounds, with $w_2(SU(2n_f))=w_2(TM)$ flux and turn on a background flux for the $\IZ _2^{(1)}$ global symmetry, $B_2$.  We can restrict to $G_{\rm total}$-bundles that lift to
\eq{
G'=\frac{Spin(2n_c)}{\IZ_2}\times \frac{SU(2n_f)\times Spin(4)}{\IZ^F_2 }~. 
}
We use the index theorem as in~\eqref{index} to compute the anomaly. The $B_2$ flux is embedded along $U(1)_{\Gamma_g}$ which is generated by 
\eq{
Q_{\Gamma_g}=\half\left(H_{n_c-1}+(-1)^{n_c}H_{n_c}+2\sum_{I=1}^{ \lceil\frac{n_c}{2}\rceil-1} H_{2I-1}\right)~.
}
So for the index~\eqref{index} we compute 
\eq{
\Tr_{\rm vec}[Q_{\Gamma_g}^2]= 2\left(\left\lfloor\frac{n_c}{2}\right\rfloor-1\right)+\begin{cases}
 \Tr_{\rm vec}\left[(H_{n_c-2}+\half H_{n_c-1}-\half H_{n_c})^2\right]& n_c\text{ odd}\\
\Tr_{\rm vec}\left[\left(\half H_{n_c-1}+\half H_{n_c}\right)^2\right]& n_c\text{ even}
 \end{cases}
}
This can be computed via an $Spin(6)\subset Spin(2n_c)_{\rm gauge}$ in terms of its Cartan matrix.  
The result for the number of fermion zero modes and the anomaly indicator for $n_c$ odd is
\eq{
\fI=2n_f n_c\quad\Longrightarrow\quad \sigma_\fI=n_f ~{\rm mod}_2~.
}
For $n_c$ even the index is $\fI=2n_f(n_c-1)$ and the anomaly indicator is  again $\sigma_\fI=n_f~{\rm mod}_2$. 
We conclude that $Spin(2n_c)$ gauge theory with $2n_f$ fermions, for all $n_c$, has an anomaly 
\eq{\label{spin2nanomaly}
\CA\supset \pi i \, n_f\int B_2\cup  w_3(TM)~.
}

We can additionally show that there are no other anomalies when we consider $Spin_{SU(2n_f)}(4)$ structure. The allowed, independent anomalies (without activating $w_2(\IZ_2^C)$ are 
\eq{
\CA\subset\pi i \,n_f\int B_2\cup w_3(TM)+\pi i \kappa_{T,T}\int w_2(\IZ_2^F)\cup w_3(\IZ_2^F)~.
}
Again, there is no independent $\kappa_{\Gamma_g,\Gamma_g}$ anomaly because it is redundat with the $B_2w_3$-anomaly. Now using the fact that $w_2(\IZ_2^F)=w_2(SU(2n_f))=w_2(TM)$, we see that the only additional anomaly is the purely gravitational anomaly for the $Spin_{SU(2n_f)}(4)$ structure. We can compute $\kappa_{T,T}$ by putting the the theory on $\IC\IP^2$, with flux along $w_2(SU(2n_f))=w_2(TM)$. In this background, the fermion charges are 
\eq{
2n_c\otimes 2n_f\longmapsto \Big(\half,-\half\Big)^{\oplus 2n_c n_f}~. 
}
The Atiyah Singer index theorem then shows that there are no fermion zero-modes. Therefore, there is no anomaly in this background and $\kappa_{T,T}=0$.

Now let us consider the anomalies for $Spin_{Spin(2n_c)}(4)$ backgrounds. First take the case where $n_c$ is even. For these bundles we can again turn on fluxes to activate the anomaly \eqref{spin2nanomaly} -- the computation follows directly as above. The allowed, independent anomalies are 
\eq{
\CA\subset\pi i \,n_f\int B_2\cup w_3(TM)+\pi i \kappa_{T,T}^\prime \int w_2(\IZ_2^C)\cup w_3(\IZ_2^C)~,
}
which is the same as the $Spin_{SU(2n_f)}(4)$ case except that  $\IZ_2^C$ has replaced the role of $\IZ_2^F$. Again, we can identify the $w_2(\IZ_2^C)w_3(\IZ_2^C)$ anomaly as the purely gravitational anomaly since $w_2(\IZ_2^C)=w_2(TM)$. We can compute $\kappa_{T,T}^\prime$ directly by  putting the theory on $\IC\IP^2$ and turning on 1-form flux along 
generated by
\eq{Q=&
\half \sum_{I=1}^{ \frac{n_c}{2}} H_{2I-1}~.
}
Now the fermions decompose into $U(1)_{Q}$ representations as $2n_c\otimes 2n_f\to \Big(\half,-\half\Big)^{\oplus 2n_c n_f}$, so the Atiyah Singer index theorem leads to $\fI=0$ zero-modes and hence $\kappa_{T,T}^\prime=0$.

For odd $n_c$, the $Spin_{Spin(2n_c)}(4)$ structure does not permit discrete flux in $SU(2n_f)$. Additionally, the fact that $\Gamma_g^{(1)}$ forms a 2-group with $(-1)^F$ forces the anomalies to be that of a $\IZ_4$-valued discrete flux  where we cannot use our mod-2 index to probe the anomalies. For completeness, we compute these anomalies using the fractionalization technique \cite{Brennan:2022tyl} in Appendix \ref{app:fractionalization} and show that there is no anomaly for the $Spin_{SU(2n_c)}(4)$ background. 

For all $n_c$, we find for these theories that there is an anomaly 
\eq{
\CA\subset\pi i \,n_f\int B_2\cup w_3(\IZ_2^F)~. 
}

\subsection{$Spin(N_c)_{\rm gauge}$ with $2n_f$ Weyl 
Fermions in the Adjoint Representation}
\label{sec:spinaqcd}

The fields couple to $G_{\rm total}$-bundles with (suppressing the $\IZ _{4h^\vee _{G_{\rm gauge}}n_f}\subset U(1)_A$)
\eq{
G_{\rm total}=\frac{Spin(N_c)}{\Gamma_g}\times \frac{SU(2n_f)\times Spin(4)}{\IZ_2}~,
}
where the one-form symmetry group $\Gamma _g^{(1)}$ is the center symmetry
\eq{
\Gamma_g=\begin{cases}
\IZ_2&N_c=2n_c+1\\
\IZ_2\times \IZ_2&N_c=4n_c\\
\IZ_4&N_c=4n_c+2~.
\end{cases}
}
Again, our $\widehat \varphi$ flux background computation can only activate a $\IZ_2\subset \Gamma _g$ flux. As discussed in Section \ref{sec:w2w3gen} the anomaly for $Spin(4n_c+2)$ adjoint QCD vanishes due to the fact that the center of $Spin(4n_c+2)$ is $\IZ_4$. We check this by the use of fractionalization techniques developed in \cite{Brennan:2022tyl} in Appendix \ref{app:fractionalization} and indeed find that there is no additional anomaly associated with $\IZ _4$ flux. The $\IZ _2$ flux obstructs lifting $SO(N_c)$ to $Spin(N_c)$ and can be embedded along $U(1)_{\Gamma_g}$ which is generated by 
\eq{
Q_{\Gamma_g}=\begin{cases}
\half H_{n_c}\quad, \quad  N_c=2n_c+1~,\\
\frac{1}{2}\left(H_{n_c-1}-H_{n_c}+2\sum_{I=1}^{ \frac{n_c-1}{2}} H_{2I-1}\right)&N_c=4n_c+2~,
\end{cases}
}
and, for $N_c=4n_c$, we choose to embed the $\IZ _2$ flux along $U(1)_{\Gamma_g}^L$ which is generated by $Q_{\Gamma_g}^L$ where 
\eq{
Q_{\Gamma_g}^L=\half \sum_{I=1}^{ \frac{n_c}{2}} H_{2I-1}\quad, \quad Q_{\Gamma_g}^R=\half H_{n_c}+\half\sum_{I=1}^{\frac{n_c}{2}-1}H_{2I-1}~.
}  
We can write the above charges in the fundamental representation and then use $\Tr_{adj}=(N_c-2)\Tr_{vec}$ in computing the index for the adjoint fermions.  The Atiyah Singer index theorem then gives 
\eq{
\fI=N_f\Tr_{adj}[Q_{\Gamma_g}^2]=\begin{cases}
2n_f(N_c-2) & N_c=2n_c+1\\
2n_f(4n_c-2)(2n_c-1)&N_c=4n_c\\ 
2n_f(4n_c)(2n_c+1)&N_c=4n_c+2 
\end{cases}
}
The anomaly indicator $\sigma _\fI = \half \fI ~{\rm mod 2}$ coefficient thus gives for the anomaly theory 
\eq{\label{spinadjointnodd}
\CA\subset\begin{cases} n_f\pi i \int B_2\cup w_3(TM)\quad, \quad &N_c=2n_c+1~,\\
0, \quad \quad &N_c=2n_c~.
\end{cases}
} 

We can also check the anomalies for $Spin(N_c)$ gauge theory by reducing to the WWW anomaly in a  $SU(2)_{\rm diagonal}\subset SU(2)_{\rm gauge}\times SU(2)_{\rm global}$, where we choose a restriction of  $SU(2n_f)\to SU(2)_{\rm global}$ and $Spin(N_c)\to SU(2)_{\rm gauge}$ that ideally does not trivialize the anomaly. For 
$SU(2n_f)\to SU(2)_{\rm global}$, we choose the embedding  where $\mathbf{2n_f}\mapsto \mathbf{2}^{\oplus n_f}$.  We will here only discuss the case of $N_c=2n_c+1$ odd; similar considerations verify that there is no anomaly for $N_c$ even, in agreement with~\eqref{spinadjointnodd}.   We can take the $Spin(2n_c+1)\to SU(2)_{\rm gauge}$ restriction where $\mathbf{N_c}\to \mathbf{N_c}$, since for $N_c=2n_c+1$ this restriction preserves the center.  The trivialization of the anomaly is controlled by the index of embedding which is here 
\eq{
\mu 
=\frac{N_c(N_c^2-1)}{24}=\frac{n_c(n_c+1)(2n_c+1)}{6}\quad, \quad N_c=2n_c+1~.
}
So the anomaly is trivialized for $N_c=1,7~{\rm mod}_8$ and detectable for $N_c=3,5~{\rm mod}_8$.  For the trivialized case, we take another embedding below.  The adjoint representation of $Spin(N_c)$ is the anti-symmetric tensor product ($\widehat\otimes $) of vector representations, so it decomposes under the 
 $\mathbf{N_c}\to \mathbf{N_c}$ embedding of  $Spin(N_c)\to SU(2)_{\rm gauge}$ as
 \eq{
{\rm Adjoint} = \mathbf{N_c}\widehat\otimes \mathbf{N_c}\longmapsto  
j=1\oplus 3\oplus...\oplus N_c-2  
}
Using the results from Section \ref{sec:su2janomaly}, we indeed find that the theory has the anomaly for  $N_c=3,5~{\rm mod}_8$, matching the result \eqref{spinadjointnodd} for those cases where the index of the embedding does not trivialize the anomaly. 

For the cases $N_c=1,7~{\rm mod}_8$, we can exhibit the anomaly  \eqref{spinadjointnodd} by choosing different embeddings, to avoid trivializing the anomaly.  For  $N_c=8n_c+1$, we can pick the $SU(2)_{\rm gauge}\subset Spin(N_c)$ sub-bundle where $\mathbf{8n_c+1}\to (\mathbf{8n_c-3})\oplus \mathbf{1}^{\oplus 4}$, which has odd index of embedding 
$\mu=\frac{(2n_c-1)(4n_c-1)(8n_c-3)}{3}\in 2\IZ+1~$.  Decomposing the adjoint representation, we verified that for $n_f$ odd there are indeed an odd number of $SU(2)_{\rm gauge}$ representations that contribute to the anomaly, reproducing the  result \eqref{spinadjointnodd}.  Similarly, for $N_c=8n_c+7$, we can choose the embedding $SU(2)_{\rm gauge}\subset Spin(N_c)$ where $\mathbf{8n_c-1}\mapsto (\mathbf{8n_c-5})+1^{\oplus 4}$, which has odd index of embedding, 
$\mu=\frac{(2n_c-1)(4n_c-3)(8n_c-5)}{3}\in 2\IZ+1~,$
so the anomaly is not trivialized.  Again, we verify that the decomposition of the adjoint yields $SU(2)_{\rm gauge}$ representations with anomaly matching  the results of \eqref{spinadjointnodd}.  

\subsection{Other $G_{\rm gauge}$ Theories with $ 2n_f$ Weyl Fermions in the Adjoint Representation}
\label{sec:otherexamples}

The theories can be put on $M_4=\ICP ^2$ using the $Spin_{SU(2n_f)}(4)$ structure, with $w_2(SU(2n_f))=w_2(TM)$.  The gauge groups with a $\IZ _2^{(1)}\subset \Gamma _g^{(1)}$ subgroup of the one-form center symmetry can have the 't Hooft anomaly under discussion.  In particular, $G_{\rm gauge}=Sp(N_c)$, for $N_c=1,2$ mod$_4$, and $G_{\rm gauge}=E_7$ have the anomaly 
\eq{\label{adjGanomaly}
\CA\supset \pi i \, n_f\int B_2\cup  w_3(TN)~.
}
We can obtain these results for $G_{\rm gauge}=Sp(N_c)$ by computing the Index of the Dirac operator on $\ICP^2$ with a background $B_2$ gauge field turned on for the $\IZ_2^{(1)}$ center symmetry. Applying the Atiyah-Singer Index theorem, one finds 
\eq{
\fI=n_f N_c(N_c+1)\quad \Longrightarrow \quad \sigma_\fI=\begin{cases}
n_f ~{\rm mod}_2&N_c=1,2~{\rm mod}_4\\
0&N_c=0,3~{\rm mod}_4
\end{cases}
}
reproducing the anomaly above. This matches the results from Sections \ref{sec:su2janomaly} and \ref{sec:spinaqcd} for the case of $Sp(1)=SU(2)$ and $Sp(2)=Spin(5)$.  

For $E_7$, we can take the restriction $E_7\to SU(2)$ where
\eq{
\mathbf{56}\longmapsto \mathbf{2}^{\oplus 5}+\mathbf{4}^{\oplus 7}+\mathbf{6}^{\oplus 3}~,
}
which maps the center $\IZ_2=Z(E_7)\mapsto \IZ_2=Z(SU(2))$. This restriction has an index of embeddeing $\mu_{E7}=15$. Under this restriction, the adjoint representation restricts
\eq{
\mathbf{133}\longmapsto \mathbf{1}^{\oplus 3}+\mathbf{3}^{\oplus 15}+\mathbf{5}^{\oplus 10}+\mathbf{7}^{\oplus 5}~. 
}
Using our results from Section \ref{sec:su2janomaly}, we see that this does indeed have an anomaly \eqref{adjGanomaly}.

\section{$\CC$ Symmetry, Higgsing, and Abelian Gauge Theory}
\label{sec:abelian}

The  $w_2w_3$ 't Hooft anomalies must of course match upon continuous, symmetry-preserving deformations.  For example, we can preserve the generalized $Spin_{G}(4)$ structure via added scalars in real representations of $G_{\rm gauge}$.  Giving the added scalars masses RG flows to our original theories, with the added scalars decoupled.  Alternatively, the added scalars can have a potential that leads them to have non-zero expectation values, Higgsing the gauge group.  The 't Hooft anomalies must be matched in the Higgsed theory.   For scalars in the adjoint representation of the gauge group, the deformed theory can flow to an abelian gauge theory in the IR.    Such deformations can also be used to give masses to the fermions by adding Yukawa couplings to the scalar field that gets condenses.  We will here consider cases without such Yukawa couplings, with massless fermions in the low-energy theory. 

\subsection{Cases where $\widehat \varphi$ Includes Charge Conjugation Symmetry $\CC$}

\label{sec:AnomalyCSymmetry}

The WWW method~\cite{Wang:2018qoy} of determining the $w_2w_3$ type anomaly involved  the classical global symmetry $\widehat\varphi= \varphi _{c.c.} \circ W$, with e.g. $W\in SU(2)$ to preserve the flux background.  In the theory with added adjoint scalar field(s), we can Higgs $SU(2)\to U(1)$ and as discussed in~\cite{Wang:2018qoy} the anomaly of the $SU(2)$ theory then connects to that of all fermion electrodynamics.  Upon Higgsing $SU(2)\to U(1)$, the element $W\notin U(1)$ and instead it is similar to charge conjugation.  We will here discuss some general aspects about cases when the analog of the $\widehat \varphi$ map needs to incorporate an element that is not a Weyl symmetry but, instead,  the outer automorphism of charge conjugation, $\CC$, to obtain a symmetry of the bosonic flux background.  We will apply the discussion to cases where the gauge group is Higgsed to an Abelian subgroup.  We thus  consider $\widehat\varphi=\varphi _{c.c.} \circ  W_{\rm global}\circ \CC$ that preserves discrete flux backgrounds 
for  $G_{\rm total}$-bundles:
\eq{
G_{\rm total}=\frac{\frac{G_{\rm gauge}}{\Gamma_g}\times G_{\rm global}\times Spin(4)}{\Gamma}~.
} 

Suppose e.g. that the fermions transform as $R_{\rm gauge}\otimes R_{\rm global}$ and are pseudo-real and $R_{\rm gauge}$ is real.  
To probe for an anomaly involving $B_2\cup w_3(TM)$ we  turn on $B_2$ flux background in addition to the flux background needed for the $Spin_{G_{\rm global}}$ structure.  For 
the mapping torus construction, $\widehat\varphi$ is given by $\widehat \varphi = \varphi _{c.c.}\circ W_{\rm gauge}\circ W_{\rm global}$, where the $W_{\rm global}$ is needed to preserve the fluxes associated with the $Spin _{G_{\rm global}}(4)$ structure, and $W_{\rm gauge}$ is needed to preserve the $\Gamma _g$ flux background.  Now we suppose that the gauge group does not include such a $W_{\rm gauge}$ element and we instead use charge conjugation to preserve the $\Gamma _g$ gauge flux.  We then consider the mapping torus based on $\widehat\varphi=\varphi _{c.c.}\circ  W_{\rm global}\circ \CC$,  which can preserve the flux background, and satisfy $\varphi _{c.c.}^2=W_{\rm global}^2=-1$ and $\CC ^2=1$, so $\widehat \varphi ^2=1$.

Now, however, the mapping torus construction for $\widehat\varphi=\varphi _{c.c.}\circ  W_{\rm global}\circ \CC$, probes a $5d$ SPT phase that includes flux associated with an $\CC$ background gauge field: 
the anomaly measured by the variation of the action is given by 
\eq{\label{w1anomaly}
\CA\supset\pi i \sigma_\fI\int w_1^\CC\cup w_2(TM)\cup B_2~, 
} 
where $w_1^\CC$ is a $\IZ_2$-valued gauge field for $\IZ_2^\CC$ symmetry.  Here, since we are using the action of $\CC$, we are coupling to $G_{\rm total}\rtimes \IZ_2^{\CC}$-bundles which allows for the non-trivial $\IZ_2^\CC$ gauge fields.  Since $\CC$ acts non-trivially on $\Gamma_g$ (we assume it acts trivially on $G_{\rm global}\times Spin(4)$), the cohomology of the $\Gamma_g$-fluxes must be modified in the presence of $\CC$-background gauge fields: the $\Gamma_g$ cohomology is valued in a $\IZ_2^\CC$-twisted cohomology which takes values in (torsion subgroups of) $\IZ^\CC$ which is an associated $\IZ_2^\CC$-bundle over spacetime where $\CC$ acts on the $\IZ$ fiber with the natural $\IZ_2$ involution, see e.g. \cite{Kapustin:2014gma,Hatcher}. 
The $\IZ_2^\CC$ bundle is characterized by a connection $w_1^\CC\in H^1(M;\IZ_2)$ which is a background field for the $\IZ_2^\CC$ symmetry. We will summarize some of the effects of turning on a non-trivial $w_1^\CC$; see \cite{Kapustin:2014gma,Hatcher}for more details.

Let us restrict our discussion to the case where $\Gamma_g=\IZ_{2N}$. 
Since $\IZ_2\subset \Gamma_g$ is invariant under the action of $\IZ_2^\CC$, the action of the $\IZ_2^\CC$ twisted Bockstein map $\tilde \beta _2$ is related that of the ordinary Bockstein map by  
\eq{\label{twistedbocksteintext}
\tilde\beta_2(x)=w_1^\CC\cup x+\beta_2(x)\quad, \quad x\in H^\ast(M;\IZ_2)~,
}
where $\beta_2$ is the $\IZ_2$ Bockstein map, similar to the structure of $PSO(n)$ gauge bundles as in ~\cite{Hsin:2020nts}.    
This implies that, for $\Gamma_g=\IZ_{2N}$, that the anomaly can only be of the form
\eq{
\CA&=\pi i \, \sigma_{\fI}\int w_2(TM) \cup \Big(\beta_2(B_2)+w_1^\CC\cup B_2\Big)~,
}
where  the two terms correlated and $B_2$ is the flux for $\IZ_2\subset \IZ_{2N}$.

In summary, if the $G_{\rm gauge}$ matter representations are real  and the path integral is anomalous under $\widehat\varphi=\varphi\circ W_{\rm global}\circ \CC$, then the theory has the $w_1^\CC\cup B_2\cup w_2(TM)$ anomaly term \eqref{w1anomaly}. The relations \eqref{twistedbocksteintext}, then imply tha tthe theory also the correlated
$\IZ_2$-valued $B_2\cup w_3$ anomaly. The same statement holds for the case when $\Gamma_g$ is a product of such finite abelian groups.  

Now consider the case where the $G_{\rm gauge}$ representation $R_{\rm gauge}$ is pseudoreal/symplectic.  After Higgsing to $U(1)$, $W_{\rm gauge}$ is matched by   $\CC \circ U$ where $U$ generates a $\IZ _4$ chiral symmetry with $U^2=(-1)^F$.  Since the fermions transform in a symplectic representation of the gauge group, the $U^2$ operation is part of the gauge group.  Again, due to the twisting of the $U(1)_{\rm gauge}$ cohomology by $\widehat\varphi$ (i.e. $O(2)=U(1)_{\rm gauge}\rtimes \IZ_2^{\widehat\varphi}$ still acts on fermion zero modes), the 5d anomaly theory associated with the $\widehat \varphi = \varphi _{c.c.}\circ \CC \circ U$ mapping torus is of the form
\eq{
\CA \subset  i \pi \,\sigma_\fI\int w_2(TM)\cup (\beta_2B_2+w_1^{\widehat\varphi}\cup B_2)~.
}
In such theories, there can additionally be a seperate anomaly for the axial symmetry.

\subsection{Anomalies in Abelian Gauge Theory}

Because anomalies are independent of continuous deformations,  one can probe the anomalies involving 1-form global symmetries in non-abelian QCD-like theories by deforming the theory so that it flows to pure abelian gauge theory  and studying the charges of the IR line operators.
This was discussed in e.g. \cite{Brennan:2022tyl,Wang:2018qoy}.  The anomalies can then be matched by identifying the UV 1-form global symmetries with discrete subgroups of  the electric- and magnetic- 1-form global symmetries of the IR theory which have their own mixed anomaly
\eq{
\CA\subset\frac{i}{2\pi}\sum_{I=1}^r\int B_{2,I}^{(e)}\wedge d B_{2,I}^{(m)}~. 
} 
Here we will show how we can use our discussion of $\CC$-symmetry to compute the anomalies in the abelian gauge theory complementary to the fractionalization technique. 

 Consider the same setup as earlier: $G_{\rm gauge}$ theory with $N_f$ fermions in representations $R_{\rm gauge}$ so that the fields couple to $G_{\rm total}$-bundles of the form 
\eq{\label{Gfermstarting}
G_{\rm total}=\frac{G_{\rm gauge}}{\Gamma_g}\times \frac{G_{UV}\times Spin(4)}{\IZ_2}~,
}
where here $G_{UV}$ is the 0-form global symmetry group of the fermions and $\Gamma_g$ is the UV 1-form center symmetry.  We now add an adjoint-valued scalar field $\Phi$, and consider how the anomalies are matched in the low-energy theory where $\langle \Phi \rangle \sim v \neq 0$, breaking $G_{\rm gauge}\to U(1)^r$.  One option is to include a Yukawa interaction with coupling $\lambda_Y \Phi \psi \psi$ to the fermions, so the fermions get a mass $m_\psi \sim \lambda v$ in the low-energy theory.  In that case,  the anomalies of the UV theory can be matched by fractionalization of the  line operators as in \cite{Brennan:2022tyl}.   We will here instead discuss how the anomalies are matched in the theory with $\lambda_Y\to 0$, where the IR theory is a $U(1)^r$ gauge theory with a collection of massless fermions $\{\psi_i\}^{\oplus N_f}$ with charge vectors $\vec{q}_i$.  We will take the spectrum to be invariant under 
\eq{
\CC:\{\vec{q}_i\}^{\oplus N_f}\longmapsto \{\vec{\tilde{q}}_i=-\vec{q}_i\}^{\oplus N_f}~,
}
 up to permutation. As a consequence of $\CC$-symmetry there is no gauge anomaly which means:
 \eq{
 \sum_i q_{i,a}q_{i,b}q_{i,c}=0~, ~ \forall a,b,c\quad, \quad \vec{q}_i=(q_{i,1},q_{i,2},...,q_{i,r})~. 
 } 
We can couple the fermions of the IR theory to $G_{\rm total}$-bundles where 
\eq{
G_{\rm total}=\frac{G_{\rm gauge}}{\Gamma_g}\times \frac{G_{UV}\times Spin(4)}{\IZ_2}~.
}
Here, the theory again has a $\Gamma_g^{(1)}$ electric 1-form global symmetry which turns on $\Gamma_g$ flux along $U(1)_{\Gamma_g}\subset U(1)^r$ which is generated by $Q_{\Gamma_g}$. Additionally, we will restrict our attention to (in general sub-group of) the 0-form global symmetry group $G_{UV}$ which matches that of the $UV$ global symmetry group. Because of the identification of $(-1)^F\sim (-\mathds{1}_{UV})$, we can consistently put the theory on a non-spin manifold. 

Now we can proceed as before by putting the theory on $\ICP^2$ and turning on  $\IZ_2\subset \Gamma_g$ $B_2$ and $w_2(G_{UV})$ flux. Let us write the charge pairing of $\psi_i$ with the total fractional flux along $U(1)_{\Gamma_g}$ and $SU(2)_{UV}$ as $Q_i$. We can then apply the Atiyah Singer index theorem to determine the number of zero modes $\fI$ of the Dirac operator in the flux background to be
\eq{
\fI=\sum_i \frac{4Q_i^2-1}{8}~.
}
The map $\widehat\varphi$ that preserves the bosonic background and acts non-trivially on the fermion zero-modes is $\widehat\varphi=\varphi_{c.c.}\circ W_{UV}\circ \CC$ or $\widehat\varphi=\varphi_{c.c.}\circ W_{UV}\circ \CC\circ U$ as we have previously discussed, where $W_{UV}$ is the Weyl transformation in an embedded $SU(2)_{\rm global}\subset G_{UV}$ that inverts the discrete flux $w_2(\Gamma)$. As before, $\widehat\varphi$ acts on the path integral measure as $D[\Psi] \stackrel{\widehat{\varphi}}{\mapsto}D[\Psi]\,(-1)^{\fI/2}$  
So the anomaly indicator is the integer $\sigma_\fI=\frac{\fI}{2}~{\rm mod}_2$.

As discussed in the previous subsection, non-trivial action of $\widehat\varphi$ on the path integral indicates that the anomaly contains a term 
\eq{
\CA_1=\pi i \sigma_\fI\int w_1^{\widehat\varphi}\cup w_2(TM)\cup {B_2} ~, 
}
which  implies the presence of an anomaly
\eq{
\CA_0= \pi i \sigma_\fI\int {B_2}\cup w_3(TM)~. 
}

\subsubsection{Example: $SU(2)$ with 2$n_f$ Weyl Fermions in the Adjoint Representation}
\label{sec:su2U1}

The fields, including the added adjoint scalar, couple to the total bundle 
\eq{
G_{\rm total}=\frac{SU(2)_{\rm gauge}}{\IZ_2}\times \frac{SU(2n_f)_{UV}\times Spin(4)}{\IZ_2}~.
}
Giving an expectation value to the added scalar Higgses $SU(2)_{\rm gauge}\mapsto U(1)_{\rm gauge}$ and the theory RG flows to QED in the IR.  The adjoint fermions decompose as 
$\mathbf{3}\otimes \mathbf{2}\longmapsto \left(2\oplus 0\oplus (-2)\right)\otimes \mathbf{2}$. 
The $\IZ_2^{(1)}$ electric 1-form symmetry can be coupled to a background flux $B_2$ and the $Spin_{SU(2n_f)}(4)$ structure leads to fractional flux  along $U(1)_{\rm gauge}$ and $SU(2)_\text{global}\subset SU(2n_f)_\text{global}$ with $Q_i$ given 
$\{Q_i\}=\left\{\pm \frac{3}{2},\pm \half,\pm \half\right\}^{\oplus n_f}$.  The calculation of the index of the Dirac operator and the anomaly indicator are the same as before:  
$\fI=\sum_i\frac{4Q_i^2-1}{8}=2n_f$ and $\sigma_\fI=n_f$.  The mapping torus has $\widehat\varphi=\varphi\circ W_{UV}\circ \CC$ where $\varphi$ is charge conjugation on $\ICP^2$ and $W_{UV}\in SU(2n_f)_{UV}$, with the anomaly now given by 
\eq{
\CA\subset\pi i \,n_f\int B_2\cup  (w_3(TN)+w_1^\CC\cup w_2(TN)).
	}

\subsubsection{Example: $SU(2)$ with Weyl Fermion in ${\bf 4}$-Representation}

Now let us consider the example of $G_{\rm gauge}=SU(2)_{\rm gauge}$ with a single Weyl fermion in the $j=3/2$ representation. The fermion in this theory couples to the total bundle 
\eq{
G_{\rm total}=\frac{SU(2)_{\rm gauge}\times Spin(4)\times \IZ_{10}
}{\IZ_2\times \IZ_2}~,
}
where here we suppress the $\IZ_{10}$ discrete chiral symmetry. 
We deform this theory by coupling to an adjoint scalar field and condensing it so that gauge symmetry is broken $SU(2)_{\rm gauge}\to U(1)_{\rm gauge}$ 
and we flow to an abelian gauge theory in the IR. There the chiral symmetry is enhanced 
\eq{
G_{\rm total}\longrightarrow \frac{SU(2)_{\rm gauge}\times Spin(4)\times \IZ_{20}}{\IZ_2\times \IZ_2}~.
}
With respect to $U(1)_{\rm gauge}$ 
the fermion decomposes into four Weyl fermions $\psi_{q_i}$ with charges 
\eq{
\mathbf{4}\longmapsto q_i=\pm 3,\, \pm 1~. 
}
In the IR, there is a $Spin_\IC(4)$ structure and the theory can be put on $\ICP^2$ by turning on a fractional flux along $U(1)_{\rm gauge}$.  
Additionally, the map $\widehat\varphi$ of the un-Higgsed theory which is given by $\widehat\varphi_{UV}=\varphi_{c.c.}\circ W_{\rm gauge}\circ W_{UV}$ is matched in the Higgsed theory $\widehat\varphi_{IR}=\varphi_{c.c.}\circ W\circ \CC\circ U\circ g$  
and $\CC,U,g$ act as 
\eq{
\CC:\psi_{q_i}\longmapsto \psi_{-q_i}\quad, \quad 
U= 
e^{\frac{2\pi i}{4}} \mathds{1}\quad, \quad g= 
e^{\frac{2\pi i}{4}Q_{\rm gauge}}~,
}
where $U$ generates a $\IZ_4\subset \IZ_{20}$ chiral rotation and $g$ generates a $\IZ_4\subset U(1)_{\rm gauge}$ gauge transformation.

We can now compute the index in this background
\eq{
\fI=\sum_i \frac{4 Q_i^2-1}{8}=2\quad\Longrightarrow \quad \sigma_\fI=1~,
}
and see that there is an anomaly 
\eq{
\CA\supset \pi i \int w_2(TM)\cup \left(w_3+w_1^{\widehat\varphi}\cup w_2(TM)\right)~. 
}
Note that  the chiral symmetry $\IZ_{20}$ has additional `t Hooft anomalies both with itself and with the $Spin_{\IC}(4)$ structure. 

%\subsection{Trivialization of $\IZ_{4N}$ Anomaly}
%
%As discussed in Section \ref{sec:w2w3gen}, the $w_2w_3$ anomaly we consider in this paper is $\IZ_2$-valued. In general, such $\IZ_2$ anomalies can exist for any $\IZ_{N}^{(1)}$ global symmetry as long as $N$ is even. In this section, we will show that the $w_2w_3$ anomalies we study in this paper are only non-zero when $N$ is of the form $N=4n+2$ so that $\IZ_N=\IZ_2\times \IZ_{2n+1}$. 
%
%Let us consider a $G_{\rm gauge}$ gauge theory with fermionic matter that transform under a $SU(2)_f^{(0)}\subset G_f^{(0)}$ global symmetry so that all of the fields in the theory couple to bundles of the form 
%\eq{
%G_{\rm total}=\frac{G_{\rm gauge}}{\IZ_{4n}}\times \frac{SU(2)_f\times Spin(4)}{\IZ_2}~. 
%}
%Let us consider coupling to an adjoint scalar field $\Phi$ with potential so that in the IR the theory flows to an abelian gauge theory. Let us restrict to a single $U(1)_0\subset G_{IR}\cong U(1)^r$ for which the $\IZ_2\subset \IZ_{4N}\subset G_{\rm gauge}$ UV 1-form center symmetry is also a subgroup $\IZ_{2}\subset U(1)_0$. 
%
%Following our previous discussion, there is a possible $w_2w_3$ anomaly of the UV theory that can be matched in the IR by a $w_1^\CC\cup w_2\cup w_2$ anomaly. 

\subsection{Anomalies and Higgsing with Adjoints Fermions}
  
The anomalies must additionally match if we partially Higgs the gauge group by an adjoint Higgs field, 
\eq{
G_{\rm gauge}\to \frac{G_{\rm gauge}^\prime \times U(1)}{\CZ}~. 
}
The $Z(G_{\rm gauge})^{(1)}$ global symmetry is unbroken by the vev and arises in the IR theory as 
\eq{
Z(G_{\rm gauge})\subset Z(G_{\rm gauge}^\prime)\times U(1)~. }
If there is no Yukawa coupling, the anomaly matching is as discussed in the previous subsection.   With Yukawa couplings the IR theory can have a $U(1)^{(1)}$ emergent 1-form global symmetry and there are two cases. 
If $Z(G_{\rm gauge})$ is in the $U(1)$ factor, then $\Gamma_g^{(1)}$ embeds into the emergent $U(1)^{(1)}$ global symmetry and the IR $U(1)^{(1)}$ symmetry matches the anomaly of $\Gamma_g^{(1)}$. 
If $Z(G_{\rm gauge})\to Z(G_{\rm gauge}^\prime)$ then the anomaly of $\Gamma_g^{(1)}$ will instead be matched by an anomaly of $\CZ^{(1)}$, the 1-form center symmetry of the IR model.  

As an example, consider $SU(N_c)$ gauge theory with $2n_f$ adjoint fermions, where 
\eq{
G_{\rm total}^{(UV)}=\frac{SU(N_c)}{\IZ_{N_c}}\times \frac{SU(2n_f)\times Spin(4)}{\IZ_2}~. 
}
Now we can consider Higgsing the theory by an adjoint Higgs field  which breaks the gauge group 
\eq{
SU(N_c)\mapsto \frac{SU(N_c-1)\times U(1)}{\IZ_{N_c-1}}=U(N_c-1)~.
}
If we additionally couple the fermions to the Higgs field via a Yukawa coupling, we break $SU(2n_f)\to Sp(n_f)$ and the 
resulting IR theory will have $2n_f$ fermions in the adjoint representation of $U(N_c-1)$ which transform faithfully under the group 
\eq{ G_{\rm total}^{(IR)}=\frac{U(N_c-1)}{U(1)}\times \frac{Sp(n_f)\times Spin(4)}{\IZ_2}~.
}
Here we see that the IR theory has an emergent $U(1)^{(1)}$ 1-form global symmetry where $\IZ_{N_c}^{(1)}=\IZ _{N_c}^{(1)}\subset U(1)^{(1)}$. 
For the case of $N_c$ odd, since $Z(SU(N_c))=\IZ_{N_c}$, there is no mixed anomaly with $\IZ_2^{(1)}$ along the entire RG flow. However, since the IR has an emergent $U(1)^{(1)}$ global symmetry which contains $U(1)^{(1)}\supset \IZ_2^{(1)}$, there is the possibility for this emergent symmetry to be anomalous. Indeed, as we know from the previous section, $SU(N_c-1)$ QCD does have a mixed anomaly when $N_c=4n_c+3$.  
For the case of  $N_c$ is even, we know that there is an anomaly for $N_c=4n_c+2$.  Again, the IR theory has an emergent $U(1)^{(1)}$ global symmetry, which contains $\IZ_{4n_c+2}^{(1)}$ as a subgroup. The $\kappa _{\Gamma _{g}T}$ anomaly of the UV theory is then matched by $\kappa _{\Gamma _{g}T}$ of the IR theory by a  mixed $U(1)^{(1)}$-gravitational anomaly.

\section{Symmetry Matching in the SSB phase: $N_f$ Adjoint Weyl Fermion QCD}
\label{sec:CP1model}

Four-dimensional gauge theories have a rich variety of possible IR phases, including IR-free electric (e.g. if not UV asymptotically free), interacting conformal field theory, exotic IR-free duals, gapped (including TQFTs), and spontaneous symmetry breaking (SSB).  While it is an unsolved problem to analytically determine the IR phase of general 4d gauge theories (see e.g.~\cite{Intriligator:1995au} for examples and discussion in the context of supersymmetric gauge theories), the IR phase is partially constrained by symmetries, 't Hooft anomaly matching, and the $a$-theorem.  For example, non-zero 't Hooft anomalies can rule out IR gapped phases\footnote{Some anomalies rule out symmetry preserving gapped TQFTs, and others do not; see~\cite{Cordova:2019bsd,Cordova:2019jqi} and  therein.} and SSB is constrained by the theorems of~\cite{Vafa:1983tf}.  

In this section, we will discuss some aspects of symmetries, 't Hooft anomalies, and anomaly matching in the IR SSB nonlinear sigma model phase. We will recall that 't Hooft anomalies must also match for the spontaneously broken symmetries and how this condition determines the coefficient of the Wess-Zumino-Witten (WZW)~\cite{Wess:1971yu, Witten:1983tw} interaction in the low-energy effective theory.  We  will illustrate anomaly matching for both standard anomalies and those associated with $Spin_G(4)$ structure in the context of a wide class of theories.   See e.g.~\cite{Freed:2006mx, Freed:2017rlk, Cordova:2018acb, Lee:2020ojw, Lu:2022zra, Chen:2022cyw, Hsin:2022heo, Chen:2023czk, Pace:2023kyi, Pace:2023mdo} for a partial list of some recent works on aspects of nonlinear sigma models and their symmetries and defects.

The class of theories that we will discuss in this section are general $G_{\rm gauge}$ theories with $N_f$ Weyl fermions, $\lambda _{\alpha =1,2} ^{f=1\dots N_f}$, in the adjoint representation of $G_{\rm gauge}$.  The theories are asymptotically free for $N_f\leq 5$ and are expected to be in an interacting CFT phase for $N_f^*\leq N_f\leq 5$, where the lower bound $N_f^*$ of the conformal window is not known (and it could be $G_{\rm gauge}$ dependent).   It is also not known for which $N_f$ values the theories have SSB.  The theories with $N_f=1$ are  ${\cal  N}=1$ supersymmetric $G_{\rm gauge}$ pure Yang Mills and the IR phase of is well-understood from a variety of SUSY-based constraints and perspectives: there are isolated gapped vacua with $\langle\Tr \lambda \lambda\rangle \neq 0$ which leads to SSB of the (discrete) chiral symmetry.\footnote{The 't Hooft anomalies for the broken discrete symmetries are matched by the TQFTs on the domain walls between the vacua; see~\cite{Gaiotto:2014kfa,Delmastro:2020dkz} and references therein.} The theories for $N_f>1$ are non-supersymmetric and less well-understood; this class of theories has been studied for various values of $N_f$, and for various choices of $G_{\rm gauge}$ (including on the lattice, especially for $N_f=2$, i.e. one Dirac flavor, and $G_{\rm gauge}=SU(2)$). See e.g.~\cite{Bolognesi:2007ut, DelDebbio:2009fd, DeGrand:2011qd, DeGrand:2013uha, Shifman:2013yca, Basar:2013sza, Athenodorou:2014eua, Anber:2018xek, Anber:2018iof, Bi:2018xvr, Cordova:2018acb, Wan:2018djl, Poppitz:2019fnp, Cordova:2019jqi, Cordova:2019bsd, Anber:2020gig} and references therein for a partial list of the large literature on this class of theories. Some of these works propose novel, IR dual theories, and related analysis of 't Hooft anomaly matching constraints.  We will not discuss any novel IR duals.   

Here we will simply assume the spontaneous symmetry breaking IR phase associated with the expectation value $\langle {\cal O}^{(fg)}\rangle \sim \Lambda ^3  \neq 0$ of the gauge and Lorentz invariant operator ${\cal O}^{(fg)}\equiv \Tr \lambda ^{(f}_\alpha \lambda ^{g)}_\beta \epsilon ^{\alpha \beta}$, and discuss how the symmetries and 't Hooft anomalies are consistent with that possibility.  These checks neither prove nor disprove that the SSB phase is physically realized, and indeed we expect that the theory is IR-free for $N_f>5$ and that the IR CFT phase conformal window includes $N_f=5$ and perhaps also $N_f=4$ and $N_f=3$.   

The vacuum manifold of the theories, with our assumed SSB, consists of $h^\vee _{G_{\rm gauge}}$ (with adjoint index e.g. $h^\vee _{SU(N_c)}=N_c$) disconnected copies of a $G/H=SU(N_f)/SO(N_f)$ nonlinear sigma model of Nambu-Goldstone boson pions.    The $h^\vee _{G_{\rm gauge}}$ copies are associated with the  spontaneous symmetry breaking of a discrete global symmetry for simply connected $G_{\rm gauge}$ theories for all $N_f$.\footnote{E.g. for $N_f=1$ and simply connected $G_{\rm gauge}$, this gives the Witten index $\Tr (-1)^F=h^\vee _{G_{\rm gauge}}$.
Non-simply connected $G_{\rm gauge}$, e.g. $SO(3)$ (see sect.~\ref{sec:SO(3)}), can be obtained from the simply connected cases by gauging subgroups of the $\Gamma _g^{(1)}$ one-form global center symmetries.  Then the  $h^\vee _{G_{\rm gauge}}$ copies of the vacuum manifold are then generally not related by SSB and indeed are generally inequivalent, e.g. with different discrete gauge symmetries.} This was discussed in  detail in~\cite{Cordova:2019bsd} for $G_{\rm gauge}=SU(2)$ and $N_f=2$.  Our discussion will extend some of the analysis and results of~\cite{Cordova:2019bsd}   to general  $G_{\rm gauge}$ and $N_f$.

\subsection{$G_{\rm gauge}$ Theory with $N_f$ Massless, Weyl adjoints: UV symmetries}

The (simply connected \footnote{\label{footnote1}For non-simply connected $G_{\rm gauge}$, the $\IZ _{2n}$ symmetry here is reduced to a subgroup.   Non-simply connected cases (e.g. $SO(N)$) can be obtained from the simply connected ones by (e.g. $Spin(N)$) by gauging subgroups of $\Gamma _g$.   For example, as discussed in~\cite{Aharony:2013hda} for $N_f=1$ and $G=SU(2)\cong Spin(3)$, there is a $\IZ _{2N_f h^\vee _{SU(2)}}=\IZ _4$ chiral symmetry that is spontaneously broken to $\IZ _2\cong (-1)^F$, leading to two equivalent vacua which are related to each other by $\theta _{YM}\to \theta _{YM}+2\pi$.  For the $N_f=1$ non-simply connected case of $G=SO(3)$, one gauges $\Gamma _{SU(2)}^{(1)}$.  Then the  periodicity is instead $\theta _{YM}\cong \theta _{YM}+4\pi$ and the $U(1)_A$ symmetry is instead explicitly broken to $\IZ _2\cong (-1)^F$ and there is no spontaneous symmetry breaking.  There are still two vacua, but they are physically inequivalent: one has an unbroken $\IZ _2$ gauge theory and the other does not.}) $G_{\rm gauge}$ theory with $N_f$ massless Weyl fermions in the adjoint has 
\eq{\label{G4adjoints}
G_{\rm total}=\frac{G_{\rm gauge}}{{\Gamma_g}}\times \frac{SU(N_f)\times Spin(4)\times \IZ_{2n}}{\IZ_{N_f}\times \IZ_2}~.
}
$\Gamma_g^{(1)}=Z(G_{\rm gauge})$ is the 1-form center symmetry.  $\IZ^{(0)} _{2n}$, with generator $g$ and $n\equiv N_f h^\vee _{\rm Gauge}$, is the anomaly free subgroup of the ABJ-anomalous $U(1)_r$ symmetry.\footnote{$U(1)_r$ assigns charge $+1$ to each Weyl fermion, with generator ${\cal U}_\phi \ : \lambda _\alpha ^f\to e^{i\phi}\lambda _\alpha ^f$.  This symmetry is broken by instantons for general $\phi$, and the anomaly free $\IZ _{2n}$ subgroup is generated by group element $g= {\cal U}_{\phi= \pi / n}$.  The 
 ABJ anomaly implies that the $U(1)_A$ transformation $\lambda _\alpha  ^f\to e^{i\phi}\lambda _\alpha ^f$ shifts $\theta _{YM}\to \theta _{YM}+2N_f h^\vee _{G_{\rm gauge}}\phi$, where $h_{G_{\rm gauge}}^\vee$ is the dual Coxeter number (normalized to $h^\vee _{SU(N_c)}=N_c$).  Since we assume  that $G_{\rm gauge}$ is simply connected, instantons preserve $\theta \to \theta +2\pi$ (when the background gauge field for $\Gamma _g^{(1)}$ is $B_2=0$) and the unbroken $\IZ _{2n}\subset U(1)_A$ symmetry has  $n=N_f h^\vee _{G_{\rm gauge}}$, with $g={\cal U}_{\phi = \pi / N_f h^\vee _{G_{\rm gauge}}}$.  Non-simply connected cases, e.g. $SO(N_c)$ instead of $Spin(N_c)$, can be constructed from this discussion of the simply-connected cases by gauging appropriate subgroups of the one-form symmetry $\Gamma_g ^{(1)}$, the center of $G_{\rm gauge}$.}  The $\IZ _{N_f}$ quotient in~\eqref{G4adjoints} identifies the $\IZ _{N_f}$ center of $SU(N_f)$ with a subgroup of $\IZ _{2n}$ generated by $g^{2n/N_f}$; the $\IZ_2$ quotient identifies $g^{n}=(-1)^F$.        

The 't Hooft anomaly coefficients $\kappa$ can be computed in  the asymptotically free limit from the anomaly of free fermions.  The perturbative anomalies, in the presence of background gauge fields, are obtained by descent from the  anomaly polynomial $I=\Tr e^{F/2\pi}\widehat A(TM)$, along with additional anomaly terms for the torsion parts.  The anomaly theory terms include 
\eq{\label{sect5anomalies}{1\over 2\pi i}{\cal A}&\supset \kappa _{SU(N_f)^3} \int CS_5[SU(N_f)]+ \frac{1}{3}\kappa _{SU(N_f)^2r} \int \frac{A_r}{2\pi} \wedge ch_2[SU(N_f)]\\ &+\kappa _{r, TT} \int \frac{A_r}{2\pi} \wedge \frac{p_1[TM]}{24}+\frac{\kappa _{r^3}}{6} \int \frac{{\bf z}}{2n}\cup \beta {\bf z} \cup \beta {\bf z}+ \kappa _{r, \Gamma} \int {\bf z}  \cup {\cal P}(B_2)\\
&+\kappa_{\Gamma, T} \frac{1}{2}\int B_2\cup w_3(TM)+ \kappa _{T,T}\frac{1}{2} \int w_2(TM)\cup w_3(TM). }
We denote the $\IZ _{2n}\subset U(1)_r$ by $r$ e.g. its background gauge field is $A_r$ and as in~\cite{Cordova:2019bsd} we also denote $\frac{A_r}{2\pi} \to {\bf z}/2n$ and $\frac{F_r}{2\pi} \to \beta {\bf z}$ where ${\bf z}$ is a 1-cochain with values in $\IZ _{2n}$ and $\beta$ is the $\IZ _{2n}$ Bockstein map ($\beta = \delta /2n$, where $2n=2N_f h^\vee _{G_{\rm gauge}}$).  The anomaly coefficients are 

\begin{table}
\begin{center}
\begin{tabular}{|c|c|c|}
\hline
$G_{\rm gauge}$ & $\Gamma_g$&$\kappa_{r\Gamma}$\\
\hline
$SU(N_c)$&$\IZ_{N_c}$&$\frac{N_c-1}{2N_c}$\\
$Sp(N_c)$&$\IZ_2$&$\frac{N_c}{4}$\\
$E_6$&$\IZ_3$&$\frac{2}{3}$\\
$E_7$&$\IZ_2$&$\half$\\
$Spin(2n_c+1)$&$\IZ_2$&$\half$\\
$Spin(4n_c+2)$&$\IZ_4$&$\frac{2n_c+1}{8}$\\
$Spin(4n_c)$&$\IZ_2\times \IZ_2$&$\left(\frac{n_c}{4}~,~\half\right)$\\
\hline
\end{tabular}
\end{center}
\caption{ \label{tab:kapparG} In this table we give the anomaly coefficients $\kappa_{r\Gamma}$ for various choices of $G{\rm gauge}$. For $G_{\rm gauge}=Spin(4n_c)$ the two anomaly coefficients correspond to the two terms in the anomaly polynomial $\CA\supset\frac{2\pi i n_c}{4}\int {\bf z}\cup {\cal P}(B_2^L+B_2^R)+\frac{2\pi i }{2} \int {\bf z}\cup B_2^L\cup B_2^R$.
}
\end{table}

\begin{itemize}
\item   $\kappa _{SU(N_f)^3}=\Tr SU(N_f)^3=|G_{\rm gauge}|$.  This is non-torsion for $N_f\geq 3$ (since $SU(N_f\geq 3)$ has a cubic Casimir). For $N_f=2$, this 't Hooft anomaly reduces to a $\IZ _2$ torsion component corresponding to the 't Hooft anomaly version of the original $SU(2)$ anomaly~\cite{Witten:1982fp}, with coefficient $|G_{\rm gauge}|$ mod 2.  

\item  $\kappa _{SU(N_f)^2 r}=\Tr SU(N_f)^2 \IZ _{2n}=|G_{\rm gauge}|$. 

\item  $\kappa _{r}=\kappa _{r^3}=N_f |G_{\rm gauge}|$.  

\item  $\kappa _{r, \Gamma}$ is related to the anomalous phase under $\theta \to \theta +2\pi$ when $B_2\neq 0$.  The $\kappa _{r, \Gamma}$ coefficient, for all $N_f$, coincides with that given as in Table 1 of~\cite{Cordova:2019uob}, where we replace $d\theta/2\pi \to {\bf z}$.  The results are reproduced here in Table \ref{tab:kapparG}. 

\item  $\kappa _{\Gamma, T}$ is as computed in earlier sections, e.g. for $SU(N_c)$, we found $\kappa _{\Gamma, T}=\delta _{N_c, 4\IZ +2}$. 

\item  $\kappa _{T,T}=0$ for this class of theories.  

\end{itemize}

\noindent
The discrete symmetry anomaly coefficients $\kappa _{SU(N_f)^2 r}$ and $\kappa _{r}$ and $\kappa _{r^3}$ are examples of  anomalies of discrete symmetries $\IZ _{2n}$ symmetries with $g^n=(-1)^F$; such anomalies were discussed in~\cite{Hsieh:2018ifc, Guo:2018vij}.  The  $\kappa _{r}$ and $\kappa _{r^3}$ terms are packaged into anomalies associated with cobordism group $\Omega ^H_5=\IZ _a\times \IZ _b$ where the modularity $a$ and $b$ depends if $n=N_f h^\vee _{G_{\rm gauge}}$ is a multiple of 2 and /or a multiple of 3, e.g. for $n=0$ mod 6 then $a=24n$ and $b=n/6$ -- see \cite{Hsieh:2018ifc, Guo:2018vij}.

\subsection{$G_{\rm gauge}$ Theory with $N_f$ Massless, Weyl adjoints: Possible IR Phase with SSB}

For $N_f=1$, the theory is ${\cal N}=1$ SUSY $G_{\rm gauge}$ pure Yang-Mills, and it is known from SUSY exact results that the theory has SSB by the gaugino condensation expectation value of the Lorentz and gauge invariant operator ${\cal O}\sim \Tr \lambda _\alpha \lambda _\beta \epsilon ^{\alpha \beta}$, which is the bottom component of the chiral glueball superfield operator $S\sim \Tr W_\alpha W^\alpha$.  The SSB breaks $\IZ _{2 h^\vee _{G_{\rm gauge}}}\to \IZ _2$, where if  $\IZ _{2 h^\vee _{G_{\rm gauge}}}$ has generator $g$, then $g^{h^\vee _{G_{\rm gauge}}}$ is identified with $\IZ _2=(-1)^F$.  This leads to $h^\vee _{G_{\rm gauge}}$ vacua, which are physically equivalent for simply connected $G_{\rm gauge}$.  These vacua have unbroken supersymmetry and are trivially gapped for $G_{\rm gauge}$ simply connected (otherwise they can have discrete gauge theories, as mentioned in Footnote \ref{footnote1}), corresponding to $\Tr (-1)^F=h^\vee _{G_{\rm gauge}}$ in these theories.  The multiple vacua are connected by BPS domain walls with 3d TQFTs on their worldvolumes, see e.g.~\cite{Delmastro:2020dkz} and references therein.

For $N_f>1$, we consider the possibility that the IR theory is also in the SSB phase with $\langle {\cal O}^{(fg)}\rangle \sim \Lambda ^3$ for the operator ${\cal O}^{(fg)}=\Tr \lambda ^f_\alpha \lambda ^g_{\beta}\epsilon ^{\alpha\beta}$. This spontaneously breaks the continuous symmetry as $SU(N_f)\to SO(N_f)$ and spontaneously breaks the discrete symmetry\footnote{As discussed in~\cite{Cordova:2019jqi}, 't Hooft anomalies obstruct an IR phase where the discrete symmetry is unbroken and only acts nontrivially on gapped states (it acts trivially on the massless NG bosons). See also~\cite{Cox:2021vsa}.} as $\IZ _{2N_f h^\vee _{G_{\rm gauge}}}\to \IZ _2=(-1)^F$, so the vacuum manifold consists of multiple disconnected copies of the $SU(N_f)/SO(N_f)$ NG boson target space.   This breaking does not lead to $N_f h^\vee _{G_{\rm gauge}}$ copies of $SU(N_f)/SO(N_f)$: because $g^{2h^\vee _{G_{\rm gauge}}}$ is equal to the generator of the 
$\IZ _{N_f}$ center of $SU(N_f)$, it follows that $g^{2h^\vee _{G_{\rm gauge}}}$ maps a vacuum in a $SU(N_f)/SO(N_f)$ to another vacuum in the same $SU(N_f)/SO(N_f)$. The SSB to $\IZ _2$ then leads to a vacuum manifold consisting of $h^\vee _{G_{\rm gauge}}$ disconnected copies of  $SU(N_f)/SO(N_f)$ non-linear sigma model for all $N_f$:
\eq{\label{hG/Hs}{\cal M}_{\rm vac}= \frac{SU(N_f)}{SO(N_f)} ^{\oplus h^\vee _{G_{\rm gauge}}} .}  For $N_f=2$, assuming SSB, the vacuum manifold consists of $h^\vee _{G_{\rm gauge}}$ disconnected copies of $SU(2)/SO(2)\cong\ICP ^1$, as in the $h^\vee _{G_{\rm gauge}=SU(2)}=2$ case discussed in~\cite{Cordova:2019bsd}.    For $N_f=3$, assuming SSB, the space $SU(3)/SO(3)$ is the 5d Wu manifold, which is  discussed further e.g. in~\cite{Lee:2020ojw}.

The fact that the vacuum manifold~\eqref{hG/Hs} consists of  $h^\vee _{G_{\rm gauge}}$ disconnected components for all $N_f$ is nicely compatible with RG flows continuously deforming from $N_f>1$ to $N_f=1$ by adding mass terms and RG decoupling of the heavy flavors.  
In the UV gauge theory, one can add mass terms for the fermions, $\Delta {\cal L}_{UV}=m_{(f,g)}{\cal O}^{(fg)}$ where  ${\cal O}^{(fg)}=\Tr \lambda ^f_\alpha \lambda ^g_{\beta}\epsilon ^{\alpha\beta}$ is the operator that gets a vev in the SSB phase.  In the IR pion theory, non-zero $m_{(f,g)}$ lifts the $SU(N_f)/SO(N_f)$ vacuum degeneracy; some of the pions become pseudo NG-bosons.  In particular, choosing $m_{(f,g)}$ to give non-zero mass to $N_f-1$ of the Weyl fermions gives a RG flow that reduces to the $N_f=1$ case in the IR.  The $h^\vee _{G_{\rm gauge}}$ disconnected copies of the $SU(N_f)/SO(N_f)$ sigma model is indeed nicely consistent with the $h^\vee _{G_{\rm gauge}}$ isolated vacua of the $N_f=1$ case upon deformation by adding mass terms to $N_f-1$ of the Weyl fermions: doing so in the low-energy sigma model lifts the $SU(N_f)/SO(N_f)$ vacuum degeneracy and leads to an isolated vacuum for each $SU(N_f)/SO(N_f)$ sigma model theory, correctly reproducing the  $h^\vee _{G_{\rm gauge}}$ vacua of the $N_f=1$ low-energy theory.   See~\cite{Cordova:2019jqi} for details in the case discussed there. 

The continuous part of the vacuum manifold, the $G/H\equiv SU(N_f)/SO(N_f)$ sigma model, has been much-studied in connection with $SO(N_c)$ QCD with $N_f$ fundamental fermions.  The sigma model admits solitonic particles, and also lines, which were matched to 
to the baryons and flux tubes of $SO(N_c)$ in~\cite{Witten:1983tx}; see also~\cite{Benson:1994dp,Benson:1994ce}  for further consideration in connection with $SO(N_c)$ QCD flux tubes and~\cite{Auzzi:2006ns, Bolognesi:2009vm} for the flux tubes in connection with $SU(N_c)$ gauge theories with adjoint fermions.  The $\pi _3$ and $\pi _2$ homotopy groups for solitonic particles and vortices, respectively, have some $N_f$ dependence for low $N_f$:  $\pi _3({\cal M}_{N_f>3})=\IZ _2$, $\pi _3({\cal M}_{N_f=3})=\IZ _4$, $\pi _3({\cal M}_{N_f=2})=\IZ$; 
$\pi _2({\cal M}_{N_f\geq 3})=\IZ _2$,  $\pi _2(\ICP ^1)=\pi _3(\ICP ^1)=\IZ$.  See \cite{Freed:2017rlk, Cordova:2019bsd, Chen:2022cyw} for discussion and clarification of the physics and symmetry associated with the Hopf number $\pi _3(\ICP ^1)=\IZ$, related to the fact that it is not independent of the $\pi _2(\ICP ^1)$ quantum number.  

In the remainder of this subsection, we will discuss the sigma model and symmetry and anomaly matching for $N_f\geq 3$.  Unlike the case of $SO(N_c)$ with $N_f$ fundamentals, where SSB might occur for arbitrary $N_f$ by taking $N_c$ sufficiently large, for the case of $G_{\rm gauge}$ with adjoint Weyl fermions the asymptotically free cases have $N_f\leq 5$; the theories with $N_f=5$ are likely in an interacting CFT phase rather than a SSB, and that is perhaps also the case for $N_f=4$ and $N_f=3$.  We will here anyway assume SSB, so the remainder of this subsection is perhaps an academic exercise in how symmetries could be matched in IR phases that are not physically realized.  The exercise also serves as a warmup for the $N_f=2$ case in the next subsection, where SSB is a quite plausible scenario.

The  $G/H \equiv SU(N_f)/SO(N_f)$ sigma model for $N_f\geq 3$ admits a WZW interaction~\cite{Wess:1971yu, Witten:1983tw} in the low-energy effective theory for the NG boson pions with coefficient $k$ determined by matching the $\kappa _{SU(N_f)^3}=|G_{\rm gauge}|$ 't Hooft anomaly:
\eq{\label{WZWmatching}e^{-S_{WZW}}=e^{2\pi i \Delta \kappa _{G^3} \int _{N_5} \frac{1}{4} \phi ^*y_5}, \qquad \hbox{so} \qquad  \Delta \kappa _{G^3} \oint _{Z_5} \frac{1}{4} \phi ^*y_5 \in \IZ}
with $M_4=\partial N_5$ and the quantization condition is needed for the WZW term to be well-defined under the ambiguity in $N_5\to N_5'$ with closed $Z_5=N_5-N_5'$.
The $H^5(G/H, \IZ)$ representative $y_5$ is normalized such that $\oint _{Z_5} y_5 \in \IZ$ for general closed $Z_5$, so the $\frac{1}{4}$ is interesting -- see~\cite{ Lee:2020ojw} for a nice, recent discussion.   For our present case\footnote{For the case of $SU(N_c)$ gauge theory with $N_f$ massless Dirac fundamental fermions, each flavor contributes 2 to the $\Tr SU(N_f)_{L-R}^3=2N_c$ 't Hooft anomaly.  For $N_c$ odd, the IR theory must have a spin structure, $w_2(TM)=w_2(TN_5)=0$, because e.g. the baryon associated with $\pi _3(SU(N_f))$ is a fermion.  This ensure that the quantization condition is satisfied because then  $\int _{Z_5=\rm{spin}} \phi ^*y_5\in 2\IZ$ \cite{Freed:2006mx,Lee:2020ojw}.} of $G/H=SU(N_f)/SO(N_f)$ it was argued in~\cite{ Lee:2020ojw} (in the context of $SO(N_c)$ gauge theory with fundamental fermions) that the quantization condition is satisfied for all $k$ if $M_4$ has a spin structure, $w_2(TM)=0$: they show that then $\int _{Z_5=\rm{spin}}\phi ^*(y_5)\in 4\IZ$.\footnote{As discussed in~\cite{Lee:2020ojw} it is easily argued that $\int _{Z_5}\phi ^*y_5=\int _{Z_5}w_2(TN)\cup w_3(TN)$ mod 2, so $w_2(TN)=0$ easily implies  $\int _{Z_5=\rm{spin}}\phi ^*(y_5)\in 2\IZ$; the additional factor of 2 requires further analysis in~\cite{Lee:2020ojw}.}
  Recall that $k$ affects the statistics of solitons in the IR theory~\cite{Witten:1983tw,  Witten:1983tx} and that  $p$-form symmetry charged objects of the UV theory can arise as skyrmionic solitons\footnote{Non-trivial topology of the vacuum manifold is of course not sufficient to ensure stability (e.g. a rubber band can unwind off the bottom of a wine-bottle potential).  Assumptions about higher derivative terms are needed to ensure stability and also to evade Derrick's theorem.} of the  $\phi$ field in the IR, with their codimension $p+1$ symmetry operators  associated with $\pi _{d-p-1}(G/H)$.  The more general, proper formulation of pions and WZW  terms in terms of generalized differential cohomology theories~\cite{Freed:2006mx, Gu:2012ib, Cordova:2019jnf, Cordova:2019uob, Yonekura:2020upo, Lee:2020ojw} shows that the homotopy and cohomology groups should be replaced with corresponding bordism classes.  This formulation also incorporates torsion anomalies, and generalizations of the solitons and defect operators. 

In the present context, $k\equiv \Delta \kappa _{G^3}=\kappa _{SU(N_f)^3}=|G_{\rm gauge}|$, so
\eq{\label{adjointwzw} e^{-S_{WZW}}=e^{2\pi i |G_{\rm gauge}| \int _{N_5} \frac{1}{4} \phi ^*y_5}, \qquad \hbox{so} \qquad |G_{\rm gauge}| \oint _{Z_5} \phi ^* y_5\in 4 \IZ.} 
The WZW 5-form $y_5$ in~\eqref{adjointwzw} includes, as usual, additional terms involving the background gauge fields for the $SU(N_f)$ global symmetry and also the background gauge field $A_r$ for the $\IZ _{2n}$ global symmetry.   For the theories with ordinary spin structure, we restrict to $M_4$ with spin structure, which can be extended to $N_5$  so $Z_5=N_5-N_5'=\rm{spin}$.  It then follows from the result of~\cite{Lee:2020ojw} that $\int _{Z_5=\rm{spin}}y_5\in 4\IZ$ so the quantization condition is satisfied for all $G_{\rm gauge}$.    We will discuss the $\frac{1}{4}$ again below in the case with generalized ${\text Spin}_{SU(N_f=2)}(4)$ structure.   For $N_f=2$, the $\Tr SU(N_f)^3$ 't Hooft anomaly reduces to the Witten $SU(2)$ anomaly (see below).

For $G_{\rm gauge}$ theories with a non-trivial center $\Gamma _g$, we expect that the one-form center symmetry $\Gamma_g^{(1)}$ is not spontaneously broken, so the Wilson loops with non-trivial $\Gamma _g^{(1)}$ representations will have area (rather than perimeter) law.  For the case of $N_f=1$, the IR theory consists of the $h^\vee _{G_{\rm gauge}}$ isolated (supersymmetric) vacua with mass gap, and the  $\Gamma_g$ symmetry is unbroken in each of the vacua.  See e.g.~\cite{Dierigl:2014xta,Gaiotto:2014kfa} and references therein for some discussion of the interplay between the vortices (quanta of the line operators) and the domain walls in the $N_f=1$ case.   For $N_f=1$, the vortices do not arise as solitons of the IR theory.

For the $N_f>1$ theories, on the other hand, the IR theory in the SSB phase has a one-form global symmetry associated with $\pi _2(SU(N_f)/SO(N_f))$; this symmetry is independent of $G_{\rm gauge}$, unlike the $\Gamma _g^{(1)}$ center symmetry.  For $N_f=2$, the $SU(2)/SO(2)\cong \ICP ^1$ theory has $\pi _2(\ICP ^1)=\IZ$, corresponding to a $U(1)^{(1)}$ one-form global symmetry of the IR theory~\cite{Gaiotto:2014kfa}.  The $U(1)^{(1)}$ IR symmetry could arise as an enhanced version of the $\Gamma _g^{(1)}$ symmetry of the UV theory in cases\footnote{For the case of $G_{\rm gauge}= SO(4n_c)$, where $\Gamma _g=\IZ _2\times \IZ _2$, one of the $\IZ _2$ factors can be embedded in the $U(1)_{IR}^{(1)}$ and the other could require being added separately to the IR description.} where the center symmetry is $\Gamma_g = \IZ _N$.  Assuming that the $\IZ _N$ center of the UV theory indeed maps to a subgroup of  $U(1)^{(1)}$, we can ask if the embedding is to identify  the basic, $\IZ _N$ charge 1 string of the UV theory with the basic $U(1)$ charge 1 string of the IR theory.  
We will assume that this is the case.  In other words, we assume that the background gauge field $B_2^{IR}$ that couples to $n^\ast(\omega)$ in the IR is identified with the integer lift of the UV $B_2^{(N)}$ background gauge field.     For $N_f\geq 3$, the IR nonlinear sigma model has $\pi _2(SU(N_f)/SO(N_f))=\IZ _2$ one-form global symmetry, so it can only directly match the UV center symmetry $\Gamma _g$ for $G_{\rm gauge}$ with $\Gamma _g= \IZ _2$.\footnote{For the case where the UV theory is instead $Spin(N_c)$ with $N_f$ fermions in the fundamental, the UV theory instead has a $\IZ _2^{(1)}$, which nicely matches the solitonic symmetry associated with  $\pi _2(SU(N_f)/SO(N_f))=\IZ _2$ solitonic symmetry for $N_f>2$. This theory is generally not bosonic, and then the spin / statistics of lines is generally not a well-defined,  scheme-independent quantity.}  

For the particle-like solitons, associated with $\pi _3(G/H)$, we recall that $\pi _3(SU(3)/SO(3))=\IZ _4$ and $\pi _3(SU(N_f>3)/SO(N_f>3))=\IZ _2$.  In the context of $SO(N_c)$ QCD, this $\IZ _2$ was identified with baryon number~\cite{Witten:1983tw, Witten:1983tx}, and this nicely fits with how the coefficient of the WZW term affects the spin / statistics of the solitons: the $\kappa _{SU(N_f)^3}=N_c$, WZW term implies that the $\pi _3(G/H)$ soliton has $(-1)^F=(-1)^{N_c}$, befitting its identification with a baryon $B\sim \psi ^{N_c}$.  In the context of $G_{\rm gauge}$ theory with adjoints, on the other hand, the WZW term  $\kappa _{SU(N_f)^3}=|G_{\rm gauge}|$ suggests that the $\pi _3(G/H)$ soliton has $(-1)^F=(-1)^{|G_{\rm gauge}|}$ and the identification of such a soliton with a composite from the $G_{\rm gauge}$ gauge fields and adjoint fermions of the UV theory is less clear.   It was suggested in~\cite{Auzzi:2006ns, Auzzi:2008hu} that the $\IZ _2$ in the case of adjoint fermions has an exotic interpretation that we will mention further in the next section.

\subsection{Symmetry Matching for $N_f=2$ in the $\ICP ^1$ Sigma Model (on Spin Manifolds)}

For $N_f=2$ massless Weyl fermions in the adjoint representation, the fields couple to 
\eq{
G_{\rm total}=\frac{G_{\rm gauge}}{{\Gamma_g}}\times \frac{SU(2)_R\times Spin(4)\times \IZ _{2n}}{\IZ_2\times \IZ_2}~,
}
where we denote $SU(N_f=2)_{\rm global}\equiv SU(2)_R$.   The $SU(2)_{\rm gauge}$ case was discussed in~\cite{Anber:2018iof,Cordova:2018acb}, and we will review some of their discussion and extend to general  (simply connected) $G_{\rm gauge}$.   In this subsection, we consider the theory on $M_4$ with spin structure on $M_4$ with $w_2(TM)=0$.  The case with $Spin_{SU(2)_R}(4)$ structure on $M_4=\ICP ^2$ will be discussed in the next subsection.   

As in~\cite{Cordova:2018acb}, one can gain some intuition about the IR phase by starting with the ${\cal N}=2$ supersymmetric pure Yang-Mills theory, which also has $N_f=2$ Weyl fermions in the adjoint (but also with an additional complex scalar $\Phi$ in the adjoint representation), and the deform the ${\cal N}=2$ theory by a supersymmetry-breaking mass term for the (unwanted) scalars, $\Delta V\sim M^2 \Tr |\Phi |^2$.    For the $SU(2)_{\rm gauge}$ case this analysis in~\cite{Cordova:2018acb} nicely leads to unbroken $\Gamma _g^{(1)}=\IZ _2^{(1)}$ (along with confinement of the photon) from monopole / dyon condensation~\cite{Seiberg:1994rs} with $h^\vee _{SU(2)}=2$ copies of the $SU(2)_R/U(1)_R\cong \ICP ^1$ sigma model arising via a linear sigma model with the photon, with the sigma model scalars coming from the monopole hypermultiplet.  The detailed analysis for analogously deforming from $\CN =2$ for more general $G_{\rm gauge}=SU(N_c)$ is more subtle and required a deeper understanding of the Coulomb branch solution of $\CN =2$ theories for $N_c>2$~\cite{DHoker:2020qlp, DHoker:2022loi} and will be discussed in~\cite{DHoker:2023}.  
We will here simply assume the confinement and chiral symmetry breaking scenario in the IR for $N_f=2$ Weyl adjoints for all $G_{\rm gauge}$, and discuss how the anomalies are then matched in the IR nonlinear sigma model.  We will review the matching  for the $G_{\rm gauge}=SU(2)$ case, following~\cite{Cordova:2018acb}, and then generalize the discussion to general $G_{gauge}$ theories.  

The $\kappa _{SU(N_f)^3}$ 't Hooft anomaly, which for $N_f\geq 3$ is matched by the WZW term~\eqref{adjointwzw} in the SSB phase, reduces to the Witten anomaly for $N_f\to 2$:  $\kappa _{SU(2)^3}\to  |G_{\rm gauge}|$ mod 2.  Correspondingly, the WZW term~\eqref{adjointwzw} reduces for $N_f\to 2$ to a  topological term related to $\pi _4(\ICP ^1)=\IZ _2$~\cite{Witten:1983tx,Cordova:2018acb}; in fact, it is given by the bordism class $\tilde \Omega _4^{\rm spin}(S^2)=\IZ _2$ see~\cite{Freed:2006mx,Lee:2020ojw}.   The $N_f\geq 3$ WZW term that matches the 't Hooft anomaly leads by transgression~~\cite{Freed:2006mx, Lee:2020ojw} to the $N_f=2$ case of the $\pi _4(SU(2))=\IZ _2$ matching anomaly.  In particular, for the $\ICP ^1$ sigma model in the context of $G_{\rm gauge}=SU(2)$ with adjoints~\cite{Cordova:2018acb}, and likewise for $SO(N_c)$ with fundamentals~\cite{Lee:2020ojw}, the $SU(2)_R$ Witten anomaly is matched in the $\ICP ^1$ sigma model by the coefficient of a $\IZ _2$ valued theta term corresponding to $\pi _4(\ICP ^1)=\IZ _2$.  For the case of general $G_{\rm gauge}$ with $N_f=2$ Weyl fermions in the adjoint, we write the  WZW term as
\eq{\label{adjointwzw2} e^{-S_{WZW}}=(-1)^{|G_{\rm gauge}|[\phi : M_4\to S^2]}}
which matches for the IR theory the $SU(2)_R$ Witten $\kappa _{\rm Witten}=|G_{\rm gauge}|$ (mod 2)  't Hooft anomaly of the UV theory.  As in~\cite{Lee:2020ojw}, the notation $[\phi : M_4\to S^2]$ denotes the reduced bordism class in $\widetilde \Omega _4^{\rm spin}(S^2)=\IZ _2$, which is integral valued so the term in~\eqref{adjointwzw2} is only non-trivial for $|G_{\rm gauge}|$ odd, reflecting the fact that the Witten anomaly is $\IZ _2$ valued. 

For QCD in the SSB IR phase,  particle-like solitons of the IR non-linear sigma model, associated with $\pi _3(G/H)$, can be skyrmionic realizations of the baryons of the UV theory, with $\pi _3(G/H)$ charge identified with $U(1)_B$ (or a $\IZ _2$ version for $SO(N_c)$ QCD with $N_f$ fundamentals)~\cite{Witten:1983tx}.  For the present case of $G/H=SU(2)/U(1)\cong \ICP ^1$, the IR sigma-model is essentially the Fadeev-Hopf model (which has quartic terms to ensure soliton stability), whose Hopfion solitons are toroidal loops or knots~\cite{Faddeev:1996zj}.   The fact that the solitons have Hopf winding number\footnote{It is measured by the Chern-Simons number $\int a da/4\pi ^2$ if one obtains $SU(2)/U(1)$ via a $U(1)$ linear sigma model with gauge field $a$.}  $\pi _3(\ICP ^1)=\IZ$ initially suggests an associated $U(1)_H$ Hopf number global symmetry, which is not present in the UV theory.   As discussed in~\cite{Freed:2017rlk} in the related context of the 3d $\theta$ angle, and in 4d in ~\cite{Cordova:2018acb,Chen:2022cyw} there is actually no $U(1)_H$ global symmetry of the IR theory either:  only a $\IZ _2\subset U(1)_H$ Hopfion number is conserved.  As discussed in~\cite{Chen:2022cyw}, this agrees with the spin bordism class $\widetilde \Omega ^{\rm spin}_3(\ICP ^1)=\IZ _2$, and the $\IZ _2$ is part of a larger,  non-invertible global symmetry.

The spin / statistics of the Hopf soliton has been much-studied in the literature, see e.g.~\cite{Wilczek:1983cy, Krusch:2005bn, Bolognesi:2009vm} and references therein.  Although there is no standard WZW to determine the spin / statistics of the solitons as in the case of the baryonic skyrmions~\cite{Witten:1983tw, Witten:1983tx},  the $\pi _4(S^2)$ Hopf-WZW term~\eqref{adjointwzw2} associated with matching the $SU(2)_R$ Witten anomaly is essentially a WZW term (and related by transgression).  This Hopf-WZW term 
suggests that the basic $\IZ _2$-charged Hopfion soliton has $(-1)^F= (-1)^{|G_{\rm gauge}|}$, i.e. it is a fermion for $|G_{\rm gauge}|$ odd and a boson for $|G_{\rm gauge}|$ even.   This was discussed for $G_{\rm gauge}=SU(2)$ case in~\cite{Cordova:2018acb}.   
The works~\cite{Auzzi:2006ns, Bolognesi:2007ut, Auzzi:2008hu, Bolognesi:2009vm}, on the other hand, emphasized that Hopfions admit an alternative,  fermionic quantization~\cite{Finkelstein:1968hy, Krusch:2005bn} which makes use of the noncontractible $\pi _1$ configuration space of the Hopfion solutions (the $\pi _3(S^2)$ Hopf fibration solitons are essentially loops of the $\pi _2(S^2)$ vortices).     These works have a speculative interpretation of the $\IZ _2$ Hopfion 0-form global symmetry as being associated with a conjectural  exotic sector of the Hilbert space in which the spin-charge relation is violated, rather than identifying the $\IZ _2$ with $(-1)^F$.   Solutions of equations can of course be in a non-trivial representation of the underlying symmetry, e.g. the Hopfion toroidal solutions vs the full rotational symmetry.  But we do not expect that the IR $\IZ^{(0)} _2$ global symmetry is related to spin-charge violation.   For example, consistency of the Donaldson-Witten twist of ${\cal N}=2$ SYM relies on non-violation of the generalized spin structure.  As far as we are aware,  the well-studied BPS soliton spectrum of ${\cal N}=2$ supersymmetric gauge theories does not exhibit violation of the spin-charge relation.\footnote{The spin-charge relation for ${\cal N}=2$ involves both the gauge charge and the spin charge. E.g. for ${\cal N}=2$ supersymmetric $SU(2)_{\rm gauge}$ QCD,  the generalized spin structure is $Spin_{SU(2)_R\times SU(2)_{\rm gauge}}$ and the spin charge relation is $(-1)^F=(-1)^Q(-1)^{F_R}$ where $Q$ is the electric or magnetic charge of the state and $F_R$ is the charge under $U(1)\subset SU(2)_R$.}  

The $\ICP ^1$ sigma model has a $U(1)^{(1)}$ one-form global symmetry~\cite{Gaiotto:2014kfa} that assigns charge to solitonic strings (these are the ANO vortices if $\ICP ^1$ is realized via a linear sigma model of a $U(1)$ gauge theory).  The associated closed 2-form is $*J^{(2)}=\phi ^*(\omega) $, where the 2-form $\omega$ is the $\ICP ^1$ K\"ahler form, and the charge of the solitonic winding configurations is the $\pi _2(\ICP ^1)=\IZ$ winding number.  The UV theory has the one-form global symmetry $\Gamma _g^{(1)}$, where $\Gamma _g$ is the $G_{\rm gauge}$ center symmetry, and if  $\Gamma _g ^{(1)}= \IZ _N^{(1)}$, it can be embedded in the apparent $U(1)^{(1)}$ of the IR theory.   With this identification, the background gauge field $B_2^{(N)}$ for the UV symmetry $\Gamma _g^{(1)}=\IZ_N^{(1)}$ is expected to\footnote{In the case of  $U(1)_B$ in QCD, the analogous coupling between the background gauge field and the $G/H$ soliton the IR is dictated by matching the $\Tr U(1)_BG^2$ 't Hooft anomaly.} couple in the IR $\ICP ^1$ sigma model's functional integral as  
\eq{\label{Bncoupling}
\exp \left( i \int _{M_4} B_2^{IR}\wedge \phi ^*(\omega)\right) \to \exp \left( \frac{2\pi i }{N} \int _{M_4} B_2^{(N)}\cup \phi ^*(\omega)\right)~.
}
Here, much as in the $SU(2)$ discussion in~\cite{Cordova:2018acb},  $B_2^{(N)}\in H^2({M}_4, \IZ _N)$ is the $\IZ _N^{(1)}$ background gauge field of the UV theory that can be embedded in $U(1)^{(1)}$ via $B_2^{IR}=\frac{2\pi}{N} B_2^{(N)}$. 
The  K\"ahler 2-form of the $\ICP ^1$ in the presence of the $SU(2)_R$ background gauge field $A_{SU(2)_R}\equiv A$ is~\cite{Cordova:2018acb} 
\eq{\label{nis}
\phi ^\ast(\omega)\equiv n^\ast (\omega) =\frac{1}{8\pi}(\epsilon_{IJK}n^Id_R n^Jd_R n^K-2n^I F^I)~,
}
where  the $\ICP ^1\cong S^2$ sigma model is written in terms of a unit vector\footnote{For general $G/H$, we prefer $\phi$ for the pions, but for $\ICP ^1$ we take $\phi \to n$ to match the notation of~\cite{Cordova:2018acb}.} $n^I$ transforming in the $\bf 3$ of $SU(2)_R$, with covariant $d_R n^I=d n^I+\epsilon_{IJK}A^Jn^K$ and $F^I=dA^I+\half \epsilon_{IJK}A^J\wedge A^K$.

The IR theory also admits a $\theta$ term, which multiplies the functional integral by~\cite{Cordova:2018acb}
\eq{\label{cptheta}\exp \left(\frac{i\theta}{2} \int _{{M}_4}n ^*(\omega)\wedge n ^*(\omega)\right),}
which by the time-reversal symmetry of the UV theory could take the values $\theta =0$ or $\theta =\pi$. For $G_{\rm gauge}=SU(2)$, it was noted in~\cite{Cordova:2018acb} that RG flow from ${\cal N}=2$ SYM suggests that $\theta =\pi$.

\subsection{Symmetry Matching on $M_4=\ICP ^2$ with Twisted, $Spin_{SU(2)_R}(4)$ Structure}

The $G_{\rm gauge}$ theory with adjoint fermions is not a bosonic theory, e.g. $\Tr (\lambda _\alpha ^f F_{\mu \nu})$ is a gauge invariant fermionic operator.  As emphasized in~\cite{Brennan:2022tyl}, in fermionic theories neither the spin nor the statistic of line operators are scheme-independent.   For even $N_f$, on the other hand, the identification $(-1)^F\sim -{\mathds 1}_{SU(N_f)}$ allows assigning a scheme independent combined $(-1)^F\circ (-{\mathds 1}_{SU(N_f)})$ charge to line operators in addition to admitting the twisted generalized $Spin_{SU(N_f)}(4)$ structure.  As mention in the previous subsection, we expect that the $Spin_{SU(N_f)}(4)$ structure is preserved along the RG flow for all $G_{\rm gauge}$, and that all of these theories can thus be placed on $M_4=\ICP ^2$ with $w_2(SU(N_f))=w_2(TM)$.  In  this subsection, we discuss how the IR $\ICP ^1$ sigma model can match the 
symmetries, and the $\kappa _{\Gamma, T}$ and $\kappa _{T, T}$ 't Hooft anomaly coefficients in~\eqref{sect5anomalies} that we have computed in previous sections,  when the theories are formulated with twisted, $Spin_{SU(2)_R}(4)$ structure on non-spin manifolds. The case $G_{\rm gauge}=SU(2)$ was discussed in~\cite{Cordova:2018acb} and we will here consider general $G_{\rm gauge}$.

As discussed in~\cite{Cordova:2018acb}, the expression~\eqref{nis} for $n^\ast (\omega)(A_R^{(1)})$ in the presence of $SU(2)_R$ background gauge fields, e.g. in the vacuum configuration with $n^I=\delta ^{I3}$, shows that 
$2n^\ast(\omega)$ is an integral cohomology class that, with the $Spin_{SU(2)_R}(4)$ structure, satisfies (mod 2): $2n^\ast \cong w_2(SU(2)_R)\cong w_2(TM)$.  In the IR theory, this is a constraint on the functional integral over the pion fields. The coupling~\eqref{Bncoupling} of $n^\ast(\omega)$ to the background gauge field $B_2^{(N)}$ for the one-form global symmetry can be modified by a counterterm, replacing $n^\ast(\omega)$ with $(n ^*(\omega) + \half \tilde w _2(TM))$, which is integral valued as needed so that the coupling is invariant under  $\IZ _N^{(1)}$ background gauge transformations of $B_2^{(N)}$~\cite{Cordova:2018acb}:
\eq{\label{newBncoupling} 
\exp \left( \frac{2\pi i }{N} \int _{{M}_4} B_2^{(N)}\cup n ^*(\omega)\right)\to  \exp \left( \frac{2\pi i }{N} \int _{{M}_4} B_2^{(N)}\cup (n ^*(\omega) + \half \tilde w _2(TM)) \right)~, 
}
where $\tilde w_2(TM)$ is an integral lift of $w_2(TM)$.  The $\theta$ term~\eqref{cptheta} also requires a modification, with an additional term $-i\frac{\theta}{8}\sigma ({M}_4)$ involving the signature of $M_4$~\cite{Cordova:2018acb}.   For $G_{\rm gauge}=SU(2)$, the appearance of $w _2(TM)$ in~\eqref{newBncoupling} correctly reproduces the $\kappa _{\Gamma , T}=1$ 't Hooft anomaly~\cite{Cordova:2018acb}. 

For general $\Gamma _g^{(1)}=\IZ _N^{(1)}$ of UV theories, the coupling~\eqref{newBncoupling}, which relies on identifying the UV $\IZ _N^{(1)}$ in an enhanced IR $U(1)^{(1)}$ of the $\ICP ^1$ sigma model, naturally leads to anomaly $\kappa _{\Gamma, T}=1$.   Recall that a $\kappa _{\Gamma, T}$ anomaly implies a variation of the partition function under either shifts of integral lifts $\tildeB_2^{(N)}\to \tildeB_2^{(N)}+Nx$ or $\tilde w_2(TM) \to \tilde w_2(TM) +2y$:
\eq{\label{anomchoices}
Z\longmapsto Z\times (-1)^{\kappa _{\Gamma, T} \int x \cup w_2(TM)} \qquad\hbox{or}\qquad Z\longmapsto Z\times e^{\frac{2\pi i \kappa _{\Gamma, T}}{N} \int B_2^{(N)} \cup y}~,
}
where $x$ and $y$ are integral cohomology classes or cochains.  These two options differ by a counterterm,  which is the latter term in~\eqref{newBncoupling}; it gives the anomalous variation as the latter one in~\eqref{anomchoices}, under 
 $\tilde w_2(TM) \to \tilde w_2(TM) +2y$ shifts.  When $N$ is odd, all anomalous variation can be cancelled by the local counter term $S_{c.t.}=\frac{2\pi i n}{N}\int \tildeB_2^{(N)}\cup \tilde w_2(TM)$ where $2n=1~{\rm mod}_N$.  As we have discussed, $\kappa _{\Gamma , T}$ is associated with flux in a $\IZ _2^{(1)}\subset \Gamma _g^{(1)}$, which requires even $N$. 
 
 In summary, the IR analysis readily leads to $\kappa _{\Gamma, T}=1$ for $N$ even.  On the other hand, we have computed the $\kappa _{\Gamma, T}$ in the UV description and found  that $\kappa _{\Gamma, T}=1$ only if $\Gamma_g=\IZ _N$ with $N\in 4\IZ +2$.  
 The { apparent mismatch} for $N\in  4\IZ $ requires a compensating effect.  One option is the addition of a TQFT in the IR sigma model in those cases~\cite{Cordova:2019bsd,Brennan:2023kpo}. Alternatively, it could be that matching of the center symmetry $\Gamma_g^{(1)}$ of the UV theory to the enhanced $U(1)^{(1)}$ symmetry of the IR sigma model requires modification.

\section{ Symmetric Mass Generation}
\label{sec:SMG}

A chiral gauge theory has symmetric mass generation if it generates an IR mass gap despite not admitting quadratic mass terms; this can happen if 't Hooft anomalies do not rule out the IR mass-gapped phase.  In some classes of examples, the IR mass gap can be generated by e.g. 4-fermion interactions \cite{Razamat:2020kyf,Tong:2021phe,Smith:2021vbf}, see also \cite{Fidkowski:2009dba,FidowskiKitaev, Zeng:2022grc, You:2017ltx,Wang:2022ucy} for earlier work and references.   

We will here consider the $Spin_G$ 't Hooft anomalies are non-zero in variants of these theories that admit a generalized spin structure.    A class theories that have been proposed to lead to symmetric mass generation are the symmetric with fundamentals examples discussed in~\cite{Tong:2021phe}.  One starts with $SU(N)$ gauge theory with a single fermion $\lambda$ in the 2-index symmetric representation $\Sym^2\mathbf{N}$, and cancels the $\Tr SU(N)^3$ gauge anomaly via $N+4$ fermions $\psi$ in the $\mathbf{\overline{N}}$ representation. The global $SU(N+4)$ symmetry can be made anomaly free by coupling to $\half(N+4)(N+3)$ uncharged fermions that transform under $SU(N+4)$ in the $\Lambda^2\mathbf{\overline{N+4}}$ (two-index antisymmetric). In all, the fermion content is given by 
\begin{center}
\begin{tabular}{c|cccc}
&$SU(N)_g$ & $SU(N+4)_R^{(0)}$ & $U(1)_r^{(0)}$\\
\hline
$\lambda$ & $\Sym^2\mathbf{N}$ & 1 & $-(N+4)$\\
$\psi$ & $\mathbf{\overline{N}}$ & $\mathbf{N+4}$ & $N+2$\\
$\chi$ & 1 & $\Lambda^2\mathbf{\overline{N+4}}$ & $-N$
\end{tabular}
\end{center}
\noindent Here $U(1)_r$ is also a non-anomalous global symmetry.  The 't Hooft anomalies for these symmetries all-vanish, and they and are all preserved by the 4-fermion interaction term
\eq{\label{fourfermion}
\CL_{\rm int}=\frac{1}{\Lambda^2}\psi \chi\psi \lambda+c.c. 
}
The theory with this interaction has no obstruction to an IR mass gap, and the proposal of symmetric mass generation is that the term~\eqref{fourfermion} indeed generates a mass gap. This can be intuitively understood if 
at some energy scale $\Lambda_{QCD}$, $SU(N)_g$ confines and IR phase is described by $\chi$ together with the  $SU(N)_g$  neutral composite  fermion 
\eq{
\Psi=\frac{1}{\Lambda_{QCD}^3}\psi\chi\psi~,
}
that transforms under the $\Lambda^2(\mathbf{N+4})$ of $SU(N+4)_R^{(0)}$. The 4-fermion interaction $\CL_{int}$ would then RG flow to a mass term for $\Psi,\chi$, leading to the conjectured trivially gapped IR phase. 

We here consider variants of the above theory, with $K$ generations, i.e. $K$ copies, of the  $\lambda$, $\psi$, and $\chi$ matter content.  The $SU(N)_g$ gauge coupling is asymptotically free if $K(N+3)<{11\over 2}N$.  For $K$ sufficiently close to the upper limit of the bound will presumably put the theory in a conformal  window, where it RG flows to an IR CFT.   We are here more interested in a possible confining gapped IR phase, with $K$ below the conformal window.  We will in particular focus on $K=2$ and can further disfavor a possible IR CFT phase by taking $N$ sufficiently large.  The $K=2$ generations leads to a global $SU(2)^3$ symmetry acting on each matter field, and the four-fermion interaction~\eqref{fourfermion} can be chosen to preserve only a diagonal $SU(2)_f$.  We will use $SU(2)_f$ to have a $Spin_{SU(2)_f}(4)$ structure.  We thus consider 
\begin{center}
\begin{tabular}{c|cccc}
&$SU(N)_g$ & $SU(N+4)_R^{(0)}$ &$SU(2)_f$ & $U(1)_r^{(0)}$\\
\hline
$\lambda$ &  $\Sym^2\mathbf{N}$ & 1 &$\mathbf{2}$&$-(N+4)$\\
$\psi$ & $\mathbf{\overline{N}}$ & $\mathbf{N+4}$ &$\mathbf{2} $ & $N+2$\\
$\chi$ & 1 & $\Lambda^2\mathbf{\overline{N+4}}$ &$\mathbf{2} $ & $-N$
\end{tabular}
\end{center}

\noindent Note that $SU(2)_f$ has no Witten anomaly, no WWW $SU(2)$ anomaly, and no mixed anomaly with $U(1)_r^{(0)}$.  Since all anomalies vanish, one might conjecture an IR gapped phase here too.  

For $N=2n$ there is a common $\IZ_2$ shared between $SU(2n)_g\times SU(2n+4)_R$, $SU(2)_f$, and $(-1)^F$.  The fields couple to $G_{\rm total}$-bundles with 
\eq{
G_{\rm total}=\frac{SU(2n)_g\times SU(2n+4)_R}{\IZ_2}\times \frac{SU(2)_f\times Spin(4)}{\IZ_2}~. 
}
This allows a generalized $Spin_{SU(2)_f}(4)$ structure and a background $B_2$ field with\footnote{Note that here we are turning on a correlated flux in the dynamical gauge field and background gauge field for $SU(2n+4)_R$ which does not correspond to a 1-form global symmetry. The interpretation of such flux backgrounds has been recently discussed in \cite{Brennan:2023tae,Cherman:2017tey,Brennan:2023ynm}.}
\eq{
w_2(TM)=w_2(SU(2)_f)\quad, \quad B_2=w_2(g)=w_2(R)~. 
} 
We can now compute the $w_2w_3$-type 't Hooft anomaly of this theory on $\ICP^2$ by computing the index of the Dirac operator in an appropriate flux background.  E.g. for $N=2n=2$, considering a diagonal $U(1)_Q\subset U(1)_\Gamma\times U(1)_{\Gamma_g}$, the fermions have the charges 
\eq{
q_\lambda&=(1,0,-1)\otimes\left(\half,-\half\right)= \left(\frac{3}{2},-\frac{3}{2}\right)\oplus \left(\half,-\half\right)^{\oplus 2}~, \\
q_{\psi}&=\left(\half,-\half\right)\otimes \left(\half,\half,\half,-\half,-\half,-\half\right)\otimes \left(\half,-\half\right)=\left(\frac{3}{2},-\frac{3}{2}\right)^{\oplus 3}\oplus \left(\half,-\half\right)^{\oplus  9}~,\\
q_\chi&=\left(\half,-\half\right)\otimes \Big((1,-1)^{\oplus 3}\oplus 0^{\oplus 9}\Big)=\left(\frac{3}{2},-\frac{3}{2}\right)^{\oplus 3}\oplus \left(\half,-\half\right)^{\oplus 12}
} 
This leads to contributions to the index $\fI_\lambda=2$, $\fI_\psi=6$, $\fI_\chi=6$ and anomaly indicator 
\eq{
\sigma_\fI=\frac{\fI_\lambda+\fI_\psi+\fI_\chi}{2}~{\rm mod}_2=1~.
} 
Therefore, we see that this theory has a non-perturbative anomaly $\kappa _{\Gamma _g, T}=1${
\eq{
\CA\subset\pi i \int B_2\cup w_3(TM)~. 
}
Therefore the two-generation theory with symmetry preserving 4-fermion operator cannot be trivially gapped. For $N=2$ this two-generation case is close to the asymptotic freedom bound, so it might be expected to be in the conformal window rather than trivially gapped.  
Two-generation cases with larger $N$ are farther from the asymptotic freedom bound, so perhaps not in the conformal window, and similarly can have $\kappa _{\Gamma, T}\neq 0$, preventing a trivially gapped IR phase. Of course, it is still possible that the one-generation case is trivially gapped. 

\section*{Acknowledgements}

We thank Greg Moore, Misha Shifman, and Juven Wang for helpful comments, and especially Clay C\'ordova and Thomas Dumitrescu for helpful discussions and related collaborations. DB and KI are supported in part by the Simons Collaboration on Global Categorical Symmetries, and by Simons Foundation award 568420.  KI is also supported by DOE award DE-SC0009919. 

\appendix

\section{Fractionalization and the Example of $Spin(4n_c+2)$ Vector-QCD}

\label{app:fractionalization}

In this appendix we will briefly describe the fractionalization technique \cite{Wang:2018qoy, Brennan:2022tyl,Delmastro:2022pfo} to compute discrete anomalies, and detail a few computations that we quoted in the main text. 

Consider $G_{\rm gauge}$ gauge theory with $N_f$ fermions in the representation $R_{\rm gauge}$ such that $G_{\rm gauge}$ is simply connected and  the fields couple to $G_{\rm total}$-bundles of the form  
\eq{
G_{\rm total}=\frac{\frac{G_{\rm gauge}}{\Gamma_g}\times G_{\rm global}\times Spin(4)}{\Gamma}~,
}
where $\Gamma_g=\IZ_n$. This theory has a 1-form global symmetry corresponding to $\Gamma_g$ and $Spin_{G_{\rm global}}$ structure corresponding to $\Gamma$.   We now consider adding an adjoint-valued scalar field that also couples to the fermions via a Yukawa interaction. A vev for the adjoint scalar generically breaks the gauge group $G_{\rm gauge}\mapsto U(1)^r$, where $r$ is the rank of $G_{\rm gauge}$, and gives the fermions a mass. The IR abelian theory has an emergent $G_{IR}^{(1)}=\left(U(1)^{(1)}_e\times U(1)^{(1)}_m\right)^r$ 1-form global symmetry with mixed 't Hooft anomalies as in~\cite{Gaiotto:2014kfa}
\eq{\label{mixedanom1form}	
\CA\subset\itwopi \sum_{I=1}^r\int B_{2,I}^{(e)}\wedge d B_{2,I}^{(m)}~. 
}
The 1-form and $Spin_G$ symmetries of the UV theory are preserved along this RG flow and can be tracked to the IR abelian gauge theory by identifying the quantum numbers of the emergent line operators.  The discrete fluxes for $\Gamma,\Gamma_g$, which we denote $w_2(\Gamma)$ and $B_2^{UV}$ respectively,  can activate fluxes for subgroups of the emergent $G_{IR}^{(1)}$.  This can be used to exhibit the existence of $w_2w_3$ and other 't Hooft anomalies of  the UV theory. 

Let us explain how to track the $\Gamma,\Gamma_g$ fluxes along the RG flow from $G_{\rm gauge}\to U(1)^r$. We first discuss how they activate  the IR electric 1-form fluxes. Note that it is clear how $\Gamma_g,\Gamma\subset G_{\rm gauge}$ embed into $U(1)^r\subset G_{\rm gauge}$ since $U(1)^r$ is identified as the Cartan torus of $G_{\rm gauge}$.   Let us focus on a single $U(1)_{\rm gauge}$ factor, and consider a unit charge Wilson line $W_1[\gamma]$. This Wilson line is charged under the corresponding 1-form electric symmetry $U(1)^{(1)}_e$. Let us introduce a 2-form background gauge field for $U(1)^{(1)}_e$: $B_2^{(e)}$. The statement that $W_1[\gamma]$ is charged under $U(1)_e^{(1)}$ means that it transforms under $B_2$ gauge transformations as 
\eq{
\delta B_2^{(e)}=d\Lambda_1^{(e)} \quad, \quad W_1[\gamma]\longmapsto W_1[\gamma]\times e^{ i \int_\gamma \Lambda_1^{(e)}}~.
}

Along the RG flow, the Wilson line can experience \emph{charge fractionalization}. This means that, if we introduce a $\Gamma$-background gauge field $B_2^{(\Gamma)}$, the Wilson line will transform under the $B_2^{(\Gamma)}$-background gauge transformation as
\eq{
\delta B_2^{(\Gamma)}=d\Lambda_1^{(\Gamma)}\quad, \quad W_1[\gamma]\longmapsto W_1[\gamma]\times e^{i \int_\gamma \Lambda_1^{(\Gamma)}}~. 
}
The only way this can be matched is if we enact the $\Gamma,\Gamma_g$ transformations with the  $U(1)_e^{(1)}$ transformation. This means that if the UV operator that flows to the IR Wilson line is charged under $\Gamma,\Gamma_g$, then $\Gamma,\Gamma_g$ flows (at least in part) to $\Gamma,\Gamma_g\subset U(1)_e^{(1)}$. So turning on a background gauge field $B_2^{(\Gamma)},B_2^{(\Gamma_g)}$ turns on a $B_{2}^{(e)}$ depending on the transformation property of $W_1[\gamma]$.  Now we wish to determine the charge fractionalization of the Wilson line. This differs between the two cases:
\begin{itemize}
\item $\Gamma_g$-non-trivial: In this case the fundamental Wilson line of $G_{\rm gauge}$ maps to the minimal Wilson line of the IR theory. This means that $\widehat\Gamma$ populates a diagonal subgroup $\widehat\Gamma\subset (U(1)^{(1)}_e)^r$ where $\widehat\Gamma$ is the extension
\eq{
1\longrightarrow \Gamma \longrightarrow \widehat\Gamma \longrightarrow \Gamma_g\longrightarrow 1~.
}
This is the case as discussed in Section \ref{sec:SpinvecQCD} where $\Gamma_g$ forms a 2-group with the $G_{\rm global}\times Spin(4)$. 

\item $\Gamma_g$ trivial: Here the fundamental Wilson line of the IR theory is described by the world volume of the fundamental fermion in the UV theory. When $\Gamma$ does not act on $G_{\rm gauge}$, the Wilson line will not have any fractionalization and $w_2(\Gamma)$ does not activate $B_2^{(e)}$. On the other hand, when $\Gamma$ acts on $G_{\rm gauge}$, it identifies part of the center of $G_{\rm gauge}$ with part of  $Z(G_{\rm global}\times Spin(4))$. Then, since the center of the gauge group flows to the diagonal subgroup of the 1-form electric symmetry in the IR, the subgroup $Z(G_{\rm global})^{(1)}\subset G_{IR}^{(1)}$ would be identified with part of  $Z(G_{\rm global}\times Spin(4))$ and $w_2(\Gamma)$ activates $B_2^{(e)}$. However, if $\Gamma$ acts trivially on $G_{\rm gauge}$, then there can be no such identification. 

\end{itemize}

\noindent Consider the special case where $\Gamma_g=\IZ_N$ and the $\Gamma=\IZ_2$ acts trivially on $G_{\rm gauge}$. Let us pick $U(1)_{\Gamma_g}\subset U(1)^r\subset G_{\rm gauge}$ where ${\Gamma_g}$ embeds into $U(1)_{\Gamma_g}$ and let $Q_{\Gamma_g}$ be the generator of $U(1)_{\Gamma_g}$ which further decomposes as 
\eq{
Q_{\Gamma_g}=\sum_I Q_{\Gamma_g}^{(I)}H_I~,
}
where $H_I$ are simple coroots of $G_{\rm gauge}$ that generate the IR $U(1)^r$.  Then the minimal UV Wilson line will flow to a minimal IR Wilson line and we can identify the IR 1-form electric fluxes as
\eq{
\frac{B_{2,I}^{(e)}}{2\pi}=\frac{1}{N}B_2^{UV}\,Q_{{\Gamma_g}}^{(I)}\quad \Longrightarrow\quad\frac{ B_{2}^{(e)}}{2\pi}=\frac{1}{N}B_2^{UV}Q_{\Gamma_g}~,
}
where again $B_2^{UV}$ is the background gauge field for the UV 1-form global symmetry. 

Now let us turn to the IR 1-form magnetic symmetries. The line operators that are charged under the 1-form magnetic symmetry are the magnetic line operators. Since we assume that $G_{\rm gauge}$ is simply connected, the magnetic line operators in the IR come from smooth magnetic monopoles that arise when the adjoint scalar field condenses. Unlike the Wilson line operators, the magnetic line operators acquire charge fractionalization from the fermion zero-modes that  localize on its world volume. 

The magnetic 1-form global symmetries are generated by weights that are dual to the generators of the $U(1)$ factors. The background magnetic gauge field to $U(1)_{\Gamma_g}$ is valued 
\eq{ 
\frac{B_2^{(m)}}{2\pi}=\frac{\sigma_f}{2}w_2(\Gamma)\,Q_{\Gamma_g}^\vee\quad, \quad Q_{\Gamma_g}^\vee=\sum_I (Q_{\Gamma_g}^\vee)_I\lambda^I\quad, \quad (H_I,\lambda^J)=\delta_I^{~J}~, 
}
where $Q_{\Gamma_g}^\vee$ is the dual weight to $Q_{\Gamma_g}$. In general, $Q_{\Gamma_g}$ which can be expressed as a (fractional) sum of the dual weights, $\lambda^I$ and $(~,~)$ is the natural pairing co-roots and dual weights. Here $\sigma_f=0,1$ is the spin indicator as in \cite{Brennan:2022tyl} which indicates whether the monopole is a fermion ($\sigma_f=1$) or a boson ($\sigma_f=0$). 

We can compute the fermion zero-modes as follows. Consider the monopole $M_{Q_{\Gamma_g}}[\gamma]$ with magnetic charge $Q_{\Gamma_g}$.  The fermions decompose into representations under $U(1)_{Q_{\Gamma_g}}$ with respect to their magnetic charge pairing with $Q_{\Gamma_g}$:
\eq{
\mathbf{R}\longmapsto \bigoplus_i p_i\quad, \quad \psi\longmapsto \sum_i \psi_i v_i~,
}
where again the $v_i$ are the basis vectors of the representation $R$ which have corresponding charges $ Q_{\Gamma_g}\cdot v_i=p_i\,v_i$. When there are $N_f$ fermions in the representation $R$, which transforms under a subgroup of $G_{\rm global}\subset SU(N_f)$, the number of fermion zero-modes given by the Callias Index Theorem \cite{Callias:1977kg} is:
\eq{\label{CalliasIndex}
\CI(\mathbf{R}\otimes \mathbf{N_f})=\{\text{\# of $\IR$ fermion z.m.}\}=N_f\sum_{i\,|\,p_i>0}p_i~. 
}
These zero-modes transform under $ G_{\rm global}\times Spin(3)_{\rm rot}$, where $Spin(3)_{\rm rot}\subset Spin(4)$ is the rotation group preserved by the monopole line configuration,  as \cite{Moore:2014jfa,Brennan:2022tyl}
\eq{
\mathbf{N_f}\otimes \bigoplus_{i\,|\,p_i>0}\mathbf{p_i}~.
}

These fermion zero-modes give rise to a quantum mechanics that is localized on the monopole which transforms under the $G_{\rm global}\times Spin(3)_{\rm rot}$. The charge fractionalization of the monopole is given by the representation of the  Hilbert space associated to this quantum mechanics: $\CH_{mono}$. The charge of the monopole with respect to $\Gamma$ is determined by whether or not $\CH_{mono}$ transforms projectively under the $G_{\rm sub}\subset G_{\rm total}$ subgroup 
\eq{G_{\rm sub}=\frac{G_{\rm global}\times Spin(3)_{\rm rot}}{\IZ_2}~.}
To any projective representation $R_{\rm proj}$ of $G$, we can associate a cohomology class $w_2[R_{\rm proj}]\in H^2(BG_{\rm sub};\IZ_2)$. When $\CH_{mono}$ transforms projectively with representation $R_{\rm proj}$ under $G_{\rm sub}$, we can identify a cohomology class $w_2[R_{\rm proj}]\in H^2(BG;\Gamma)$ which corresponds to a 1-form magnetic flux
\eq{
\frac{B_{2}^{(m)}}{2\pi}=\frac{1}{2}w_2[R_{\rm proj}]Q_{\Gamma_g}^\vee~.
}
For our case where $\Gamma=\IZ_2$ acts trivially on $G_{\rm gauge}$,    there is a $\IZ_2$ possibility of charge fractionalization corresponding to whether or not $\CH_{mono}$ transform projectively under $G_{\rm sub}$. Physically this means the monopole  is a fermion, or is in a projective $G_{\rm global}/\IZ_2$ multiplet. If we introduce a spin/fractionalization indicator 
\eq{
\sigma_f=\begin{cases}
1&\text{$\CH_{mono}$ is proj. rep. of $G_{\rm sub}$}\\
0&else
\end{cases}
}
then we can succinctly write the 
magnetic background flux as 
\eq{
\frac{B_{2}^{(m)}}{2\pi}=\frac{\sigma_f}{2}w_2(TM)Q_{\Gamma_g}^\vee~.
}
Plugging these into the anomaly formula \eqref{mixedanom1form}, we find the resulting anomaly  
\eq{
\CA\subset\itwopi \int B_2^{(e)}\wedge dB_{2}^{(m)}=\pi i\, \sigma_f \int B_2^{UV}\cup w_3(TM)~. 
}  
See \cite{Brennan:2022tyl} for more details on this approach to computing non-perturbative anomalies.

\subsection{Anomaly for $Spin(4n_c+2)$ Vector QCD}

Now let us compute the anomaly in $Spin(4n_c+2)$ QCD with $2n_f$ vector fermions to complete the discussion in Section \ref{sec:SpinvecQCD}. The fields couple to $G_{\rm total}$-bundles:
\eq{
G_{\rm total}= \frac{\frac{Spin(4n_c+2)}{\IZ_2}\times SU(2n_f)\times Spin(4)}{\IZ_2^F\times \IZ_2^C}~, 
}
where $\IZ_2^F$ identifies $(-1)^F\sim -\mathds{1}_{SU(2n_f)}$ and $\IZ_2^C$ identifies $(-1)^F\sim -\mathds{1}_{Spin(4n_c+2)}$. Here we will focus on the case of $Spin_{G_{\rm gauge}}(4)$. In this case we can activate the obstruction class for lifting $PSO(4n_c+2)$ to $Spin(4n_c+2)$-bundles which is $\IZ_4$-valued, but we are not allowed to activate the flux for $w_2(\IZ_2^F)$. The 2 independent fluxes for $B_2$ and $w_2(\IZ_2^F)=w_2(TM)$ then parametrize the $\IZ_4$ flux as 
\eq{
w_2(\IZ_4)=2 B_2+w_2(TM)~. 
}
To compute the anomaly, we will deform the theory by adding an adjoint valued scalar field with Yukawa coupling -- which breaks $SU(2n_f)\to Sp(n_f)$ -- and condense the scalar field so that we flow to pure abelian gauge theory. 
First, since the fermions do not screen Wilson lines in the spin representation of $Spin(4n_c+2)$, we can identify the minimal Wilson lines of the IR theory with the Wilson lines of the UV theory. These lines are charged under $w_2(\IZ_4)$ so that we can identify 
\eq{
\frac{B_2^{(e)}}{2\pi}=\frac{1}{4}\left(w_2(TM)+w_2(\Gamma)+2 w_2(\Gamma_g)\right)Q_{\Gamma_g}~,
}
where 
\eq{
Q_{\Gamma_g}= \left(2\sum_{I=1}^{ n_c-1} H_{2I-1}-H_{2n_c+1}+H_{2n_c}\right)~.
}
Since $\frac{1}{4}Q_{\Gamma_g}$ generates $Z(Spin(2n_c))$, we know that every fermion coming from the vector representation has charge $\pm 2$ under $Q_{\Gamma_g}$. This means that in the presence of the minimal monopole of $U(1)_{Q_{\Gamma_g}}$ the fermion zero-modes transform as $\mathbf{2}\otimes \mathbf{2n_f}^{\oplus 2n_c+1}$ 
under $Spin(3)\times Sp(n_f)$. This implies that the monopole Hilbert space decomposes as 
\eq{
\CH_{mono}=\left(\CH_{fund}\right)^{\otimes 2n_c+1}~.
}
Since there are an odd number of factors and we are looking for $\IZ_2$-valued fractionalization, we see that $\CH_{mono}$ has the same fractionalization as $\CH_{fund}$, which is the Dirac spinor representation of $Spin(4n_f)$, 
$\CH_{fund}=\mathbf{S_0}\oplus \mathbf{S_1}$,  where $S_{0,1}$ has an even (odd) number of fermion mode operators respectively (here $S_0$ is the bare monopole and $S_1$ is the dyon as in \cite{Brennan:2022tyl}). 
 When $n_f$ is odd (even) the fermionic (bosonic) components of $\CH_{fund}$ transform as a real (pseudo-real) representation of $Sp(n_f)$ which implies that there is (possible) charge fractionalization. 

Now we can use the fact that the fermion zero-modes transform as spin-$\half$ under $Spin(3)$. This means that if the Fock-vacuum of $\CH_{fund}$ has charge $q$ under $T_3\subset Spin(3)$, then the maximum state has charge $2n_f+q$. If we then tune the UV $\theta$-angle so that the IR theory is $\CT$-symmetric, then we see that 
\eq{
-q=2n_f+q\quad\Rightarrow \quad q=-n_f~, 
}
which implies that $S_0$ is a fermion (boson) if $n_f$ is odd (even) which implies that 
\eq{
\frac{B_2^{(m)}}{2\pi}=\frac{n_f}{2}w_2(TM)\, Q_{\Gamma_g}^\vee~. 
}
This is consistent with the observation \cite{Brennan:2022tyl} that the monopole Hilbert space decomposes as 
\eq{
\CH_{mono}=\CH_{\rm mono}\oplus \CH_{\rm dyon}~,
}
since the fractionalization class for $\CH_{\rm dyon}$ is related to $B_2^{(m)}$ by two times $B_2^{(e)}$.  The anomaly is thus
\eq{
\CA\subset&\frac{2\pi i n_f}{8}\int w_2(TM)\wedge d(2B_2+w_2(TM))=\frac{2\pi i n_f}{8}\int w_2(TM)\wedge (4 dB_2)\\
&=2\pi i\, n_f\int B_2\cup w_3(TM)=0~{\rm mod}_\IZ~,
}
where here we used the 2-group identity $dw_2(TM)=2 dB_2$ and the identity in Footnote \ref{WuRelation}. The upshot is thus that the anomaly vanishes, $\kappa _{\Gamma _g, T}=0$, for this case.

\subsection{Anomaly for $Spin(4n_c+2)$ Adjoint QCD}

For $Spin(4n_c+2)$ adjoint QCD with $2n_f$ fermions, the fields couple to the total bundle  
\eq{
G_{\rm total}=\frac{Spin(4n_c+2)}{\IZ_4}\times \frac{SU(2n_f)\times Spin(4)}{\IZ_2}~.
}  
Using the fractionalization method, we consider coupling to an adjoint fermion with Yukawa couplings (breaks $SU(2n_f)\to Sp(n)_f)$ and Higgsing $Spin(4n_c+2)\to U(1)^{2n_c+1}$.  The $\IZ_4^{(1)}$ UV center symmetry is matched to a $\IZ_4^{(1)}\subset (U(1)^{(1)})^{2n_c+1}$ of the enhanced, IR 1-form electric symmetries. The IR and UV background gauge fields are thus matched as 
\eq{
\frac{B_2^{(e)}}{2\pi}=\frac{1}{4}B_2^{UV}Q_{\Gamma_g}^\vee~, 
}
where $\frac{1}{4}Q_{\Gamma_g}$ generates the center of $Spin(4n_c+2)$. 
The fermion zero-modes of a minimal $U(1)_{\Gamma_g}$ monopole are of the form 
\eq{
\mathbf{4}\otimes \mathbf{2n_f}^{\oplus 2n_c+1}~.
}
 Again, the monopole Hilbert space decomposes into $N$ tensor factors 
\eq{
\CH_{mono}=\left(\CH_{fund}\right)^{\otimes 2n_c+1}~,
}
where $\CH_{fund}$ is the Dirac spin representation of $Spin(8n_f)$, $\CH_{fund}=\mathbf{S_b}\oplus \mathbf{S_f}$, 
($S_{b,f}$ is the bosonic, fermionic component). 
 For all $n_f$ that the fermion spin representations are real so that the fermionic (bosonic) components of $\CH_{fund}$ transform as a pseudo-real (real) representation of $Sp(n_f)$ and hence there is no fractionalization. We therefore see that adjoint QCD does not carry the center symmetry anomaly for $N_c=4n_c+2$.

This can be checked explicitly in $Spin(6)\cong SU(4)$ gauge theory. The fields couple to 
\eq{
G_{\rm total}=\frac{SU(4)}{\IZ_4}\times \frac{SU(2)_f\times Spin(4)}{\IZ_2}~,
}
bundles, where $\IZ_4\subset U(1)_{\Gamma_g}$ is generated by 
\eq{
\frac{1}{4}Q_{\Gamma_g}=\half H_2+\frac{1}{4}(H_1-H_3)~. 
}
Since this theory has  $\IZ_4^{(1)}$ 1-form global symmetry, we can identify the $U(1)_{\Gamma_g}$ Wilson line as the fundamental Wilson line of the UV theory and hence 
\eq{
\frac{B_{2}^{(e)}}{2\pi}=\frac{1}{4} B_2^{UV}Q_{\Gamma_g}~. 
}
We now compute the monopole fractionalization. %Under $U(1)^3_{\rm gauge}$, an adjoint fermion decomposes as 
%\eq{
%\mathbf{15}\longmapsto &\pm (1,0,1)\oplus\pm (1,1,-1)\oplus \pm (1,-1,-1)\oplus (0,-1,2)\oplus \pm (2,-1,0)\oplus \\
%&\oplus\pm(-1,2,-1)\oplus (0)^{\oplus 3}~. 
%}
Upon decomposing the adjoint fermions under the  $U(1)^3_{\rm gauge}$ Cartan, the charge coupling has $Q_{\Gamma_g}$  $\mathbf{15}\longmapsto  \pm 4^{\oplus 3}\oplus 0^{\oplus 9}$.  
So each adjoint fermion provides $12$ real fermion zero-modes. As a representation of $Spin(3)_{\rm rot}\times SU(2)_{\rm global}$ these zero-modes transform as 
$(\mathbf{4},\mathbf{2})^{\oplus 3}$.  Upon quantization of these zero modes, the monopole Hilbert space decomposes under $SU(2)_{\rm global}\times Spin(3)_{\rm rot}$ as:
\eq{
\CH_{mono}=\Big((\mathbf{4},\mathbf{2})\oplus (\mathbf{1},\mathbf{3})\oplus (\mathbf{5},\mathbf{1})
\Big)^{\otimes 3}~.
}
So the monopole Hilbert space is a faithful representation of $\frac{SU(2)_{\rm global}\times Spin(3)_{rot}}{\IZ_2}$ and therefore $\frac{B_{2}^{(m)}}{2\pi}=0$ and there is no anomaly.

\bibliographystyle{utphys}
\bibliography{1FormAnomaliesBib}

\end{document}